\documentclass[12pt]{article}
\usepackage{amsfonts}
\usepackage{graphicx}
\usepackage{epstopdf}
\usepackage{algorithmic}
\usepackage{amssymb}
\usepackage{kbordermatrix}
\usepackage{amsmath}
\usepackage{natbib, booktabs, blkarray}
\usepackage{algorithm}
\usepackage{subfigure}
\usepackage{soul}
\usepackage{comment}
\usepackage{lineno}

\newtheorem{corollary}{Corollary}
\newtheorem{example}{Example}

\newtheorem{remark}{Remark}
\newtheorem{theorem}{Theorem}

\newcommand{\bx}{\textbf{x}}

\newcommand{\bh}{\textbf{h}}
\newcommand{\bo}{\textbf{o}}

\newcommand{\by}{\textbf{y}}

\newcommand{\bw}{\textbf{w}}

\newcommand{\balpha}{\mbox{\boldmath${\alpha}$} }

\newcommand{\bgamma}{\mbox{\boldmath${\gamma}$} }

\newcommand{\bdelta}{\mbox{\boldmath${\delta}$} }

\newcommand{\bm}[1]{\mbox{\boldmath$ #1 $\unboldmath}}

\def\TT{{\mbox{\tiny T}}}

\usepackage{xcolor}

\newcommand{\bt}[1]{{#1}}

\newcommand{\rr}[1]{{#1}}
\newcommand{\rb}[1]{{#1}}

\newcommand{\rf}[1]{{#1}}

\begin{document}
\begin{center}
{\bf\Large Modeling and Active Learning for Experiments with Quantitative-Sequence Factors}
\vskip 25pt
{Qian Xiao$^{a}$, Yaping Wang$^{b}$, Abhyuday Mandal$^{a}$ and Xinwei Deng$^{c}$\footnote{Address for correspondence: Xinwei Deng, Professor, Department of Statistics, Virginia Tech, Blacksburg, VA, 24061 (xdeng@vt.edu).}}

$^{a}$Department of Statistics, University of Georgia, Athens, GA

$^{b}$School of Statistics, East China Normal University, Shanghai, China

$^{c}$Department of Statistics, Virginia Tech, Blacksburg, VA
\end{center}

\begin{quote}{\bf Abstract:}
A new type of experiment that aims to determine the optimal quantities of a sequence of factors is eliciting considerable attention in medical science, bioengineering, and many other disciplines. Such studies require the simultaneous optimization of both quantities and the sequence orders of several components which are called quantitative-sequence (QS) factors. Given the large and semi-discrete solution spaces in such experiments, efficiently identifying optimal or near-optimal solutions by using a small number of experimental trials is a nontrivial task. To address this challenge, we propose a novel active learning approach, called QS-learning, to enable effective modeling and efficient optimization for experiments with QS factors. QS-learning consists of three parts: a novel mapping-based additive Gaussian process (MaGP) model, an efficient global optimization scheme (QS-EGO),
and a new class of optimal designs (QS-design).
The theoretical properties of the proposed method are investigated, and optimization techniques using analytical gradients are developed. The performance of the proposed method is demonstrated via a real drug experiment on lymphoma treatment and several simulation studies.
\end{quote}

\begin{quote}
\end{quote}

\section{Introduction}
 \label{intro}
 
In modern scientific areas, nontraditional experiments that consider the quantities and sequences for arranging components, called quantitative sequence (QS) factors, are being conducted. For example, both the doses and order-of-addition for multiple drug components as a combination showed significant impacts on the efficacy of cancer treatments \citep{ding2015optimized, wang2019}.
In nanocellulose (NC) gel production, a pre-treatment process involved swelling agents, different acids and enzymes to release hemicellulose. The sequence in which the pretreatment components were added, along with their quantities, was to be optimized for the NC size \citep{bh2015}. In the bio-plastics industry, the order in which the compatibilizer and scavenger were mixed with resin, along with their quantities, can cause a significant difference between catfish algae plastic and Solix microalgae plastic. Such QS factors are also used in physical or simulation experiments (a.k.a. computer experiments) in biochemistry \citep{shinohara1998stimulation}, food science \citep{jourdain2009mixed} and management science \citep{panwalkar1973sequencing}.

To illustrate the characteristics of experiments with QS factors, Table~\ref{drugsample} presents three runs from an in vitro drug combination experiment \citep{wang2019}. Three antitumor drugs (A, B, and C) were added every 6 hours in a sequence at different doses. The percentage of tumor inhibition was measured as the response 6 hours after administering the last drug. As indicated in Table~\ref{drugsample}, different drug doses (comparing Runs 1 and 2) and sequences of adding drugs (comparing Runs 1 and 3) lead to varying responses. Thus, to identify the best drug combination, the doses and sequence for administering drugs should be optimized simultaneously. Notably, this experiment is different from crossover trials \citep{jones2014design}. A crossover trial measures all responses after the addition of every drug, and each drug exerts a fixed effect that may be carried over to the next period but does not depend on its order-of-addition. By contrast, only the end point efficacy after adding all the drugs will be measured as the response in a QS experiment, and drug effects are assumed to be dependent on the order-of-addition.

\begin{table}[ht]
\centering
\caption{Illustration of drug data involving QS factors.}
\begin{tabular}{cccccccc}
\toprule
\multicolumn{1}{c}{Run} & \multicolumn{2}{c}{Drug $A$} & \multicolumn{2}{c}{Drug $B$} & \multicolumn{2}{c}{Drug $C$} & \multicolumn{1}{c}{Response} \\
\hline
& dosage & order & dosage & order & dosage & order & \\
\hline
 1 & 3.75 $\mu$M & 1 & 95 nM & 2 & 0.16 $\mu$M & 3  & 39.91\\
 2 & 2.80 $\mu$M & 1 & 70 nM & 2 & 0.16 $\mu$M & 3  & 30.00\\
 3 & 3.75 $\mu$M & 3 & 95 nM & 1 & 0.16 $\mu$M & 2  & 34.68\\
\bottomrule
\end{tabular}%
\label{drugsample}%
\end{table}%

For experiments with QS factors, one of the key tasks is finding the optimal settings of quantities and sequences for arranging components to optimize experimental outcomes. In the current literature, researchers frequently enumerate all possible sequences and apply factorial designs to determine the quantities for each sequence \citep{wang2019}. However, when the number of components is moderate or large, such a strategy may require a prohibitively large number of runs. It may also miss the optimal setting unless a wide range of levels is adopted. To the best of our knowledge, very few studies have been conducted on how to optimize the settings of QS factors via efficient modeling and experimental design. This problem is new and challenging, because QS factors are neither purely continuous nor categorical. To fill this gap, we propose a novel active learning approach, called QS-learning, which can identify good solutions by using only a few sequential experimental trials.

Active learning has attracted considerable attention in recent years \citep{cohn1996active, settles.tr09, deng2009active}.
It sequentially queries the next data point on the basis of what it has learned from the current ones. Different methods for formulating query strategies have been proposed in the literature, including uncertainty sampling \citep{lewis1994sequential}, query-by-committee \citep{seung1992query}, expected model change \citep{settles2007multiple} and variance reductions \citep{cohn1996active}. Refer to \cite{settles.tr09} for a survey.
From the experimentation perspective, active learning overlaps with optimal design \citep{burnaev2015adaptive} and  Bayesian optimization \citep{frazier2018tutorial}. It allocates runs in an adaptive manner, efficiently improving the decision for designing the next experimental trial as more information is acquired over time. Active learning frequently outperforms one-shot experimental designs for optimization when the solution space is large and complex \citep{kapoor2007active, burnaev2015adaptive, frazier2018tutorial}. It involves three major parts: (1) a method for statistical modeling and inference, (2) optimization of some acquisition functions for choosing the next design point, and (3) an initial design for exploring input space. Here, the acquisition function is typically a function that measures the ``utility" of the run that will be evaluated next. It often considers ``exploration/exploitation" trade-off, such that balance is achieved between focusing on alternatives that appear to be good and experimenting with little known alternatives. Common choices of acquisition functions include expected improvement \citep{jones1998efficient}, knowledge gradient \citep{frazier2009knowledge}, and entropy search \citep{hennig2012entropy}.

\begin{figure}[ht]
\centering
\includegraphics[width=0.7\linewidth]{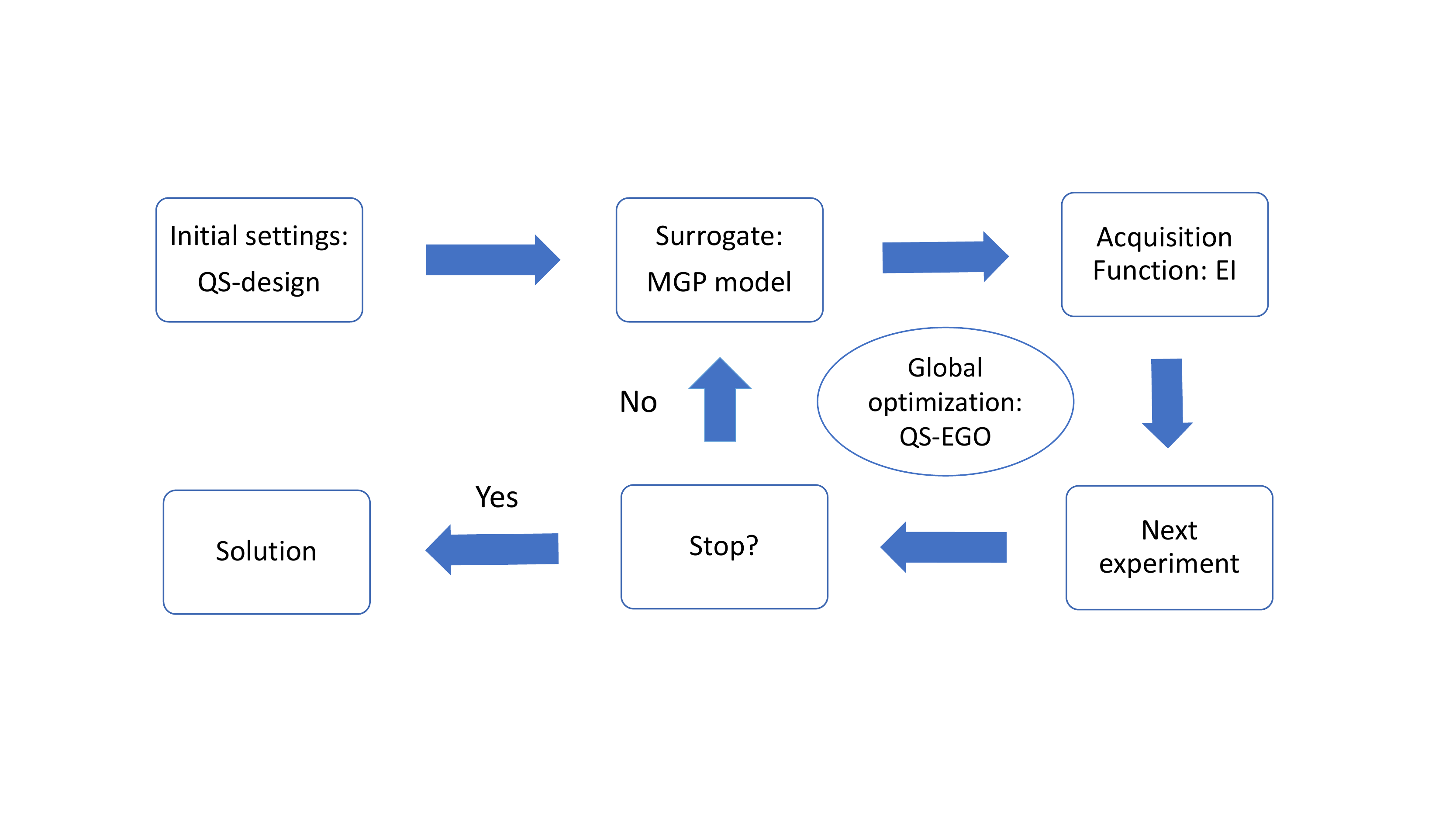}
\vspace{-.3in}
\caption{A flow chart of QS-learning for optimization.}
\label{fig:act}
\end{figure}

In this work, we propose an active learning approach (QS-learning) for experiments with QS factors. It includes a novel mapping-based additive Gaussian process (MaGP) model for prediction and uncertainty quantification, a sequential scheme that uses efficient global optimization algorithms (QS-EGO), and a new class of optimal experimental designs (QS-design) for collecting initial data points. A flowchart of the proposed QS-learning  method is presented in Figure~\ref{fig:act}. The proposed method targets experiments with budget constraints in which practitioners prefer fewer runs. For cases with large data, we develop a variant of QS-learning for computational efficiency.

The key contributions of this work are summarized as follows. First, our proposed MaGP model enables the use of the Gaussian process (GP) in analyzing quantitative and sequence factors, providing desirable predictions and uncertainty quantification. Notably, the classic GP method has been widely used for modeling data with only quantitative factors \citep{Rasmussen2006, kleijnen2009}. Some recent developments have enabled it for both quantitative and qualitative factors \citep{qian2008, deng2017additive, zhang2018latent, xiao2021ezgp}. However, this method cannot be easily adapted for modeling data with QS factors due to the semi-discrete nature of sequence input. Second, we develop a new algorithm (QS-EGO) for efficiently optimizing the expected improvement (EI) acquisition function \citep{jones1998efficient}, which is nontrivial for QS factors. To address a complex solution space with both continuous and semi-discrete characteristics, the proposed QS-EGO combines a genetic algorithm and a new space-filling threshold-accepting (SFTA) algorithm. We derive analytical gradients for model estimation and acquisition function optimization to facilitate computation. Third, we develop a new class of optimal designs (QS-designs) for collecting initial data in the proposed active learning. QS-designs can reduce the number of required sequential runs while simultaneously improving performance. New design criteria are established to search for QS-designs with flexible sizes. We also develop an algebraic construction for QS-designs with certain sizes and prove their desirable properties. In the current experimental design literature, researchers have focused on either quantitative \citep{wu2011experiments, joseph2016space} or sequence \citep{robert2018, voelkel2019design, lin2019order, yang2018} factors, while the proposed QS-designs consider both factors simultaneously.

The remainder of this article is organized as follows. In Section~\ref{lr}, we review several related methods in the literature. In Section~\ref{secuk}, we describe the formulation and estimation of the proposed MaGP model in detail. In In Section~\ref{qslearn}, we discuss the proposed sequential scheme along with QS-EGO. In Section~\ref{sec:design}, we illustrate the construction of a new class of optimal designs (i.e., QS-designs). A case study of a drug combination experiment on lymphoma is reported in Section~\ref{realdat}, and a simulation study on the traveling salesman problem (TSP) is presented in Section~\ref{tsp}. We conclude this work with discussions in in Section~\ref{dis}. All proofs, technical details, convergence results, and additional numerical studies are included in the Supplementary Materials.

\section{Brief Literature Review}
\label{lr}
QS factors are commonly observed in drug combination studies \citep{wang2019}. However, conventional methods often consider only the effects of drug doses (quantitative input), e.g., the Hill model \citep{chou2006theoretical}, polynomial model \citep{jaynes2013application}, Hill-based model \citep{ning2014}, and Kriging or GP model \citep{xiao2019}. Some recent studies have shown that if several drugs with fixed doses are added in desirable sequences, then such drug combinations will have enhanced efficacy \citep{ding2015optimized}. To model drug sequences with fixed doses, two types of linear models are proposed: the pairwise ordering (PWO) model \citep{van1995design, voelkel2019design, robert2018} and component-position (CP) model \citep{yang2018}. We first review the two models and then generalize them for QS factors.

Let us consider a drug combination experiment with $n$ runs and $k$ drugs. For its $i^{th}$ run, let $\bx_i = (x_{i,1}, \ldots, x_{i,k})^\TT$ be a vector containing the doses of $k$ drugs and \bt{$\balpha_i = (\alpha_{i,1}, \ldots, \alpha_{i,k})^\TT$ be a vector containing the sequence of applying them. For example, if we add Drug $B$ first, then $C$ and finally $A$ ($k=3$), then the sequence of adding drugs ($B,C,A$) is represented by the vector $\balpha_i = (2,3,1)^\TT$.}
\bt{In the PWO model, the features of $\balpha_i$ are represented by the precedence patterns between all $k \choose 2$ pairs of drugs. Explicitly, let $S$ be the set of all pairs $(p, q)$ for $1 \leq p < q \leq k$ and define the PWO indicator between $p$ and $q$ for any $(p,q) \in S$ as
$$
z_{p,q}(\balpha_i)=
\left\{
\begin{array}{clc}
				1  & \text{if $p$ precedes $q$ in $\balpha_i$}, \\
				-1 & \text{if $q$ precedes $p$ in $\balpha_i$}.
\end{array}
\right.
$$
As an illustration, for $\balpha_i = (2,3,1)^\TT$, we have $S = \{(1,2), (1,3), (2,3) \}$, $z_{1,2}(\balpha_i) = -1$, $z_{1,3}(\balpha_i) = -1$, and $z_{2,3}(\balpha_i) = 1$. Based on PWO indicators, the PWO model \citep{van1995design, voelkel2019design} is defined as}
\begin{equation}
\label{opwo}
f(\bx_i^\TT, \balpha_i^\TT) = \beta_0 + \sum_{(p,q) \in S}z_{p,q}(\balpha_i)\beta_{p,q} + \epsilon,
\end{equation}
where the residual $\epsilon$ follows the standard assumptions for linear models and parameters $\beta$ can be estimated via the least squares method. To further capture the two-factor interactions between PWO indicators, \cite{robert2018} proposed the triplet PWO model.
In this work, we consider the PWO model, which often suffices in practice. The triplet PWO model includes as many as $1+k+k(k-1)/2+k(k-1)(k-2)/3$ parameters, which often exceed the total number of runs in sequential experiment considered in this work.

Another class of linear models is CP model \citep{yang2018}, which is defined as
\begin{equation}
\label{pem}
f(\bx_i^\TT, \balpha_i^\TT) = \beta_0 + \sum_{j=1}^{k-1}\sum_{c=1}^{k-1}x_{i,c}^{(j)}\beta_{j,c} + \epsilon_i,	
\end{equation}
where $x_{i,c}^{(j)}$ equals 1 if $\alpha_{i,j} = c$ and 0 otherwise. That is, $x_{i,c}^{(j)}$ is an indicator of whether Drug $c$ is used in the $i$th run at the $j$th position. Simply put, CP is a multivariate linear regression model treating each position as a factor with $k$ levels.

Both the PWO and CP models in the current literature work only for sequence factors. In order to establish some benchmark models, we generalize the PWO and CP models via adding covariates for quantitative factors (e.g., doses), such that they can work for QS factors. Both the generalized PWO and CP models can be represented as
\begin{equation}
g(\bx_i^\TT, \balpha_i^\TT)  = \sum_{j=1}^{k}\beta^{'}_{j}x_{i,j} + f(\bx_i^\TT, \balpha_i^\TT) ,
\end{equation}
where $\beta_j^{'}$ denotes the coefficients for quantitative factors, and $f(\bx_i^\TT, \balpha_i^\TT)$ can be either the PWO model in \eqref{opwo} or the CP model in \eqref{pem}.

Linear models may work well under one-shot experimental designs for prediction purposes, but they are less popular in active learning for optimization.
Compared with GP models, linear models often perform worse in uncertainty quantification  \citep{smith2013uncertainty, burnaev2015adaptive}.
The GP model, where responses are represented by random variables whose probability distributions characterize the beliefs of experimenters about the unknown values, provides a good probabilistic framework for active learning \citep{Rasmussen2006, kapoor2007active, frazier2018tutorial}. It enables the predictive distribution \rf{of} the outcome of the next experiment and the selection of the best one by maximizing an acquisition function.

\section{Mapping based Additive GP Model}
\label{secuk}

\subsection{Model Formulation}
Let us consider a QS experiment with $n$ runs and $k$ components (i.e., $c_1, c_2, \ldots, c_k$), where the $i^{th}$ input is denoted as $\bw_i = (\bx_i^\TT, \bo_i^\TT)^\TT$ and the corresponding output is denoted as $y_{i}$.
Here, $\bx_i = (x_{i,1}, \ldots, x_{i,k})^\TT$ is a vector of quantitative values for $k$ components, and $\bo_i = (o_{i,1}, \ldots, o_{i,k})^\TT$ is a vector that contains the orders of components in the arrangement sequence.
Notably, $x_{i,h}$ and $o_{i,h}$ ($h=1,\ldots,k$ and $i=1,\ldots,n$) are the quantitative and sequence parts of the $h^{th}$ component $c_h$, respectively.
Without loss of generality, we assume that $\bo$ is a permutation of the integers 1 to $k$. As an illustration, the third run in Table~\ref{drugsample}, $\bw_3^\TT = (\bx_3^\TT, \bo_3^\TT)$, has $\bx_3 = (3.75, 95, 0.16)^\TT$ and $\bo_3 = (3,1,2)^\TT$. The vector $\bo_3$ represents that Drug A is added in the third place, B is added in the first place, and C is added in the second place, i.e., $B \rightarrow C \rightarrow A$. Notably, the $\bo$ defined here contains the index orders of the  corresponding elements in the vector $\balpha$ defined in Section \ref{lr}, and they have the same practical meaning.

The order sequence is semi-discrete in nature; hence, the relationship between output and QS input can be complicated. To model such data, we consider the adoption of the GP model because of its flexibility and promising prediction and uncertainty quantification.
For an experiment with $n$ runs and $k$ components, we model the output at $\bw = (\bx^\TT, \bo^\TT)^\TT$ as
\begin{equation}\label{eq:additiveGP}
Y(\bw) = \mu + \sum_{h=1}^{k}G_{h}(\bw) + \epsilon,
\end{equation}
where $G_{1}, \ldots, G_{k}$ are independent zero-mean GPs with stationary covariance functions, and $\epsilon \sim N(0, \tau^2)$ is a random error. The GP component $G_h$ ($h = 1, \ldots, k$) corresponds to the effect of the $h^{th}$ component $c_h$ on the output.
For physical experiments, we assume homogeneous error variances $\tau^2 >0$ which may come from measurement errors or some environmental factors.
For computer experiments, we take $\tau^2 = 0$, because computer codes  provide deterministic output \citep{fang2005}.

In GP models, distances between pairs of input are used to measure their similarities when formulating covariance functions.
For the $h^{th}$ ($h = 1, \ldots, k$) component, its sequence input $o_{i,h}$ (for any $i = 1, \ldots, n$) is ordinal.
Thus, a distance measure should be specified for sequence input to form the covariance function in the GP component $G_h$.
To address this challenge, we consider mapping the order $o_{i,h}$ ($o_{i,h} \in \{1, \ldots, k \}$) to a $t$-dimensional  latent vector $(\tilde{o}_{i,h}^{(1)}, \ldots, \tilde{o}_{i,h}^{(t)}$).
Given that the sequence input $\bo$ is an assignment of $k$ components to $k$ ``fixed" order positions, we should use the same mapping for all components $c_1, \ldots, c_k$ (corresponding to GP components $G_1, \ldots, G_k$, respectively) to quantify the effects of fixed order positions via latent variables.
In particular, the $t$-dimensional mapping ($t=1,\ldots, k-1$) for the order of any component is defined as
\begin{equation}
\label{eq:mapping-rule}
\begin{blockarray}{c}
\hbox{$c_1, \ldots, c_k$}   \\
\begin{block}{[c]}
  1  \\
  2  \\
  \vdots  \\
  k \\
\end{block}
\end{blockarray}
\rightarrow
\kbordermatrix{
    & \tilde{o}^{(1)} &  \tilde{o}^{(2)} & \dots & \tilde{o}^{(t)} \\
    &\delta^{(1)}_{1} & \delta^{(2)}_{1} & \dots & \delta^{(t)}_{1}\\
    &\delta^{(1)}_{2} & \delta^{(2)}_{2} & \dots & \delta^{(t)}_{2} \\
    &\vdots & \vdots & \vdots & \vdots \\
    &\delta^{(1)}_{k} & \delta^{(2)}_{k} & \dots & \delta^{(t)}_{k}
}_{k \times t},
\end{equation}
where $\delta^{(j)}_{l} = 0$ for all $j \ge l$ to avoid over-parametrization. \rr{The interactions among different levels (i.e., orders) can be reflected by the mapping parameters in \eqref{eq:mapping-rule}, which are estimated from the data.} As all components use the same mapping, the total number of mapping parameters is $t(t+1)/2+(k-t-1)t$. Specifically, when $t = k-1$, we call it full mapping with a total of $k(k-1)/2$ mapping parameters. When $t=2$, we call it 2d-mapping, which has $(2k-3)$ mapping parameters.

\begin{example}
For illustration, consider a QS experiment to find the optimal sequence and quantity to add for $k = 4$ operations in a single production line with four fixed locations to be assigned with different operations. We use the same mapping for all four operations ($c_1, c_2, c_3, c_4$), which quantifies the effects due to locations (i.e., position orders) to be assigned with operations:
\[
\begin{blockarray}{c}
\hbox{full mapping}   \\
\hbox{$c_1$, $c_2$, $c_3$, $c_4$}   \\
\begin{block}{[c]}
 \text{order } 1  \\
  \text{order } 2  \\
  \text{order } 3  \\
  \text{order } 4  \\
\end{block}
\end{blockarray} \ \
\rightarrow
\begin{blockarray}{c}
\hbox{  }   \\
\kbordermatrix{
    & \tilde{o}^{(1)} &  \tilde{o}^{(2)} & \tilde{o}^{(3)} \\
    & 0 & 0 & 0\\
    & \delta^{(1)}_{2} & 0 & 0 \\
    & \delta^{(1)}_{3} & \delta^{(2)}_{3} & 0 \\
    & \delta^{(1)}_{4} & \delta^{(2)}_{4} & \delta^{(3)}_{4} \\
}
\end{blockarray}, \ \  \ \
\begin{blockarray}{c}
\hbox{2d-mapping}   \\
\hbox{$c_1$, $c_2$, $c_3$, $c_4$}   \\
\begin{block}{[c]}
  \text{order } 1  \\
  \text{order } 2  \\
  \text{order } 3  \\
  \text{order } 4  \\
\end{block}
\end{blockarray} \ \
\rightarrow
\begin{blockarray}{c}
\hbox{  }   \\
\kbordermatrix{
    & \tilde{o}^{(1)} &  \tilde{o}^{(2)} \\
    & 0 & 0 \\
    & \delta^{(1)}_{2} & 0 \\
    & \delta^{(1)}_{3} & \delta^{(2)}_{3} \\
    & \delta^{(1)}_{4} & \delta^{(2)}_{4} \\
}
\end{blockarray},
\]
where $\delta^{(j)}_{l} (j < l)$ denotes parameters to be estimated via maximum likelihood estimation (MLE).
\end{example}

The prespecified tuning parameter $t$ ($t \in \{ 1,\ldots, k-1 \}$) controls the flexibility of defining similarities between pairs of order positions.
Under full mapping ($t = k-1$), all pairwise distances between order positions can be independently determined.
Then, all possible patterns in defining similarities between sequence input can be captured. By contrast, under 1d-mapping ($t=1$), the mapping  in \eqref{eq:mapping-rule} is simplified as $\text{order 1} \rightarrow 0$, $\text{order 2} \rightarrow \delta_1, \ldots, \text{order } k \rightarrow \delta_{k-1}$, or equivalently $\text{order 1} \rightarrow 0$, $\text{order 2} \rightarrow \delta_1^{'}$, $\text{order 3} \rightarrow \delta_1^{'} + \delta_2^{'}, \ldots, \text{order } k \rightarrow \sum_{i=1}^{k-1}\delta_{i}^{'}$. Evidently, only the distances between adjacent order positions are independently determined here.
For example, when $t=1$, the distance between orders 1 and 3 (determined via $\delta_1^{'} + \delta_2^{'}$) is dependent on the distance between orders 1 and 2 (determined via $\delta_1^{'}$) and the distance between orders 2 and 3 (determined via $\delta_2^{'}$).
Such restrictive mapping works for cases where only adjacent orders interact with one another. In this work, we \rf{consider} $t \ge 2$ to allow a more general pattern of interactions.

An appropriate choice of $t$ provides a trade-off between model flexibility and computational cost. When many components are involved (i.e., large k), low-dimensional mapping (e.g., 2d-mapping) is often a good choice. It will considerably reduce the number of parameters in the MaGP model, facilitating model estimation in practice. In 2d-mapping, any pairwise distance between order positions can be partially (not fully) determined by other pairwise distances. Thus, this type of mapping can provide certain flexibility to capturing possible patterns for defining similarities between sequence input.

In the $i^{th}$ run $\bw_i = (\bx_i^\TT, \bo_i^\TT)^\TT$, the elements that correspond to the $h^{th}$ component $c_{h}$ are $(x_{i,h},o_{i,h})$, where $i=1,\ldots,n$ and $h=1, \ldots, k$. From the mapping in \eqref{eq:mapping-rule}, we define the distance between the $i^{th}$ and $j^{th}$ runs that correspond to the $h^{th}$ component $c_{h}$ under the $L_2$ norm as 
\begin{equation}
\label{eq:dist}
d_{i,j}^{(h)} =\vert \vert (x_{i,h},o_{i,h}) - (x_{j,h},o_{j,h})  \vert \vert = \sqrt{\theta_h (x_{i,h} - x_{j,h})^2 + \sum_{l=1}^{t} (\tilde{o}_{i,h}^{(l)} - \tilde{o}_{j,h}^{(l)})^2},
\end{equation}
where  $\theta_h$ is the correlation parameter that scales the quantitative input of $c_h$. Here, the $t$-dimensional  latent vectors $(\tilde{o}_{i,h}^{(1)}, \ldots, \tilde{o}_{i,h}^{(t)}$) and $(\tilde{o}_{j,h}^{(1)}, \ldots, \tilde{o}_{j,h}^{(t)}$) correspond to the orders $o_{i,h}$ in $\bw_i$ and $o_{j,h}$ in $\bw_j$, respectively, and their values are determined by the mapping parameters in \eqref{eq:mapping-rule} (denoted as $\bdelta$).
Notably, there is no need to include any correlation parameters to scale latent vectors, because the mapping parameters ($\bdelta$) are estimated from the data.

Subsequently, we can describe the proposed covariance function for the $h^{th}$ GP component $G_{h}$ ($h=1,\ldots,k$) in \eqref{eq:additiveGP} as
\begin{align}\label{eq: cor-1new}
\phi_h(\bw_i, \bw_j \vert \sigma_h^2, \theta_h, \bdelta) = \phi_h((x_{i,h},o_{i,h}), (x_{j,h},o_{j,h}) \vert \sigma_h^2, \theta_h, \bdelta) =
\sigma_h^2 K(d_{i,j}^{(h)}),
\end{align}
where $\sigma_h^2$ is the variance parameter corresponding to the $h^{th}$ component $c_h$, and $K(\cdot)$ is any valid kernel function. Popular kernels include the Matern class with a smoothness parameter $\nu \in (0, \infty)$,
\begin{equation}
\label{eq:mat}
K(d_{i,j}^{(h)}) = \frac{2^{1-\nu}}{\Gamma(\nu)}(\sqrt{2} \nu d_{i,j}^{(h)})^{\nu}k_{\nu}(\sqrt{2} \nu d_{i,j}^{(h)}),
\end{equation}
where $k_{\nu}$ is a modified Bessel function of the second kind. Specifically, we focus on the case of $\nu \rightarrow \infty$, i.e. the Gaussian kernel, in this work:
\begin{equation}
\label{eq:gaus}
K(d_{i,j}^{(h)}) = \hbox{exp} (-(d_{i,j}^{(h)})^2).
\end{equation}
In \eqref{eq:gaus}, we remove a constant multiplier of $1/2$ in the exponent for re-parameterization.

By \eqref{eq:additiveGP}, \eqref{eq:mapping-rule}, \eqref{eq:dist}, \eqref{eq: cor-1new}, and \eqref{eq:gaus},
for any two input $\bw_i$ and $\bw_j$, the covariance function for the MaGP model in  \eqref{eq:additiveGP} can be specified by:
\begin{eqnarray}\label{eq:corr-fun}
\phi(\bw_i, \bw_j) & = & \hbox{Cov} ( Y(\bw_i), Y(\bw_j) ) = \sum_{h =1}^{k} \phi_h(\bw_{i},\bw_{j}| \sigma_h^2, \theta^{(h)}, \bdelta) + \tau^2 \textbf{1}(\bw_i = \bw_j) \nonumber \\
& = & \sum_{h=1}^{k}  \sigma_h^2 \hbox{exp} \big \{- \theta_h (x_{i,h} - x_{j,h})^2 -\sum_{l=1}^{t} (\tilde{o}_{i,h}^{(l)} - \tilde{o}_{j,h}^{(l)})^2 \big\} + \tau^2 \textbf{1}(\bw_i = \bw_j),
\end{eqnarray}
where $\tau^2 \geqslant 0$, and $\textbf{1}(\cdot)$ is an indicator function. 
Here, the variance parameter $\sigma_h^2$ corresponds to the effect of the $h^{th}$ component, and $\tau^2$ is the variance of the error term $\epsilon \sim N(0, \tau^2)$ in \eqref{eq:additiveGP}. This covariance function combines different dimensions via addition, and the quantity part and  sequence part in each dimension via multiplication. It cannot be decomposed into the sum or product of a covariance for purely quantitative factors and a covariance for purely sequence factors. Thus, it is not the separable covariance function \rf{as defined in} \cite{gneiting2002nonseparable}.

\rf{Given the noise variance} $\tau^2$, the MaGP model with the covariance function in \eqref{eq:corr-fun} includes $n_{par} = 1 + 2k + kt - t(t+1)/2$ parameters. Specifically, the full-MaGP ($t=k-1$) and the 2d-MaGP ($t=2$)  include $1 + k(k+3)/2$ and $4k-2$ parameters, respectively.

\begin{theorem} \label{pd}
Given $n$ input $\bw_i = (\bx_i^\TT, \bo_i^\TT)^\TT$ ($i=1,\ldots,n$), the covariance matrix of outputs $\by = (Y(\bw_1), \ldots, Y(\bw_n))^\TT$ induced by the covariance function in \eqref{eq:corr-fun} is positive semi-definite.
\end{theorem}
Theorem \ref{pd} holds for any $\bw_1, \ldots, \bw_n$, including duplicated input, and $\tau^2$ can be 0. For appropriate model inference, $\hbox{Cov}(\by)$ must be positive definite, and the following two corollaries shed some light on this aspect.

\begin{corollary}
\label{pdc}
Given $n$ input and the noise variance $\tau^2 > 0$, the covariance matrix $\hbox{Cov}(\by)$ induced by the covariance function in \eqref{eq:corr-fun} is positive definite.
\end{corollary}

\begin{corollary}
\label{pdc2}
When the noise variance $\tau^2 = 0$, if no two runs have the same quantitative input (i.e., $\bx_i \neq \bx_j$ for $i \neq j$), then the covariance matrix $\hbox{Cov}(\by)$ induced by the covariance function in \eqref{eq:corr-fun} is positive definite.
\end{corollary}

Corollary~\ref{pdc} guarantees the validity of the covariance matrix for modeling physical experiments.
For modeling computer experiments, if Latin hypercube designs \citep{lin2015latin}, orthogonal arrays \citep{hedayat2012orthogonal}, or space-filling designs \citep{wang2018} are used as the quantitative parts of design matrices (where all $\bx_i \neq \bx_j$ for $i \neq j$), then the covariance matrices in the proposed model are positive definite by Corollary~\ref{pdc2}.
In Section~\ref{sec:design}, we propose a new class of optimal designs for QS factors that satisfies the requirements in Corollary~\ref{pdc2} and has more attractive properties. 
Notably, if two runs have the same quantitative part but different sequence parts, then we need to set $\delta_{l}^{(l-1)} \neq 0$ for $l=2, \ldots, k$ in the model estimation to guarantee that the covariance matrix is positive definite.

Notably, the warping technique in the literature \citep{snelson2004warped, xiao2021mapping} can be used for ordinal factors, where an ordinal input $o_{i,h}$ \rf{is mapped to} a quantitative input $f(o_{i,h})$ via a certain transformation function $f_h(\cdot)$.
Evidently, such a technique is a special case of, and thus, more restrictive than the 1d-mapping used in the current work. The warping technique frequently considers the case of independent ordinal factors. However, the sequence factors in QS experiments are not independent, because they are required to form sequence input (i.e. permutations of $1, \ldots, k$). \cite{zhang2018latent} considered a latent approach for mapping qualitative input to some quantitative vectors in GP. Their method differs from the proposed MaGP in at least two aspects. First, they considered a single GP with a multiplicative covariance structure wherein a single variance parameter is adopted. Such a model structure may not distinguish the specific effects of different qualitative factors. By contrast, the MaGP model considers additive GPs, wherein each GP component has a specific variance parameter that measures the effect of each component. Second,  \cite{zhang2018latent} set different mapping matrices for various qualitative factors. While, the MaGP model adopts the same mapping for all components, because the order sequence is an assignment of $k$ components to $k$ ``fixed" order positions.

\subsection{Model Estimation}
\label{sec:model_est}
For parameter estimation, the proposed MaGP model in \eqref{eq:additiveGP} with the covariance function in \eqref{eq:corr-fun} contains parameters $\mu$, $\bm{\sigma^2} = (\sigma_1^2, \ldots, \sigma_k^2)^T$, $\bm{\theta} = (\theta_1, \ldots, \theta_k)^T$,  $\bm{\delta} = (\delta^{(1)}_{2}, \ldots, \delta^{(t)}_{k})^T$, and $\tau^2$.
These parameters can be estimated via the likelihood function.
The covariance matrix is denoted by $ \bm{\Phi} =  \Phi(\bm{\sigma^2}, \bm{\theta}, \bm{\delta}, \tau^2) = (\phi(\bw_i, \bw_j))_{n \times n}$, which follows the covariance function in  \eqref{eq: cor-1new}.
With some simple algebra, the negative log-likelihood function can be expressed as (up to a constant)
\begin{equation}
\label{loglike1}
\hbox{log}\vert \bm{\Phi} \vert + (\by-\mu\textbf{1})^\TT\bm{\Phi}^{-1}(\by - \mu\textbf{1}),
\end{equation}
where the response vector $\by = (Y(\bw_1), \ldots, Y(\bw_n))^\TT$, and $\textbf{1}$ is an $n \times 1$ column vector of all 1s.
For given $\bm{\sigma^2}, \bm{\theta}, \bm{\delta}$, and $\tau^2$, the MLE of $\mu$ can be obtained explicitly as
\begin{equation}
\label{mu}
\widehat{\mu} = (\textbf{1}^\TT\bm{\Phi}^{-1}\textbf{1})^{-1}\textbf{1}^\TT\bm{\Phi}^{-1}\textbf{y}.
\end{equation}
By substituting \eqref{mu} into \eqref{loglike1}, the estimation of $\bm{\sigma^2}, \bm{\theta}, \bm{\delta}$, and $\tau^2$ can be obtained by
\begin{equation}
\label{eq:likihood}
[\bm{\sigma^2}, \bm{\theta}, \bm{\delta}, \tau^2] = \textrm{argmin} \left\{ \log \vert \bm{\Phi} \vert + (\by^\TT \bm{\Phi}^{-1} \by) - (\textbf{1}^\TT\bm{\Phi}^{-1} \textbf{1})^{-1} (\textbf{1}^\TT\bm{\Phi}^{-1} \textbf{y})^2 \right\}.
\end{equation}
This minimization problem can be solved using some standard nonlinear optimization algorithms in Matlab or R. Different algorithms or the same algorithm with different initializations may lead to different parameter estimates \citep{erickson2018comparison}.
In the current work, we adopt the Broyden–Fletcher–Goldfarb–Shanno (BFGS) algorithm with random initialization \citep{liu1989limited}. It is a popular method for estimating GP models. It determines descent direction by preconditioning the gradient with curvature information. 
Numerical gradients can be used, but they are approximate and are expensive to compute. In the current study, we derive analytical gradients to facilitate fast and exact computation. We report all analytical gradients and the implementation of the optimization algorithm in Supplementary Materials Sections S1.2 and S3.1, respectively.

Given all the estimated parameters, the prediction mean and variance of the response at the target input $\bw_*$ are given by
\begin{equation}\label{eq:pred:mean}
\widehat{Y}(\bw_*) = \widehat{\mu} + \bm{\gamma}^\TT \bm{\Phi}^{-1}(\by - \widehat{\mu} \textbf{1}),
\end{equation}
\begin{equation}\label{eq:pred:var}
 s^2(\bw_*) = \phi(\bw_*,\bw_*) - \bm{\gamma}^\TT \bm{\Phi}^{-1}\bm{\gamma} +
\frac{(1 - \textbf{1}^\TT \bm{\Phi}^{-1}\bgamma)^2}{(\textbf{1}^\TT \bm{\Phi}^{-1} \textbf{1})}.
\end{equation}
Here, $\bm{\gamma}$ is the covariance vector $(\phi^{'}(\bw^{*}, \bw_i))_{n \times 1}$, where $\phi^{'}(\bw^{*}, \bw_i) = \sum_{h=1}^{k}  \sigma_h^2 \hbox{exp} \big \{- \theta_h (x_{i,h} - x_{*,h})^2 -\sum_{l=1}^{t} (\tilde{o}_{i,h}^{(l)} - \tilde{o}_{*,h}^{(l)})^2 \big\}$. Notably, the estimators in \eqref{eq:pred:mean} and \eqref{eq:pred:var} are commonly used in the literature \citep{Rasmussen2006, kleijnen2009, gramacy2020surrogates}. For an unbiased small-sample estimator of $s^2(\bw_*)$, refer to  \cite{mehdad2015classic}.

For computer experiments with $\tau^2=0$, when $\bw_*$ is the $i^{th}$ observed input $\bw_i$, $\bm{\gamma}^\TT$ is the $i^{th}$ row in $\bm{\Phi}$, \rf{and} $\bm{\gamma}^\TT\bm{\Phi}^{-1}$ is a row vector with the $i^{th}$ entry being 1 and \rf{the others being 0}. Evidently, $\widehat{Y}(\bw_*) = \widehat{Y}(\bw_i) = y_i$, and thus \eqref{eq:pred:var} yields $s^2(w^*) = 0$. Therefore, the interpolation property holds.
Note that if $\bm{\Phi}$ is ill-conditioned, then a nugget (or noise) effect may be added, and the interpolation property may not hold; refer to \cite{gramacy2012cases} for details.
For physical experiments, the interpolation property does not hold due to the presence of random errors $\epsilon$ in \eqref{eq:additiveGP}.
In practice, when no replicates are included in physical experiments, the homogeneous noise variance $\tau^2$ of random errors $\epsilon$ is often prespecified according to known background information, and different small $\tau^2$ values may not lead to a significant difference in prediction \citep{xiao2019}.  When replicates are included, $\tau^2$ should be estimated via MLE as shown in \eqref{eq:likihood}. For some basic derivations of GP model estimation, refer to \cite{Rasmussen2006} and \cite{roustant2012dicekriging} for a survey.

\section{Active Learning for Experiments with QS Factors}
\label{qslearn}

In this section, we first introduce a general active learning scheme for experiments with QS factors and then discuss its variant for computational scalability. Given the large and semi-discrete input spaces in such experiments, adapting existing methods for optimization is nontrivial. This issue motivates us to develop a tailored new optimization algorithm.

\subsection{EI Optimization with QS Factors}
\label{sec:BO}

In experimentation, active learning has received considerable attention since the expected improvement (EI) framework, which works for quantitative factors, was proposed by \cite{jones1998efficient}. In this section, we adapt the EI acquisition function to work with QS factors and develop an efficient global optimization algorithm (QS-EGO). This new algorithm adopts the proposed MaGP as the probabilistic model for the input–output relationship, under which we derive the analytical gradients for optimizing EI.

Without loss of generality, we focus on finding the optimal solution $\bw$ to minimize the ``black-box" objective function $y(\bw)$ in \rf{either physical or computer experiments}. Notably, any maximization problem can be viewed as a minimization problem to the negative objective function $-y(\bw)$.
Let the improvement function be $I(\bw) = (y_{\min }^{(n)}-y(\bw))_{+}$ for an input $\bw$, where $a_+ = \hbox{max}(a, 0)$ indicates the nonnegative part of $a$, and $y_{\min}^{(n)}$ is the minimum response of the $n$ current  observations. EI is defined as $E [ I(\bw) ] = \int I(\bw) f_n(y|\bw)dy$, where $ f_n(y|\bw)$ is the probability density function of the  predictive distribution given by the MaGP model based on the $n$ current  observations. EI at input $\bw = (\bx^T, \bo^T)^T $ can be expressed in closed form as
\begin{equation}
\label{eq:EImin}
EI = E[ I(\bw) ] = (y_{\min }^{(n)}-\widehat{Y}(\bw)) \Phi \left( \frac{ y_{\min }^{(n)}-\widehat{Y}(\bw) }{s(\bw)} \right) + s(\bw) \varphi \left( \frac{ y_{\min }^{(n)}-\widehat{Y}(\bw) }{s(\bw)} \right),
\end{equation}
where $\Phi(\cdot)$ and $\varphi(\cdot)$ denote the cumulative distribution function and probability density function of the standard normal distribution, respectively. The prediction mean $\widehat{Y}(\bw)$ and its standard error $s(\bw) = \sqrt{s^2(\bw) }$ are provided in \eqref{eq:pred:mean} and \eqref{eq:pred:var}, respectively. EI inherits a trade-off between exploitation and exploration \citep{jones1998efficient}.
The first term in \eqref{eq:EImin} is maximized by the experimental point having the smallest mean value, and thus, it can be interpreted as the exploitation part. Meanwhile, the second term is maximized by the unexplored point having the largest uncertainty, and thus, it can be interpreted as the exploration part.

\rf{As shown in Figure~\ref{fig:act}, the workflow of QS-learning includes the following four steps.}
\begin{enumerate}
	\item Construct an optimal initial design for QS factors with $n_0$ runs $\bw_1,\ldots, \bw_{n_0}$.  Compute (or simulate) their responses as $y_1, \ldots, y_{n_0}$. Then, fit the MaGP model based on these observations. Set $n = n_0$.
	\item Select the next design point $\bw_{n+1}$ that maximizes the EI acquisition function in \eqref{eq:EImin} by using the QS-EGO (shown as Algorithm~\ref{ag1}),
and then compute (or simulate) its response as $y_{n+1}$.
	\item Re-fit the MaGP model based on observations $(\bw_1, y_1),\ldots, (\bw_{n+1}, y_{n+1})$.  Set $n=n+1$.
	\item Repeat Steps 2 and 3 until the stopping criterion is satisfied.
\end{enumerate}

In Step 2, when the number of components $k$ is small,  we can enumerate $k!$ sequences (possibly with parallel computing) and identify the optimal $\bx$ given each sequence $\bo$ that can maximize the EI function.
For a large $k$, such an enumeration may become prohibitively time-consuming. To address this challenge, we propose to iteratively optimize quantitative input $\bx$ and sequence input $\bo$ given the other, as summarized in Algorithm \ref{ag1}. \rf{In both the four-step QS-learning and Algorithm~\ref{ag1}, we adopt the stopping criterion used in \cite{jones1998efficient}, i.e., that the algorithm stops when three consecutive EIs do not produce more than $\alpha$ ($ \alpha \in [0.1\%, 1\%]$) improvement over the current best output.}

\begin{algorithm}[htbp]
	\begin{algorithmic}[t]
		\caption{An efficient optimization of EI for QS factors (QS-EGO)}
		\label{ag1}
		\STATE Initialize the maximum number of rounds $N_{round}$.
		\STATE Initialize the current input $\bw_{c} = (\bx_{c}, \bo_{c})$. Set the current optimal input $\bw_{opt} = \bw_{c}$.
		\STATE Initialize an empty vector $\textbf{ei}$ and a scalar $\hbox{ei}_{opt} = 0$.
		\FOR {$i = 1$ to $N_{round}$}
		\STATE \textbf{1}. Given $\bo_{c}$, find the quantitative input $\bx$ that maximizes EI via a genetic optimization algorithm. Set $\bx_{c} = \bx$.
		\STATE \textbf{2}. Given $\bx_{c}$, find the sequence input $\bo$ that maximizes EI via the SFTA method in Algorithm~\ref{alg:ta}. Set $\bo_{c} = \bo$.
		\STATE \textbf{3}. Set $\bw_{c} = (\bx_{c}, \bo_{c})$ and evaluate $E[I(\bw_{c})]$ defined in \eqref{eq:EImin}.
		\STATE \textbf{if} $E[I(\bw_{c})] > \hbox{ei}_{opt}$,
		\STATE \ \ \ \ \textbf{then} set $\hbox{ei}_{opt} = E[I(\bw_{c})]$ and $\bw_{opt} = \bw_{c}$.
		\STATE Set $\textbf{ei}[i] = \hbox{ei}_{opt}$, where $\textbf{ei}[i]$ denotes the $i^{th}$ element of \textbf{ei}.
		\STATE \textbf{if} the stopping criterion is satisfied,
		\STATE \ \ \ \ \textbf{then} \textbf{break}.
		\ENDFOR
		\STATE Return $\bw_{opt}$.
	\end{algorithmic}
\end{algorithm}

In Algorithm \ref{ag1}, to optimize the quantitative input $\bx$ given the sequence input $\bo_{c}$, we adopt a BFGS method in a hybrid genetic optimization algorithm \citep{Walter2011}.
Numerical gradients suffice for small cases, but they can be slow for large ones. In this study, we derive the analytical gradients for EI maximization under the proposed MaGP model. These gradients are exact and fast to compute. See Supplementary Materials Section S1.3 for details.

In Algorithm~\ref{ag1}, optimizing the sequence input $\bo$ given the quantitative input $\bx_{c}$ is nontrivial, because its solution space is semi-discrete and can be extremely large. To address this issue, we propose the so-called space-filling threshold accepting (SFTA) algorithm for a large $k$. As detailed in Algorithm~ \ref{alg:ta}, the proposed SFTA algorithm includes two phases. The first phase seeks space-filling points that are far from one another to achieve robustness \citep{Johnson1990}. The second phase inherits from the classic threshold-accepting (TA) algorithm \citep{Dueck1990}, which balances exploration and exploitation.

\begin{algorithm}[htbp]
	\begin{algorithmic}[t]
		\caption{SFTA algorithm for optimizing $\bo$ given $\bx_{c}$}
		\label{alg:ta}
		\STATE Initialize $n_{step}^{(1)}$ (number of steps) in SFTA Phase I.
		\STATE Initialize $n_{seq}$ (number of iterations to compute the threshold sequence), $n_{rounds}$ (number of rounds), and $n_{steps}^{(2)}$ (number of steps) in SFTA Phase II.
		\STATE  Initialize a starting solution $\bo_0$, set current optimal $\bo_{opt} = \bo_0$, and let $O_{obs} = [\bo_0^\TT]^\TT$.
		\WHILE {$i \leqslant n_{step}^{(1)}$}
		\STATE  Generate a random sequence $\bo_{i}$.
		\STATE \textbf{if} $ H_{min}(\bo_i, O_{obs})/k > \epsilon$  ($\epsilon$ is drawn from  Unif(0,1)),
		\STATE \ \ \ \ \textbf{then} let $i = i+1$,  $O_{obs} = [O_{obs}, \bo_i^\TT]^\TT$, and $\delta = f(\bo_i)-f(\bo_{opt})$,
		\STATE \textbf{if} $\delta < 0$, \textbf{then} let $\bo_{opt}=\bo_i$.
		\ENDWHILE
		\STATE Set the current solution $\bo_c = \bo_{opt}$.
		\FOR {$j = 1$ to $n_{seq}$}
		\STATE  Generate a neighbor solution $N(\bo_c)$, and let $\Delta_j=|f(o_c)-f(N(\bo_c))|$.
		\ENDFOR
		\STATE Compute the empirical distribution of $\Delta_j$ , $j= 1,2,\ldots,n_{seq} $, denoted as $F$.
		\FOR{$r=1$ to $n_{rounds}$}
		\STATE Generate {threshold $\tau_r=F^{-1} \left(0.5 (1-r/n_{rounds}) \right)$} 
		\FOR{$i=1$ to $n_{steps}^{(2)}$}
		\STATE  Generate a neighbor solution $N(\bo_c)$, and let $\delta = f(N(\bo_c))-f(\bo_c)$.
		\STATE \textbf{if} $\delta < \tau_r$, \textbf{then} let $\bo_c=N(\bo_c)$.
		\STATE {\textbf{if} $f(\bo_c) < f(\bo_{opt})$, \textbf{then} let $\bo_{opt}=\bo_c$}.
		\ENDFOR
		\ENDFOR
		\STATE Return $\bo_{opt}$.
	\end{algorithmic}
\end{algorithm}

Specifically, Phase I of SFTA starts from the current optimal sequence input $\bo_0$ with the smallest $f(\bo)$ value in the observed data, where the \rf{objective} function $f(\bo)$ is the negative EI value for $\bo$ given $\bx$.
Then, the algorithm iteratively accepts a random sequence $\bo_i$ ($i = 1, \ldots, n^{(1)}$) for evaluation with probability $P(\bo_i) = H_{min}(\bo_i)/k$, where $H_{min}(\bo_i)$ is the minimum pairwise Hamming distance between the sequence vector $\bo_i$ and all observed sequence vectors $\bo_1, \ldots, \bo_{i-1}$. The Hamming distance is the number of positions at which corresponding symbols are different in two vectors.
If a candidate $\bo_i$ is far from the observed $\bo_1, \ldots, \bo_{i-1}$ under the Hamming distance, then it has a high probability \rf{of being} included for evaluation.
Phase II of SFTA starts from the optimal solution $\bo_c = \bo_{opt}$ with the smallest $f(\bo)$ value in Phase I.
We define the neighbor solution as $N(\bo_c)$ by randomly exchanging two elements of the sequence vector $\bo_c$ (current solution). Evidently, all possible $N(\bo_c)$'s have \rf{a Hamming distance of} 2 with $\bo_c$. The threshold values for accepting neighbor solutions are generated by empirical distributions of increments (denoted as $F$) for the objective function $f$; refer to \cite{Dueck1990} for details on the calculation of threshold values.
A neighbor solution is more likely to be accepted early in the search than later in the search, because threshold values decrease. The SFTA algorithm can avoid being trapped at local optima and focus more on exploration in the beginning.

When the number of allowed evaluations is considerably fewer than the total number of possible sequences, random initialization (and generation) of neighbor solutions may not consistently and efficiently explore space. To address this issue, we adopt space-filling samples in Phase I, which provide a good initialization for TA global optimization in Phase II. When parallel computing is available, more than one solution $\bo_{opt}$ in Phase I can be selected as multi-starting points. Refer to Supplementary Materials Section S3.2 for additional details on Algorithms~1 and 2.

\subsection{\rf{Fast QS-learning for Large Experiments}}
\label{sec:fast}

In most literature on experimentation, the costs for estimating surrogate models and assessing acquisition functions are negligible compared with the costs of conducting experiments \citep{frazier2018tutorial, gramacy2020surrogates}. In other cases, experiments may be executed rapidly, and researchers will need a fast sequential scheme for a large number of runs  \citep{gramacy2020surrogates}.

Classic GP-based active learning approaches have a computational complexity of $O(N^4)$, where $N$ is the total number of runs.
The estimation of GP models in each iteration requires $O(n^3)$ computation, where $n$ is the number of data points used.
Here, we propose a fast variant of QS-learning with $O(N^3)$ computation.
Suppose that the total budgets for the number of runs and computing time are $N$ and $T$, respectively. The proposed fast QS-learning approach includes the following four steps.
\begin{enumerate}
	\item Construct an optimal initial design for QS factors with $n_0$ runs $\bw_1,\ldots, \bw_{n_0}$, evaluate their responses  $y_1,\ldots, y_{n_0}$, and fit the MaGP model based on these observations. Set $n = n_0$. Record the time used for fitting the MaGP model as $t$. Record the time left from the budget $T$ as $T^{'}$ and the number of runs left from the budget $N$ as $N^{'}$.
	\item For the next $\lceil N^{'}t/T^{'} \rceil$ iterations, fix the estimated parameters of the MaGP model and sequentially select runs based on EI by using the fast updating technique illustrated below. Set $n = n+\lceil N^{'}t/T^{'} \rceil$ \rf{($\lceil a \rceil$ is the largest integer not exceeding $a$)}.
	\item Refit the MaGP model (including reestimating all parameters) based on observations $(\bw_1, y_1),\ldots, (\bw_{n}, y_n)$. Update the time used for fitting the model as $t$. Record the time left from the budget as $T^{'}$ and the number of runs left from the budget as $N^{'}$.
	\item Repeat Steps 2 and 3 until the stopping criterion is satisfied.
\end{enumerate}

In Step~2, we adopt fast updating of model fit in $O(n^2)$ computing time given all parameters in the MaGP model. Let $ \bm{\Phi}_n$ be the covariance of the $n$ current input.
The key is to update the model when the $(n+1)^{th}$ data point arrives via fast calculation of the covariance $\bm{\Phi}_{n+1}$ and its inverse $\bm{\Phi}_{n+1}^{-1}$.
Similar to a rank one Sherman--Morrison update \citep{sherman1950adjustment}, we have
\begin{equation*}
\bm{\Phi}_{n+1} =
\begin{blockarray}{cc}
\begin{block}{[cc]}
  \bm{\Phi}_{n} & \bm{\gamma}  \\
  \bm{\gamma}^\TT & \phi(\bw_{n+1}, \bw_{n+1}) \\
\end{block}
\end{blockarray}, \
\bm{\Phi}_{n+1}^{-1} =
\begin{blockarray}{cc}
\begin{block}{[cc]}
  \bm{\Phi}_{n}^{-1}+ \bm{g}\bm{g}^\TT v & \bm{g}  \\
  \bm{g}^\TT & v^{-1} \\
\end{block}
\end{blockarray},
\end{equation*}
where the covariance function $\phi$ is defined in \eqref{eq:corr-fun}, the covariance vector $\bm{\gamma} = (\phi^{'}(\bw_{n+1}, \bw_i))_{n \times 1}$ with $\phi^{'}(\bw_{n+1}, \bw_i) = \sum_{h=1}^{k}  \sigma_h^2 \hbox{exp} \big \{- \theta_h (x_{i,h} - x_{n+1,h})^2 -\sum_{l=1}^{t} (\tilde{o}_{i,h}^{(l)} - \tilde{o}_{n+1,h}^{(l)})^2 \big\}$ for $i=1, \ldots, n$, $v = \phi(\bw_{n+1}, \bw_{n+1}) - \bm{\gamma}^\TT \bm{\Phi}^{-1}\bm{\gamma}$ and $\bm{g} = - v^{-1} \bm{\Phi}_{n}^{-1}\bm{\gamma}$. Here, the update on the covariance inverse requires $O(n^2)$ time, and thus, the updates to all relevant quantities for each model fit are of $O(n^2)$ \citep{gramacy2020surrogates}.
The total cost of updating with sequential runs from $n =
n_0, \ldots ,N$ demands flops in $O(N^3)$.
Compared with the general QS-learning that reestimates parameters for every sequential run, the fast QS-learning method reestimates them much less frequently, and thus, saves computations for large experiments. Refer to Supplementary Materials Section~S3.3 for details on the stopping criterion and parameter tuning. 

Notably, small-sample performance is often more important and relevant than the convergence rate in experimentation, because only a small number of trials are frequently allowed in practice \citep{fang2005}. Asymptotic guarantees may provide minimal information about the practical effectiveness of the method \citep{sutton2018reinforcement}. The discussion of convergence for learning QS experiments is included in Section~S2 of the Supplementary Materials.

\section{Optimal Initial Designs for QS-learning}
\label{sec:design}

Desirable initial designs are important in active learning. They may save the total number of runs and lead to better solutions. In this section, we propose a new class of optimal designs for QS factors, called QS-design, which exhibits space-filling and pair-balanced properties. We first propose a general approach for constructing QS-designs with flexible sizes in Section~\ref{sec: general construction}, and then provide a deterministic algebraic construction for QS-designs with certain sizes in Section~\ref{sec: algebra construction}.

\subsection{General Construction}
\label{sec: general construction}
The design for QS factors is denoted as $\bm D = (\bm X, \bm O)$ where $\bm X$ is the quantitative part, and $\bm O$ is the sequence part. Both parts use components as columns. To construct a desirable design $\bm D$, we will first construct a good sequence design $\bm O$, and then construct a good quantitative design $\bm X$ in combination with $\bm O$.

Sequence designs have two equivalent representations: one with components as columns (denoted as $\bm O$), and the other with order positions as columns (denoted as $\bm O^{'}$). Designs $\bm O$ consist of runs $\bo$ defined in Section~\ref{secuk}, and designs $\bm O^{'}$ consist of runs $\balpha$ defined in Section~\ref{lr}. For illustration, the following two designs have the same practical meaning:
\renewcommand{\kbldelim}{(}
\renewcommand{\kbrdelim}{)}
\[
\bm O = \kbordermatrix{
	& A & B & C \\
	& 1 & 2 & 3 \\
	& 2 & 1 & 3 \\
	& 2 & 3 & 1 \\
}
\Leftrightarrow
\bm O^{'} = \kbordermatrix{
	& 1 & 2 & 3 \\
	& A & B & C \\
	& B & A & C \\
	& C & A & B \\
}.
\]
\rf{To identify the optimal sequence design $\bm O$, we first find the optimal $\bm O^{'}$ by minimizing the following $\nu_p$ criterion:}

\begin{equation}
\label{op}
\nu_p=\left(\rho_1 \underset{i \neq j}{\sum_{i=1}^k \sum_{j=1}^{k}} \frac{1}{(t_{i,j}+1)^p} + \rho_2 \sum_{i=2}^n \sum_{j=1}^{i-1} \frac{1}{(h_{i,j}+1)^p} \right)^\frac{1}{p},
\end{equation}
where $t_{i,j}$ is the number of appearances of the subsequence ``(i j)" in rows of $\bm O^{'}$; $h_{i,j}$ is the Hamming distance between the $i^{th}$ and $j^{th}$ rows in $\bm O^{'}$; and $\rho_1$, $\rho_2$, and $p$ are tuning parameters. A design is called pair-balanced if it has the same $t_{i,j}$ value for all subsequences (i.e., pairs) of ($i,j$), where we use $\{1,2, \ldots, \}$ to denote the levels $\{A,B, \ldots \}$. A pair-balanced design assigns equal importance to all pairwise interactions among components. It also accounts for different precedence patterns where pairs ($i,j$) and ($j,i$) are different. It is similar to the ``balance" idea in crossover designs \citep{dean2015handbook}. 
To find (near) pair-balanced designs, we propose to maximize designs' minimum $t_{i,j}$ values, which is equivalent to minimizing the term $\sum \sum 1/(t_{i,j}+1)^p$ in \eqref{op} for a sufficiently large tuning parameter $p$. In practice, $p=15$ often suffices. In the denominators, we add 1 to $t_{i,j}$ (and $h_{i,j}$) to avoid numerical problems when they are equal to 0.

\rb{The term $\sum \sum 1/(h_{i,j}+1)^p$ in \eqref{op} considers designs' space-filling properties. Here, we adopt the popular maximin distance criterion \citep{Johnson1990}, which seeks to scatter design points over the experimental domain such that the minimum pairwise distance between points is maximized. The Hamming distance is used here because the elements in $\bm O^{'}$ are categorical.
Analogous to the scalar criterion in \cite{Morris1995}, minimizing the  term $\sum \sum 1/(h_{i,j}+1)^p$ is asymptotically equivalent to the maximin Hamming distance criterion as $p$ goes toward infinity, where $p=15$ often suffices. A space-filling $\bm O^{'}$ benefits the exploration of \rf{the} response surface and is a robust choice for initial points \citep{frazier2018tutorial}. In this study, we set weights $\rho_1 = 0.2$ and $\rho_2 = 0.8$ in~\eqref{op} emphasizing more on the design's space-filling property.

\begin{example}
	\label{eg:o}
	Consider a drug combination experiment consisting of four drug components. Two designs $\bm O^{'}_A$ and $\bm O^{'}_B$ shown below are compared, and they have the same Hamming distance structure. Their $t_{i,j}$ pairs are listed in Table~\ref{method1}.
	Given the possibility of synergistic or antagonistic interactions between two drugs, the order in which they are administered is important. For example, if Drugs $A$ and $B$ exhibit a strong synergistic effect, then they should be administered in adjacent order. By contrast, if they exhibit a strong antagonistic effect, then their order of administration should be well separated in time. Furthermore, drugs may have different (i.e., immediate, delayed, or cumulative) time course effects \citep{al2009time}, and thus, their precedence patterns matter. For example, consider that Drugs $A$ and $B$ have a synergistic interaction, where the effect of $A$ is immediate, whereas the effect of $B$ appears delayed with respect to the concentration–time profile. In such case, $B$ should be added before $A$ to maximize the synergistic effect, because $B$ requires more time to fully exert its work with $A$. Evidently, subsequences $(A,B)$ and $(B,A)$ may lead to different outcomes in this study.
	Considering all of the above, it is clear that the pair-balanced design $\bm O^{'}_B$ is better than $\bm O^{'}_A$, as all possible adjacent pairs in $\bm O^{'}_B$ appear the same number of times in the experiment (i.e. all $t_{i,j}$ values are equal). Finally, we should use $\bm O_B$, the equivalent form of $\bm O^{'}_B$, to be the sequence part of the QS-design $\bm D$.
	\[
	\bm O^{'}_A =
    \kbordermatrix{
    & 1 & 2 & 3 & 4 \\
	& A & B & C & D \\
	& B & C & D & A \\
	& C & D & A & B \\
	& D & A & B & C \\
}, \
\bm O^{'}_B =
    \kbordermatrix{
    & 1 & 2 & 3 & 4 \\
	& A & B & C & D \\
	& B & D & A & C \\
	& C & A & D & B \\
	& D & C & B & A \\
}
\Leftrightarrow
\bm O_B =
    \kbordermatrix{
    & A & B & C & D \\
	& 1 & 2 & 3 & 4 \\
	& 3 & 1 & 4 & 2 \\
	& 2 & 4 & 1 & 3 \\
	& 4 & 3 & 2 & 1 \\
}.
\]
	\begin{table}[ht]
		\label{method1}
		\centering
		\caption{Comparison of designs' $t_{i,j}$ pairs.}
		\begin{tabular}{rrrrrrrrrrrrr}
			\hline
			$t_{i,j}$ pairs & $AB$ & $AC$ & $AD$ & $BA$ & $BC$ & $BD$ & $CA$ & $CB$ & $CD$ & $DA$ & $DB$ & $DC$ \\
			\hline
			$\bm O^{'}_A$ &   3 &   0 &   0 &    0 &   3 &   0 &   0 &   0 &   3 &   3 &   0 &   0 \\
			$\bm O^{'}_B$ &   1 &   1 &   1 &   1 &   1 &   1 &   1 &   1 &   1 &   1 &   1 &   1 \\
			\hline
		\end{tabular}
	\end{table}
\end{example}

To search for optimal designs $\bm O^{'}$, we adopt a standard TA algorithm \citep{Dueck1990, xiao2018} by using the criterion $\nu_p$ in \eqref{op} as the objective function. The algorithm starts with a random design $\bm O^{'}$ and defines its neighbor design $N(\bm O^{'})$ by exchanging two random levels in a random row. It can be implemented with the R package ``NMOF" \citep{Enrico}.


\rf{After obtaining the optimal $\bm O^{'}$, we propose to  construct the optimal $\bm D^{'} = (\bm X, \bm O^{'})$ minimizing the following $C_p$ criterion, which measures a design's space-filling property.}
\begin{equation}
\label{Cp}
C_p=\left(\sum_{i=2}^n \sum_{j=1}^{i-1} \frac{1}{(\rho_1^{'}d_{i,j} + \rho_2^{'}h_{i,j} + 1)^p} \right)^\frac{1}{p},
\end{equation}
where $d_{i,j} =  \sqrt{\sum_{l=1}^{k}(x_{il} - x_{jl})^2}$ is the $L_2$-distance between the rows $\bx_i$, and $\bx_j$ in $\bm X$ and  $h_{i,j}$ is the Hamming distance between the rows $\bo_i^{'}$ and $\bo_j^{'}$ in $\bm O^{'}$. Here, we adopt weights $\rho_1^{'} = \rho_2^{'} = 0.5$ and the tuning parameter $p=15$.

\rf{To search for the optimal $\bm D^{'}$ given $\bm O^{'}$}, we adopt the same TA algorithm. It starts from a design $\bm D_c^{'} = (\bm X_c, \bm O^{'})$, where $X_c$ is the maximin distance Latin hypercube design (LHD) found by the R package ``SLHD"  \citep{ba2015optimal} or ``LHD" \citep{Hongzhi2020}. Notably, the maximin distance LHD is a popular type of space-filling design, which has been proved to be robust for model misspecification and can minimize the theoretical prediction variance of fitted GP models \citep{gramacy2020surrogates}. Here, the $C_p$ criterion in \eqref{Cp} is used as the objective function, and neighbor designs $N(\bm D_c^{'})$ are defined by exchanging two randomly chosen rows of $\bm X_c$.
Finally, we convert the optimal $\bm D^{'}$ to its equivalent form, i.e., the QS-design $\bm D$.
In practice, we may need to normalize quantitative designs $\bm X$ to $[0,1]$ range.}

\subsection{\rf{Algebraic Construction Method}}\label{sec: algebra construction}
We develop an algebraic construction for QS-designs whose component sizes $k$ and run sizes $n$ are $p_r-1$, where $p_r$ is any odd prime number.
Denote the $n \times n$ good lattice point design \citep{zhou2015space} as $\bm D_{glp}$, whose $i^{th}$ row is $\bh \times i \text{ mod } p_r$, \rf{where vector $\bh = (1, \ldots, n)$ and $i = 1, \ldots, n$}. Design $\bm D_{glp}$ is a Latin square whose rows and columns are both permutations of $1, \ldots, n$. \rf{To construct $\bm D^{'} = (\bm X, \bm O^{'})$, we propose to use $\bm O^{'} = \bm D_{glp}$ and $\bm X$ as any column permutation of $\bm D_{glp}$}.
For illustration, design $\bm O_{b}^{'}$ in Example \ref{eg:o} is a $4 \times 4$ $\bm D_{glp}$, where we treat $A$ as 1, $B$ as 2, and so on.

\begin{theorem}\label{prop:Oglp}
  Let $n = k = p_r-1$, where $p_r$ is any odd prime number. Then, the n-run sequence design $\bm O^{'} = \bm D_{glp}$ has the following properties (for any $1\leq i \neq j \leq n$):\\
  (i) $\bm O^{'}$ is the maximin Hamming distance design, where all $h_{i,j} = n$; \\
  (ii) $\bm O^{'}$ is the pair-balanced design, where all $t_{i,j} = 1$;\\
  (iii)  $\bm O^{'}$ is optimal under the $\nu_p$ criterion defined in \eqref{op} for any positive weights $\rho_1$ and $\rho_2$, and it has a $\nu_p$ value of:
  \begin{equation}\label{op:Oglp}
  \nu_p (\bm O^{'})= \left\{  n(n-1) \left( \frac{\rho_1}{2(n+1)^p} + \frac{\rho_2}{2^p} \right) \right\}^\frac{1}{p}.
  \end{equation}
\end{theorem}

\begin{theorem}\label{prop:XOglp}
 Let $n = k = p_r-1$ ($p_r$ is any odd prime number), and $\bm D^{'} = (\bm X, \bm O^{'})$, where $\bm O^{'} = \bm D_{glp}$ and $\bm X$ is any column permutation of $\bm D_{glp}$.  $\bm D^{'}$ has the following properties: \\
 (i)  the minimum row-pairwise $L_2$-distance in $\bm X$ is
 $\sqrt{ n (n+1) (n+2) / 12}$; \\
(ii) \rf{it has a upper bound of $C_p$ as defined in \eqref{Cp}, i.e.,} $C_p(\bm D^{'}) \leq  n^\frac{1}{p}  C(\rho_1^{'},\rho_2^{'},p)$,
where $$
C(\rho_1^{'},\rho_2^{'},p) =\left( \frac{n/2-1}{ \left[ \rho_1^{'} \left(\frac{1}{12} n (n+1) (n+2) \right)^{\frac{1}{2}} + n \rho_2^{'} + 1\right]^p}+ \frac{1}{ 2 \left[  \rho_1^{'} \left( \frac{1}{3} n \left(n^2-1\right) \right)^{\frac{1}{2}}   + n \rho_2^{'} + 1\right]^p} \right)^\frac{1}{p}
$$
is a constant that only depends on adopted weights $\rho_1^{'}$ and $\rho_2^{'}$ in the $C_p$ criterion.
\end{theorem}

\begin{corollary}
\label{coro:l2}
Let $n = k = p_r-1$ ($p_r$ is any odd prime number), and $\bm X$ be any column permutation of $\bm D_{glp}$. Then, the minimum row-pairwise $L_2$-distance in $\bm X$, denoted as $d(\bm X)$, \rf{satisfies}
$$\frac{d(\bm X)}{d_{upper}} = \sqrt{\frac{n+2}{2n}} > \dfrac{\sqrt{2}}{2}, $$
where $ d_{upper} = n \sqrt{(n+1)/6}$ is the upper bound of $d(\bm X)$.
\end{corollary}

Theorem~\ref{prop:Oglp} shows that the proposed sequence design $\bm O^{'}$ is optimal under both the space-filling and  pair-balanced criteria. Theorem~\ref{prop:XOglp} shows that \rf{the constructed $\bm D^{'} = (\bm X, \bm O^{'})$, or equivalently the QS-design $\bm D=(\bm X, \bm O)$, has the best space-filling property, i.e., the minimized $C_p$ value}. 
Corollary~\ref{coro:l2} shows that the proposed quantitative design $\bm X$ also exhibits good space-filing property, because it has a large minimum pairwise $L_2$-distance. Corollary~\ref{coro:l2} can  easily be obtained on the basis of Theorem 3 in \cite{zhou2015space} along with the proofs for Theorems \ref{prop:Oglp} and \ref{prop:XOglp} in this work. Notably, the upper bound $ d_{upper} = n \sqrt{(n+1)/6}$ may not be achievable for all design sizes.

Although the minimum run size of an initial design in active learning can be as small as 2, many researchers have recommended using initial designs with moderate sizes \citep{jones1998efficient, loeppky2009choosing, frazier2018tutorial}.
We would remark that the run size of QS-design can be flexibly determined. One recommended run size is the number of parameters in the GP part,
e.g.  $4k-3$ for 2d-MaGP and $k(k+3)/2$ for full-MaGP. When a larger number of runs is allowed, we recommend a rule-of-thumb run size of $2 + k(k+3)/2$ for any $t$-dimensional MaGP for simplicity, which is one more than the total number of parameters in full-MaGP.
For the special case, when $k = p_r-1$ and $p_r$ is any odd prime, we find that the $k$-run QS-design proposed in this subsection performs very well, as illustrated in Sections S5 and S6 of Supplementary Materials.

\section{Case Study}
\label{realdat}
Lymphoma is cancer that begins in infection-fighting cells of the immune system, called lymphocytes. When a patient has lymphoma, lymphocytes change and grow out of control. In a recent pioneering work \citep{wang2019}, the researchers conducted a series of drug experiments on lymphoma treatment. Among them, a 24-run in vitro experiment of three Food and Drug Administration (FDA) approved chemotherapeutics, namely, paclitaxel, doxorubicin, and mitoxantrone (denoted as Drugs $A$, $B$, and $C$, respectively), was included. This experiment considered the doses and sequences of drugs. In this experiment, all six sequences of the three drugs were enumerated. For each sequence, two dose levels for $A$ (Level 0: 2.8 $\mu$M; Level 1: 3.75 $\mu$M) and $B$ (Level 0: 70 nM; Level 1: 95 nM), and a fixed dose level for $C$ (0.16 $\mu$M) were considered. The experiment was performed on Raji cells, a human lymphoma cell line. In any treatment (run), each drug was added every 6 hours in a sequence into the Raji cell culture, and the inhibition percentages (the larger, the better the response) were measured 6 hours after the addition of the last drug. The four largest responses in this experiment are 47.18, 44.87, 44.38 and 44.33.

\begin{figure}[htp]
\vspace{-.2in}
	\centering
	\subfigure []{\includegraphics[scale=0.4]{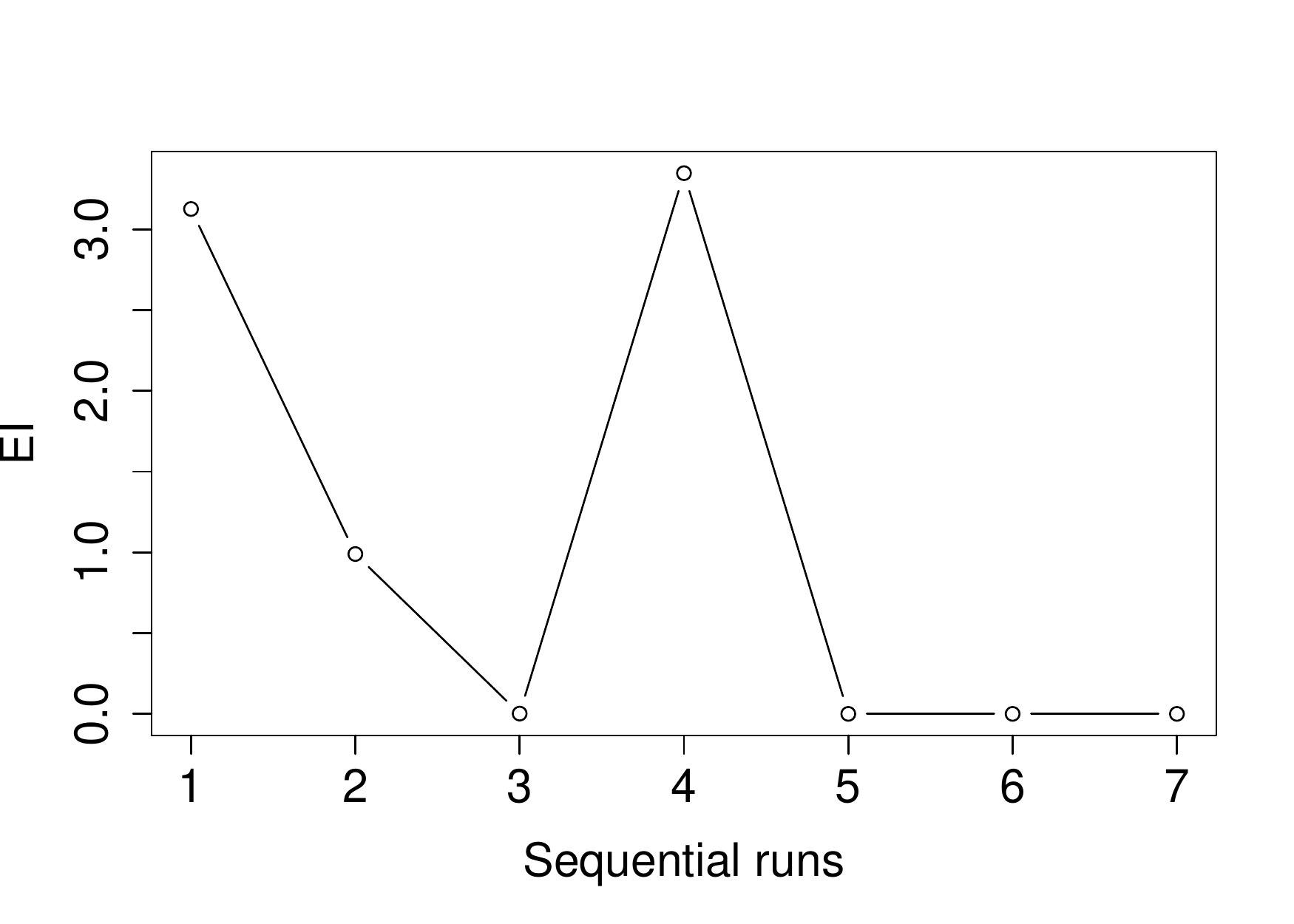}}\quad
	\subfigure [] {\includegraphics[scale=0.4]{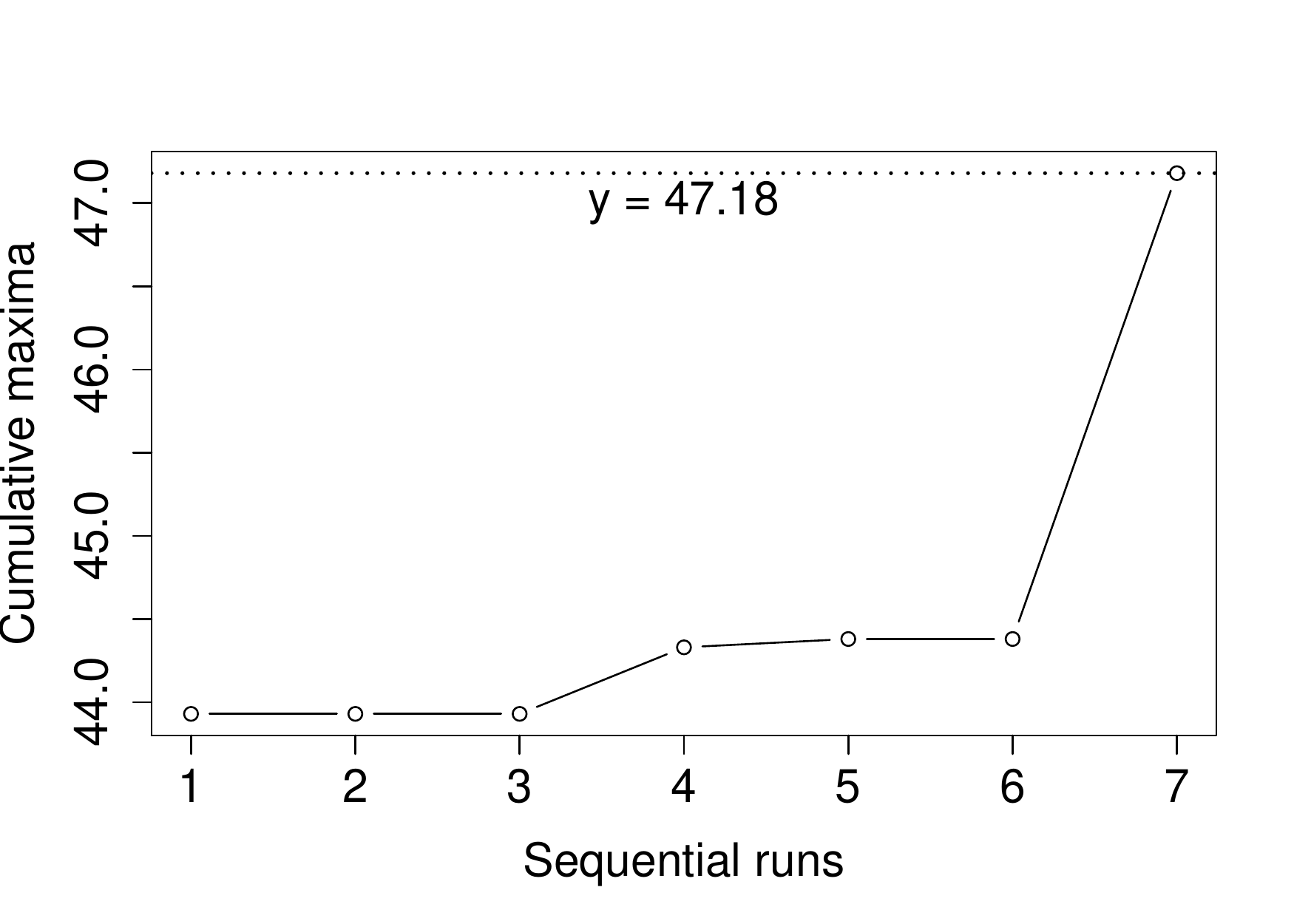}}
	\caption{Plots for (a) EIs and (b) cumulative maximum responses from the QS-learning approach in the case study.}
		\label{fig:ei}
\end{figure}

Here, we run the proposed QS-learning to determine if we can use fewer runs (compared with the original 24 runs) to identify the optimal treatment in this experiment. Notably, 2d-MaGP and full-MaGP are the same for $k = 3$ components. Given that the GP part of the model has eight parameters, we construct an eight-run QS-design to collect the initial data. The proposed QS-learning selects seven sequential runs until the stopping rule is satisfied, i.e., the last three EIs are all less than 1\% of the current best output. Figure~\ref{fig:ei} shows the plots for EIs and the cumulative maximum responses of the seven sequential runs. The true maximum response, i.e., 47.18, is found, along with the third and fourth largest responses, 44.38 and 44.33, respectively. Notably, the initial QS-design does not include settings that lead to the largest four responses. The proposed QS-learning requires 15 runs (8 initial runs plus 7 follow-up runs) to identify the optimal treatment of the original 24-run experiment, saving 37.5\% of the budget. Refer to Supplementary Materials Section S4.1 for the complete data and more details regarding the analysis.

\begin{figure}[htp]
\vspace{-.2in}
    \centering
  \subfigure []{\includegraphics[scale=0.35]{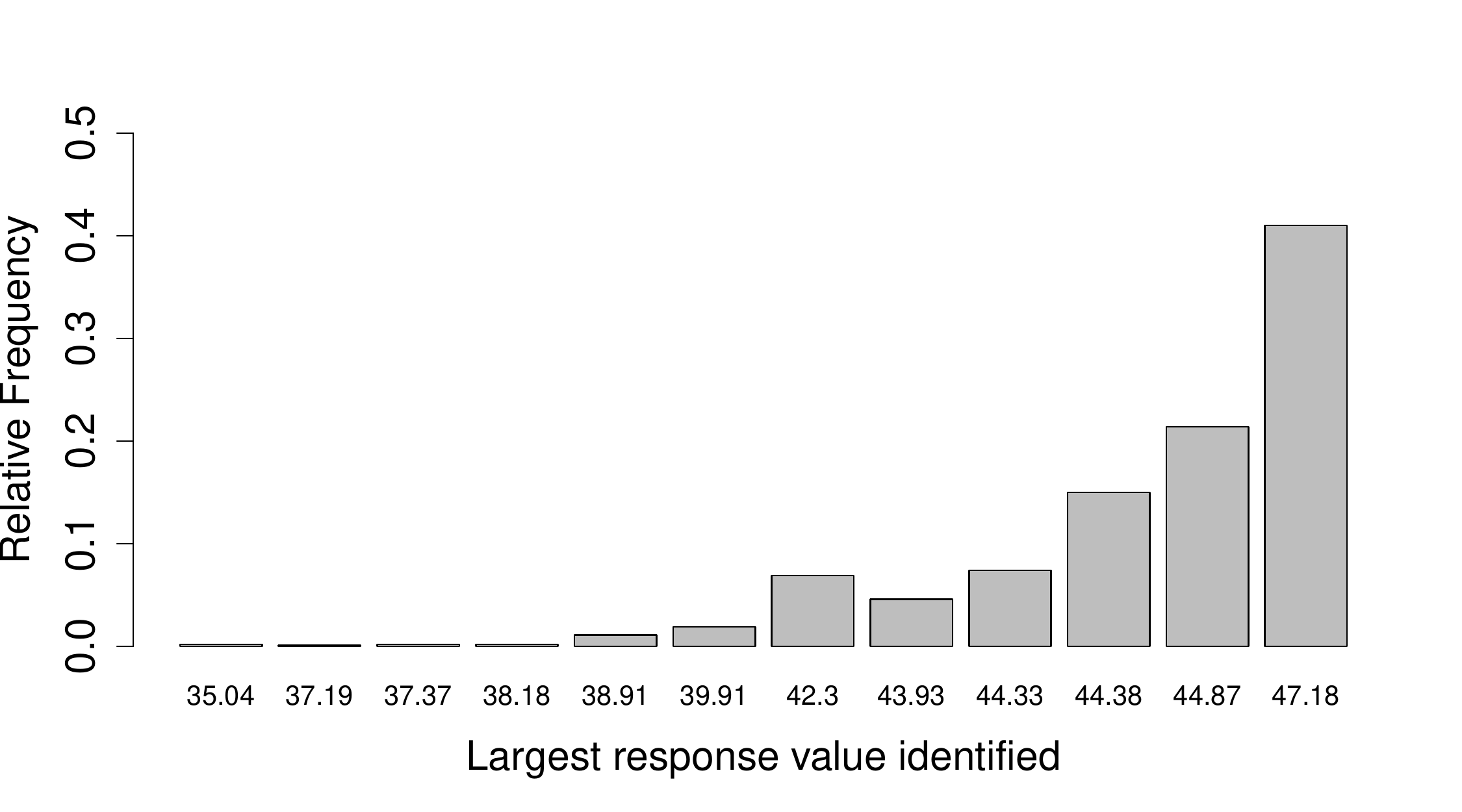}}\quad
  \subfigure [] {\includegraphics[scale=0.35]{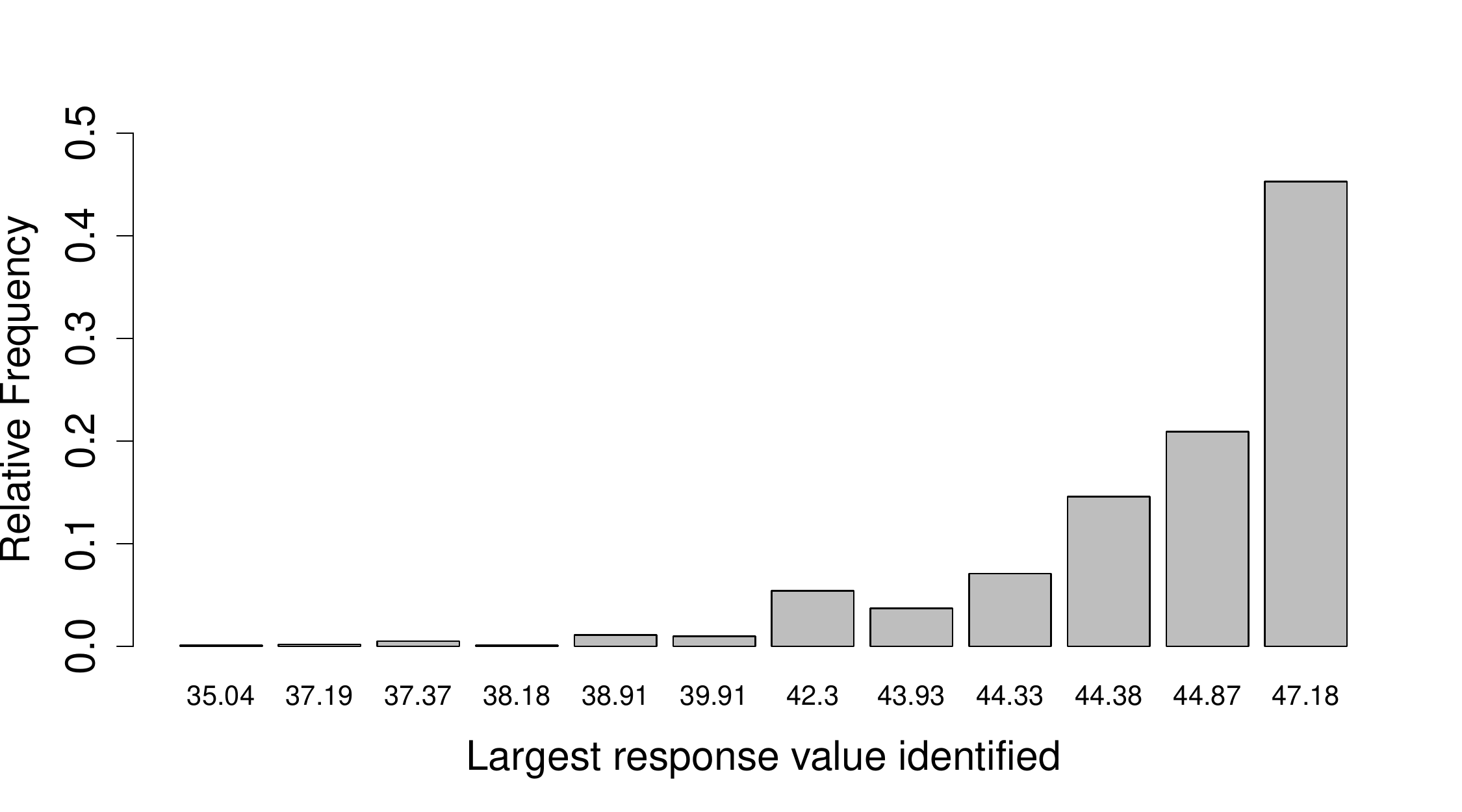}} \\
      \caption{Histograms of \rf{the} largest response values identified by (a)  sequential generalized PWO approach and (b) sequential generalized CP approach in the case study.}
          \label{fig:bar}
\end{figure}

To further evaluate the proposed QS-learning compared with other approaches, we consider three benchmark methods: the random sampling approach ($BM_{1}$),  sequential generalized PWO approach ($BM_{2}$), and sequential generalized CP approach ($BM_{3}$). Here, the $BM_{1}$ method considers a random sampling of 15 runs out of the original 24 runs, which is the same total run size used above under QS-learning. Proving that the probability of including the optimal treatment in such a random sample is only 62.5\% will be straightforward. The $BM_{2}$ and $BM_{3}$ methods consider sequential experiments based on the generalized PWO and CP approaches, respectively, introduced in Section~\ref{lr}. Both methods will start from 8 initial runs, and then choose the setting with the optimal prediction to be the next experiment trial until no further improvement is achieved. Specifically, when the $BM_{2}$ and $BM_{3}$ methods start from the QS-design, the largest response identified is only 43.93. In addition, when they start from random designs, we present the plots of their largest responses found from 1000 replications in  Figure~\ref{fig:bar}. The probability of successfully identifying the optimal solution (47.18) for $BM_{2}$ and $BM_{3}$ is less than 50\%, and many results are not good in  Figure~\ref{fig:bar}.

This 24-run real experiment \citep{wang2019} on doses and sequences is a pioneering work in the literature. It serves as a good example to demonstrate the importance of such experiments. Nevertheless, it also has limitations. First, only two dose levels for Drugs $A$ and $B$ are considered, which does not support the estimation of any curvature effect. In addition, only one dose level for Drug $C$ is used, and we cannot estimate its effect. When doses are not restricted to only a few levels and more drugs are included, QS-learning is expected to perform better.

\section{Simulation Study}
\label{tsp}
In this section, we evaluate the performance of the proposed QS-learning and its fast variant through a traveling salesman problem (TSP, \citet{applegate2006traveling}). In Sections S5 and S6 of the Supplementary Materials, we provide two additional simulations of arranging the four mathematical operations problem \citep{robert2018, yang2018} and the single machine scheduling problem \citep{allahverdi1999review, wan2013single}. These additional results illustrate the advantages of QS-designs (particularly for those from algebraic construction), the superior predictive power of the MaGP model, the difference between 2d-MaGP and full-MaGP, and the general applicability of QS-learning.

TSP is a well-known nonpolynomial-hard problem in combinatorial optimization \citep{ tan2000comparison, applegate2006traveling}. Here, we consider a modified TSP that involves the optimization of quantitative input and sequence input. We regard it as a computer experiment wherein the simulator is assumed to be black-box and expensive to evaluate \citep{fang2005}. The cost for evaluating runs is assumed to be considerably higher than that for estimating surrogate models or assessing acquisition functions.

Suppose a salesman needs to travel to $k$ cities to sell products, indexed as Cities $1,\ldots, k$. All cities are available for visiting at time zero, and the salesman must visit all cities one by one. The time to travel from City $i$ to City $j$  ($i \neq j$) is $s_{i,j}$ days, and $s_{i,j}$ can be different from $s_{j,i}$. The salesman will stay in City $i$ for $x_i$ days to sell products. He has a due date to complete the business in City $i$, denoted as $d_i$. If he misses the due date, then a penalty rate of $f$ dollars per day will be charged. After completing his business in each city, he will earn a fixed income of $a$ dollars and a variable income of $e$ dollars per day when staying in the city. During his entire trip, the expense is $b$ dollars per day. In this problem, the target is to identify the optimal traveling schedule that can maximize the profit.

Let us define $\balpha = (\alpha_1, \ldots, \alpha_k)$ as the sequence of cities visited, and the corresponding order sequence is $\bo = (o_1, \ldots, o_k)$, where City $\alpha_i$ is visited at order $o_{\alpha_i} = i$ ($i = 1, \ldots, k$). $\alpha_0 = 0$ is defined to be the starting point at time 0. The completion time of the business in City $\alpha_i$ is
$
C(\bx, \alpha_i) = \sum_{l=1}^{i}(s_{\alpha_{l-1}, \alpha_l} + x_{\alpha_l}),
$
and the tardiness (days \rf{passed} the due time) for City $\alpha_i$ is
$
T(\bx, \alpha_i) = max(0, C(\bx, \alpha_i) - d_{\alpha_i}).
$
Thus, the profit function that involves the days-staying-in-cities $\bx$ and the sequence-of-cities-visited $\bm{\alpha}$ (or equivalently $\bo$) is
$$
F(\bx, \balpha) = ka + e\sum_{i=1}^{k}x_i - bC(\bx, \alpha_k) - f\sum_{j=1}^{k}T(\bx, \alpha_j).
$$

\begin{example}\label{ex:Travel:BO}
	Consider the above TSP with $k = 8$ cities. Here, we set $a=20$, $e = 10$, $b=2$, $f=15$, due dates $(d_1,\ldots,d_8)=(26, 10, 42, 23, 25, 12, 44, 10)$, and staying days $x_i \in [1,4]$ for $i = 1, \ldots, 8$. The traveling time $s_{i,j}$ ($i<j$) is sampled from a uniform distribution $U(0.5,3)$, and set $s_{j,i} = (1+0.1\times\epsilon_{ji})s_{i,j}$, where
	$\epsilon_{ji}$ is sampled from the standard normal distribution.  Refer to Section~S4.2 of Supplementary Materials for additional details of this simulation.
\end{example}

\begin{figure}[ht]
\vspace{-.2in}
	\centering
	\subfigure []{\includegraphics[scale=0.4]{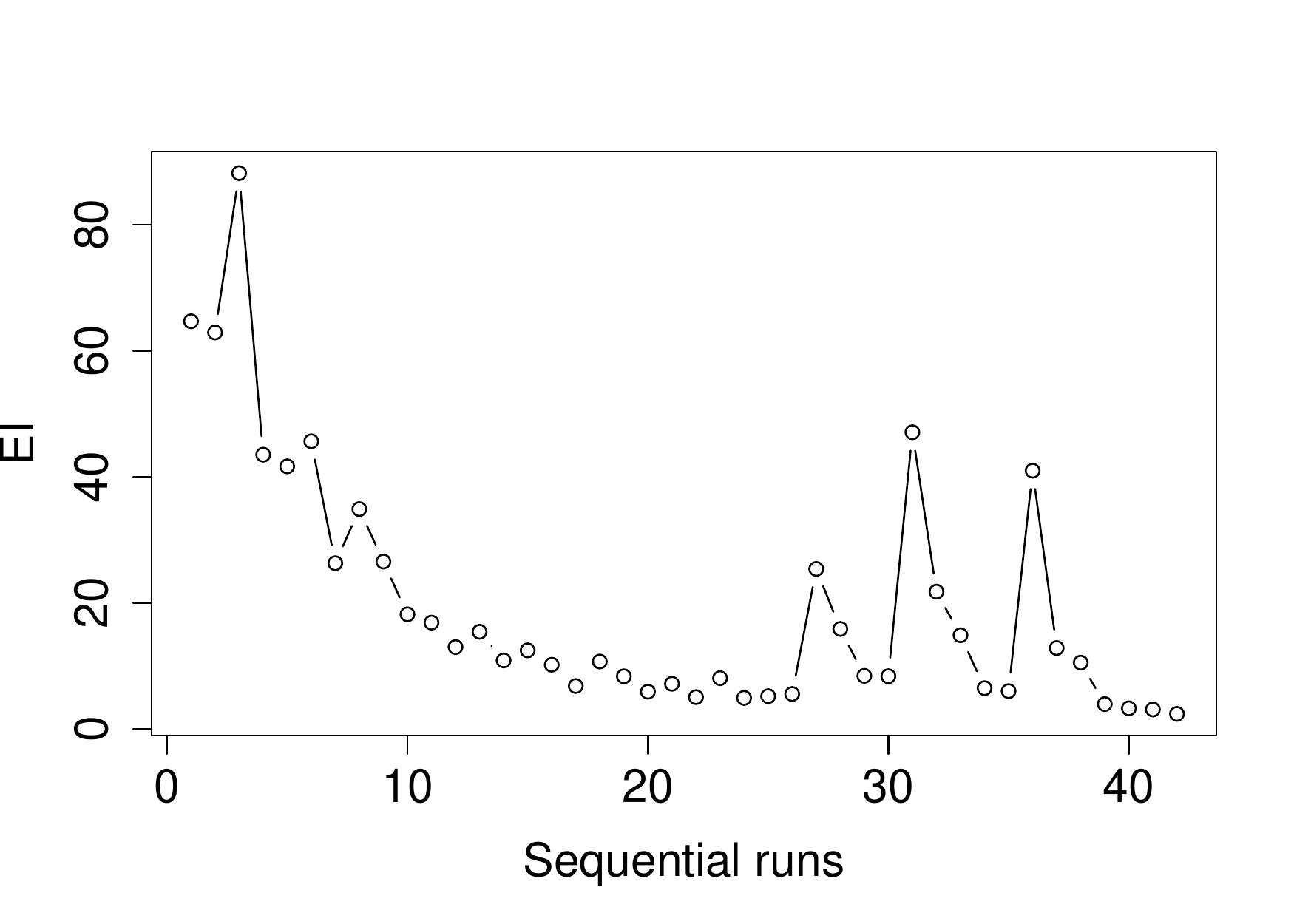}}\quad
	\subfigure [] {\includegraphics[scale=0.4]{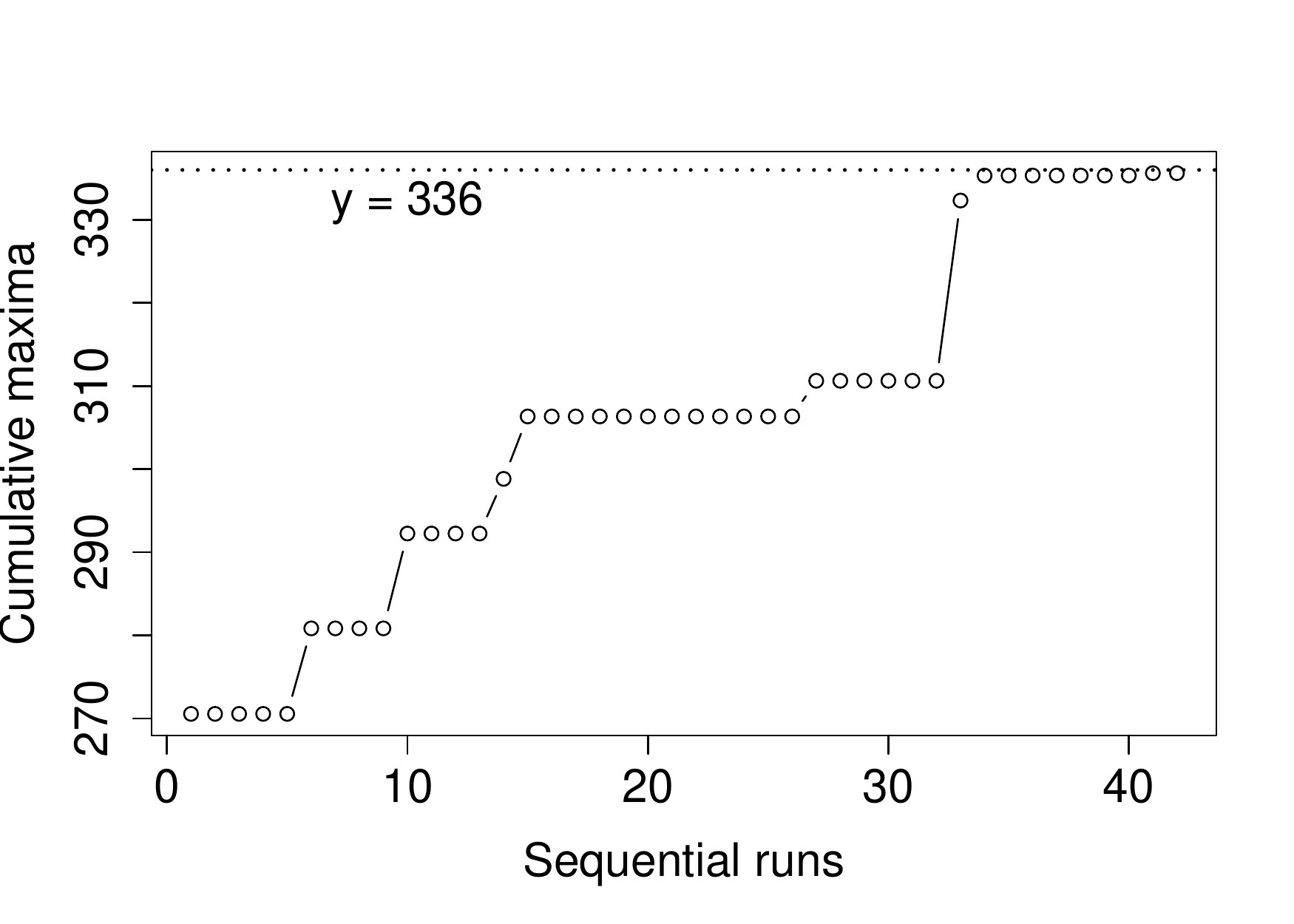}}
	\caption{Plots for (a) EIs and (b) cumulative maximum responses from the QS-learning approach under 2d-MaGP in Example \ref{ex:Travel:BO}.}
		\label{fig:ei5}
\end{figure}

Such a TSP does not have a known analytical solution. We consider the proposed QS-learning to identify \rf{the (nearly) optimal setting that will maximize} the profit function via a few experimental trials.
It starts from the 46-run (the rule-of-thumb run size illustrated in Section~\ref{sec: algebra construction}) QS-design  and selects 42 sequential runs under the  2d-MaGP model before the stopping criterion is satisfied.
The maximum response \rf{identified} is 336, which is found at the $41^{st}$ sequential run. The optimal setting includes
$\bx_* = (1.14, 3.44, 2.48, 2.86, 3.78, 4.00, 3.11, \\
4.00)$ and  
$\bo_*=(8,6,2,1,4,5,3,7)$.
Figure~\ref{fig:ei5} displays EIs and cumulative maximum responses of sequential runs.
After the $10^{th}$ sequential run, the ordinal parts in all the runs are either $\bo = (8,6,2,1,4,5,3,7)$ or $\bo = (8,6,2,1,4,5,7,3)$, both of which are good candidates.
Such an observation indicates the stability of QS-EGO (i.e., Algorithm~\ref{ag1} in Section \ref{sec:BO}).
In practice, 2d-MaGP is preferred over full-MaGP when many components are involved, because it is more computationally efficient.

To make comparisons, we first consider the random sampling approach ($BM_1$), which uses a large random sample of 4,032,000 observations, where a random Latin hypercube design is used for the quantitative part and a hundred replicates of all possible sequences are used for the sequence part.
The maximum response found is 325, which is clearly worse than that \rf{identified by the QS-learning by using only 88 runs}.
Next, we consider the sequential generalized PWO ($BM_2$) and CP ($BM_3$) approaches starting from random initial designs with \rf{the} required sizes (i.e., 37 and 38 runs, respectively). We replicate $BM_{2}$ and $BM_{3}$ 1000 times. Their average results are 254 and 264, and their best results are 321 and 324, respectively. Their performances are clearly inferior.

\begin{figure}[ht]
\vspace{-.2in}
	\centering
	\subfigure []{\includegraphics[scale=0.4]{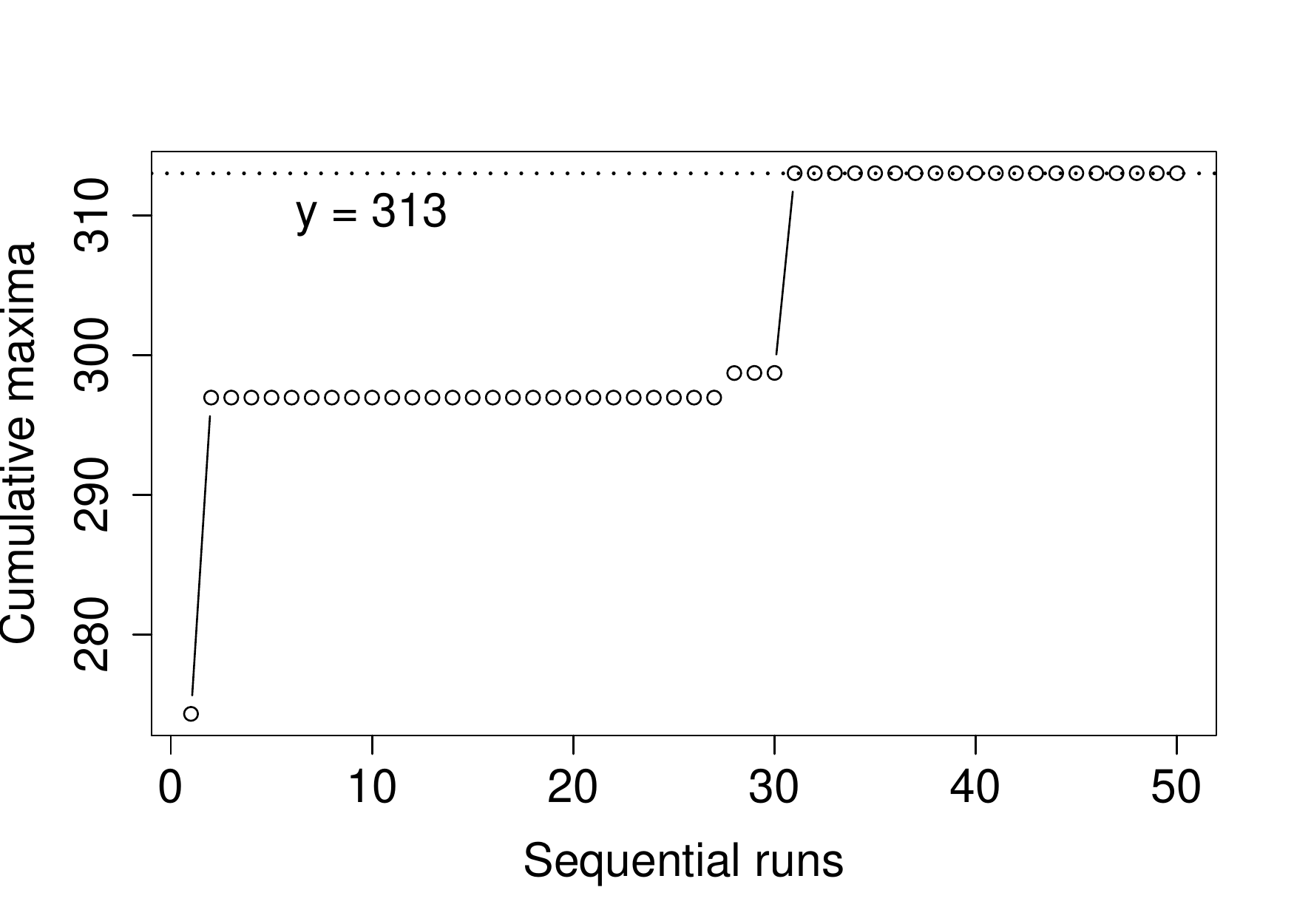}}\quad
	\subfigure [] {\includegraphics[scale=0.4]{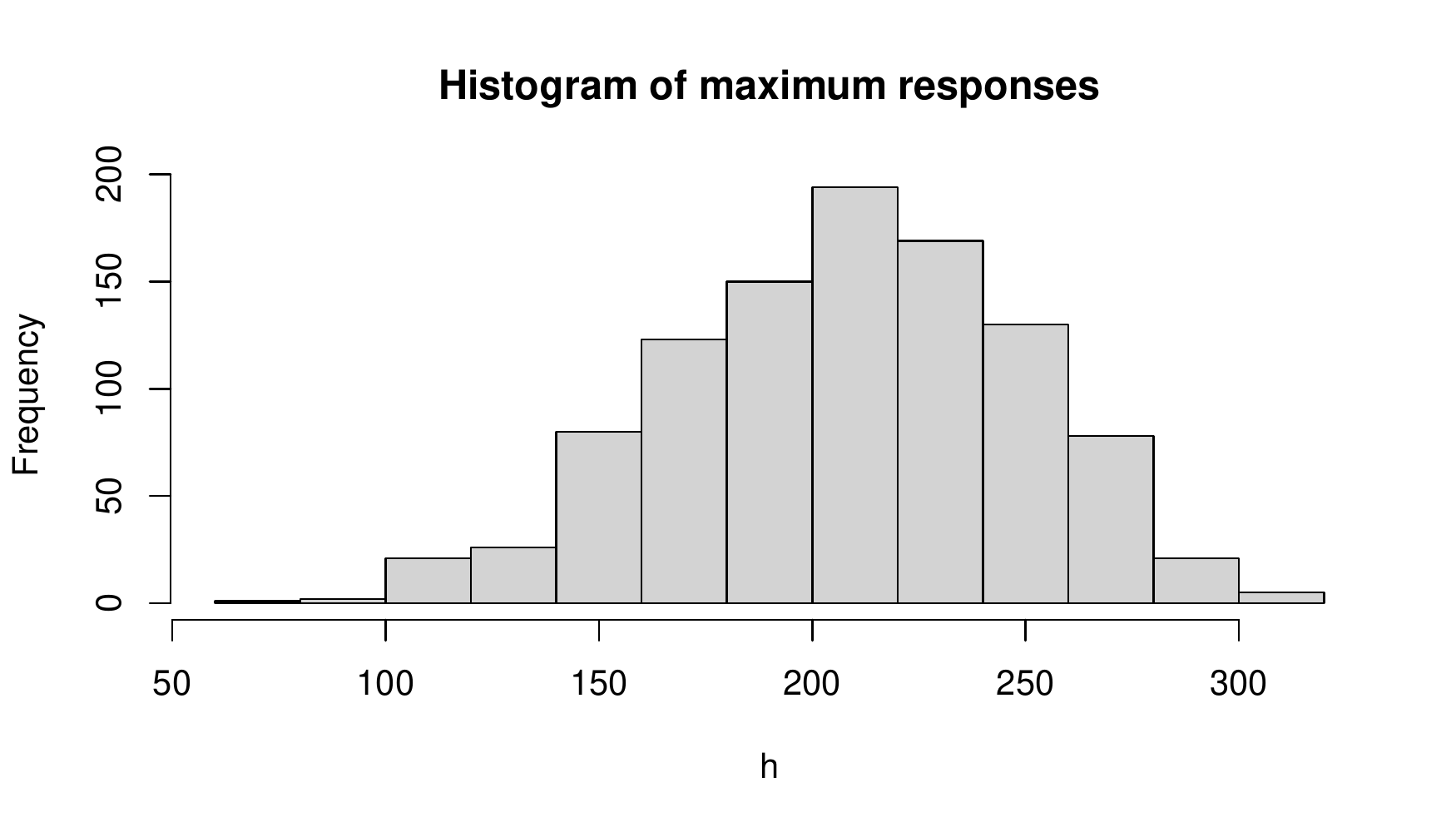}}
	\caption{Plots for (a) cumulative maximum responses from the fast QS-learning approach and (b) maximum responses in random samples of 96 \rf{trials} in Example \ref{ex:Travel:BO}.}
		\label{fig:tspfast}
\end{figure}

Finally, we evaluate the performance of the fast QS-learning approach introduced in Section~\ref{sec:fast}. It starts from the same QS-design as above, \rf{and we set a total time budget of 1 hour}. It includes 96 runs in total, where the adopted 2d-MaGP model is estimated for only 12 times. Its cumulative maximum responses are reported in Figure~\ref{fig:tspfast}(a).
On average, each sequential run takes about 1 minute, while the general QS-learning takes about 4 minutes in this example.
The maximum response found here is 313, which is still much better than the average results of $BM_{2}$ and $BM_{3}$ \rf{(i.e., 254 and 264, respectively)}.
In addition, we show the histogram of maximum responses found from 1000 random samples of 96 trials in Figure~\ref{fig:tspfast}(b), where the average value is 208 and the largest value is 309.
Evidently, the fast QS-learning approach appears to exhibit  reasonably good performance.

\section{Discussion}
\label{dis}
In this work, we propose an active learning approach to identify (nearly) optimal solutions for experiments with QS factors. Analyzing such experiments is challenging due to their semi-discrete and possibly extremely large solution spaces and complex input–output relationships. The proposed QS-learning includes a novel MaGP surrogate model, an efficient sequential scheme (QS-EGO), and a new class of optimal experimental designs (QS-designs), providing a systematic solution for analyzing QS experiments. The theoretical properties of the proposed method are investigated, and techniques for optimization by using analytical gradients are developed. A case study on lymphoma treatment and several simulations are presented to illustrate the advantages of the proposed method.

In this work, we focus on the widely used EI framework, which works well empirically. In the current literature, the upper confidence bound (UCB) is another popular framework for working with GP models in active learning, particularly for purely discrete input spaces \citep{srinivas2009gaussian, djolonga2013high, JMLR:v20:18-213, vakili2020regret}. Results on convergence rates have been established for GP-UCB and its variants. For example, \cite{srinivas2009gaussian} proved a cumulative regret bound of $n^{(\nu + d(d+1))/(2\nu+d(d+1))}$ by using a Matern kernel of smoothness $\nu$ on a $d$-dimensional space. \cite{vakili2020regret} further improved the bound to $O(n^{(d-\nu)/d})$ for $d > \nu$, $O(\log (n))$ for $d=\nu$, and some constants for $d < \nu$. All these results require finite or general compact input spaces, but the space in the QS-experiment (a joint one of \rf{continuous and sequence spaces}) does not satisfy this requirement. 
Considering QS-learning under the UCB framework and studying its convergence rate will be interesting topics for future research.

Embedding some of the currently popular non-separable covariance functions \citep{gneiting2002nonseparable} into the MaGP model is also interesting. For example, one may consider
$$
\hbox{Cov} ( Y(\bx_i, \bo_i), Y(\bw_j, \bo_j)) = \frac{\sigma^2}{\psi ( \vert \vert \bo_i - \bo_j  \vert \vert ^2)^{1/d}} \varphi \left( \frac{\vert \vert \bx_i - \bx_j \vert \vert^2}{\psi ( \vert \vert \bo_i - \bo_j  \vert \vert ^2)} \right),
$$
where $\vert \vert \cdot \vert \vert$ is the $L_2$ norm (or other norms), $\varphi(\cdot)$ is a completely monotonic function (e.g., $\varphi(t) = \exp (-ct^\gamma)$), and $\psi(t)$ is a positive function with a completely monotonic derivative (e.g. $\psi(t) = (at^{\alpha} + 1)^{\beta}$). 
Such a structure may have an interpretation for certain choices of $\varphi$ and $\psi$ \citep{haslett1989space}.
A planned future study can investigate the appropriate choices for $\varphi$ and $\psi$ and their parameters to explain QS experiments. 

In active learning and other non-sequential learning methods,
optimal designs for experiments with QS factors are important but not well addressed.
In this work, we propose criteria for QS-designs that consider designs' space-filling and pair-balanced properties.
The current literature presents various types of space-filling designs, including maximin distance designs \citep{Johnson1990, xiao2017}, minimax distance designs \citep{chen2015minimax}, uniform designs \citep{fang2003uniform}, MaxPro designs \citep{Joseph15} and uniform projection designs \citep{SunWangXu19}, which can all be used as the quantitative part in the QS-design.
For the sequence part, desirable properties beyond the pair balance can be studied analogous to component orthogonal arrays \citep{yang2018}, order-of-addition orthogonal arrays \citep{voelkel2019design} and optimal fractional PWO designs \citep{peng2017, CHEN2020108728}. Moreover, some desirable structures that connect the quantitative and sequence parts of QS-designs can be investigated \citep{deng2015design}.

\newpage

\section{Supplementary Materials}
\subsection{Technical Details}
\subsubsection{Proofs}
\textbf{Proof for Theorem ~1 and Corollaries~1 and 2.}
It is known that Gaussian kernel $K_{i,j} = \sigma^2 \exp \left(-\theta(x_i-x_j)^2\right)$ is a function of $x_i-x_j$ and is isotropic.  According to the Bochner's theorem (Theorem 4.1 in \cite{Rasmussen2006}) and Wiener-Khintchine theorem \citep{chatfield2003analysis}, showing the positive semidefiniteness of  Gaussian kernel can reduce to finding its spectral density $S(s)$ such that $K_{i,j} = \int S(s)\exp(2 \pi is (x_1-x_2)) ds$. As the Fourier transform of a Gaussian is another Gaussian, we have $S(s) = \sigma^2(\pi / \theta)^{1/2}\exp(-\pi^2s^2/\theta)$. Such a  conclusion on positive semidefiniteness can be easily generalized to any vector $\bx \in \mathrm{R}^D$ where $S(s) = \sigma^2(\pi / \theta)^{D/2}\exp(-\pi^2s^2/\theta)$.
If we view the inputs $\bw$ in the covariance function $\phi_h$ in (7) as a new input $(\bx, \tilde{o}^{(1)}, \ldots, \tilde{o}^{(t)} )$ with correlation parameters for $\tilde{o}$ being 1, the corresponding covariance matrix $\Phi_h = (\phi_h(\bw_i, \bw_j \vert \sigma_h^2, \theta_h, \bdelta))_{n \times n}$ is positive semi-definite for $h=1,\ldots,k$. As the diagonal matrix $\tau^2 \textbf{1}(\bw_i = \bw_j)$ is positive semi-definite and the sum of positive semi-definite matrices are still positive semi-definite, the covariance matrix defined by (10) is positive semi-definite, which completes the proof. 

If $\tau^2 \neq 0$, the diagonal matrix $\tau^2 \textbf{1}(\bw_i = \bw_j)$ is positive definite. Since the sum of a positive definite matrix and  some positive semi-definite matrices is  positive definite, the covariance matrix in Corollary~1 is positive definite. 

When $\tau^2 = 0$, if there are no two runs having the same quantitative inputs (that is $\bx_i \neq \bx_j$ for $i \neq j$), all matrices $\Phi_h$ ($h=1,\ldots,k$) are positive definite. Thus, the covariance matrix in Corollary~2 is positive definite.

\vskip 5pt
\noindent {\bf Proof of Theorem~2}
(i) Because $p_r$ is an odd prime number, for any $1\leq x \leq n$, the set $\{ i \times x \text{ mod } p_r : i = 1, \ldots, n \}$ equals $\{1, \ldots, n \}$. Therefore, each column of $O^{'}$ is a permutation of 1 to $n$, and thus the Hamming distance between any two distinct rows of $O^{'}$ is $n$, which is clearly the optimal.

(ii) For any $1\leq i \neq j \leq k = p_r -1$, let $\alpha = (j-i) \text{ mod } p_r$. Consider the $\alpha^{th}$ row in $O^{'}$. By the construction of $D_{glp}$, the $\alpha^{th}$ row is $\alpha \times \bh \text{ mod } p_r$ which is a permutation of 1 to $n$.
The first and last elements in the $\alpha^{th}$ row are $\alpha$ and $p_r-\alpha$, respectively. Here, $i$ cannot equal $p_r-\alpha$; otherwise, it will lead to $j=0 \text{ mod } p_r$, which is a contradiction. Thus, the sub-sequence ``(i j)" must occur in the $\alpha^{th}$ row of $O^{'}$.
Next, we show that the sub-sequence ``(i j)" can only occur in the $\alpha^{th}$  row of $O^{'}$. If it does not, suppose that ``(i j)" also occurs in the ${\alpha'}^{th}$ row with $1\leq \alpha' \neq \alpha \leq n$. Since $j$ is  adjacent to $i$ in the $\alpha'^{th}$ row,
$j = i + \alpha' = i + \alpha$, which leads to the contradiction ($\alpha' = \alpha$). Therefore,  $t_{i,j} = 1$ for any $1\leq i \neq j \leq k$, and $O^{'}$ is pair-balanced.

(iii) By (i) and (ii), it is straightforward to give the formula for $\nu_p (O^{'})$  in (19). To show that $\nu_p (O^{'})$ achieves the lower bound among all possible designs, we only need to prove that for any other design $\tilde O^{'}$,
$$
\sum_{i=2}^n \sum_{j=1}^{i-1} \frac{1}{(h_{i,j}+1)^p}  \geq  \frac{n(n-1)}{2(n+1)^p}
$$
and
$$
\underset{i \neq j}{\sum_{i=1}^k \sum_{j=1}^{k}} \frac{1}{(t_{i,j}+1)^p} \geq \frac{k(k-1)}{2^p} = \frac{n(n-1)}{2^p}.
$$
The first inequality is obvious as $h_{i,j} \leq n$ for any two rows. The second inequality follows from the fact that
$
\underset{i \neq j}{\sum_{i=1}^k \sum_{j=1}^{k}} (t_{i,j}+1) = 2k(k-1)
$
and
$$
\underset{i \neq j}{\sum_{i=1}^k \sum_{j=1}^{k}} \frac{1}{(t_{i,j}+1)^p} \geq \underset{i \neq j}{\sum_{i=1}^k \sum_{j=1}^{k}} \frac{1}{\left[\frac{2k(k-1)}{k(k-1)}\right]^p} = \underset{i \neq j}{\sum_{i=1}^k \sum_{j=1}^{k}} \frac{1}{2^p}.
$$

\vskip 5pt
\noindent {\bf Proof of Theorem~3}
(i) We first consider the pairwise $L_2$-distances between rows in $X$. Since column permutations do not change the $L_2$-distances, WLOG, we take $X = D_{glp}$.
Denote the minimum $L_2$-distance of $D$ as $d(D) = \min\{ d_{i,j} : 1\leq i < j \leq n \}$.
Based on the Theorems~1 and 4 in \cite{zhou2015space}, we can prove that
 $$ d(X) =
 \left(\sum_{i=1}^{k} \min\{i, p_r-i\}^2\right)^{\frac{1}{2}}=   \left(\frac{1}{12} k (k+1) (k+2) \right)^{\frac{1}{2}} = \left(\frac{1}{12} n (n+1) (n+2) \right)^{\frac{1}{2}},
 $$
which completes the proof.

(ii) By Theorem~2, $O^{'}$ is a Hamming equidistant design with $h_{i,j} = k$ for any $1\leq i \neq j \leq n$. To get the upper bound of $C_p(D^{'})$, the key is the  distribution of all pairwise $L_2$-distances of the design $X$.
For any $1\leq i \neq j \leq n = p_r -1$, consider the  $i^{th}$ and $j^{th}$ rows of $X$ where $\bx_i = i \times \bh \text{ mod } p_r$ and $\bx_j = j \times \bh \text{ mod } p_r$. As $p_r$ is an odd prime, there exists a unique number $l$ ($1\leq l \leq p_r-1$) such that $ j = i \times l  \text{ mod } p_r$. This means that
the $j^{th}$ row can also be written as
$ l \times (i \times \bh) \text{ mod } p_r. $
Since each row is a permutation of $(1,\ldots,n)$,
$(\bx_i^\TT, \bx_j^\TT)^\TT$ is the same as $(\bx_1^\TT, \bx_l^\TT)^\TT = (\bh^\TT, \bx_l^\TT)^\TT$ up to some column permutation.
Thus, $d_{i,j}$ equals  $d_{1,l}$. For all $n(n-1)/2$ possible distances $d_{i,j}$, it suffices to only consider the  $(n-1)$ different $d_{1,l}$ with $l=2,\ldots,n$.
In addition, for $1\leq l \leq n=p_r-1$ and $l\neq p_r-1$, there exists a unique number $1\leq l' \leq n$ ($l' \neq l$)
such that $l \times l' = 1 \text{ mod } p_r $. Thus,   $(\bx_1^\TT, \bx_l^\TT)^\TT$ is the same as $ (\bx_{l'}^\TT, \bx_1^\TT)^\TT = l'(\bx_1^\TT, \bx_l^\TT)^\TT \text{ mod } p_r $. Then, all possible $d_{i,j}$ take at most $n/2$ different values, which are in the set
$ \{ d_{1,i_1},\ldots,d_{1,i_{n/2-1}}, d_{1,n}\}$
where $i_1,\ldots,i_{n/2-1}$ are in $\{2,\ldots,n-1\}$ such that the product of any two of them does not equal $1 \text{ mod } p_r $. Among all $d_{i,j}$, there are $n$ of them take the value $d_{1,i_j}$ for $j=1,\ldots,n/2-1$ and $n/2$ of them take the value of $d_{1,n}$.
WLOG, let $i_1 = 2$, we have
$d_{1,2} =  \left(\frac{1}{12} n (n+1) (n+2) \right)^{\frac{1}{2}},$
which equals the minimum $L_2$-distance $d(X)$.
Additionally, we can show that
$$d_{1,n} =\left(\sum_{i=1}^{k} [i - (p_r-i)]^2\right)^{\frac{1}{2}} =  \left( \frac{1}{3} n \left(n^2-1\right) \right)^{\frac{1}{2}}. $$
Denote $d_{1,i_2},\ldots,d_{1,i_{n/2-1}}, d_{1,n}$ as $ d_2,\ldots, d_{n/2-1}$, respectively.
We can write the  distributions of all pairwise $L_2$-distances of $X$ as
$$
 \begin{array}{c|ccccc}\hline
   \text{distance} &  \sqrt{ \frac{1}{12} n (n+1) (n+2) } & d_2 & \ldots & d_{n/2-1} & \sqrt{\frac{1}{3} n \left(n^2-1\right) } \\
   \text{frequency} & n & n & \ldots & n & n/2  \\ \hline
 \end{array}
$$
With this distribution and by (18), we can derive an upper bound
\begin{eqnarray*}
 C_p(D^{'}) &=& \left(\sum_{i=2}^n \sum_{j=1}^{i-1} \frac{1}{(\rho_1^{'}d_{i,j} + n \rho_2^{'} + 1)^p} \right)^\frac{1}{p}\\
 & = & n^\frac{1}{p} \Bigg( \sum_{i=2}^{n/2-1} \frac{1}{(\rho_1^{'}d_{i} + n \rho_2^{'} + 1)^p}  +  \frac{1}{ \left[ \rho_1^{'} \left(\frac{1}{12} n (n+1) (n+2) \right)^{\frac{1}{2}} + n \rho_2^{'} + 1\right]^p} \\
 & & +   \frac{1}{ 2 \left[  \rho_1^{'} \left( \frac{1}{3} n \left(n^2-1\right) \right)^{\frac{1}{2}}   + n \rho_2^{'} + 1\right]^p}    \Bigg)^\frac{1}{p} \\
 & \leq & n^\frac{1}{p}  C(n,\rho_1^{'},\rho_2^{'},p) ,
\end{eqnarray*}
where $$
C(n,\rho_1^{'},\rho_2^{'},p) =\left( \frac{n/2-1}{ \left[ \rho_1^{'} \left(\frac{1}{12} n (n+1) (n+2) \right)^{\frac{1}{2}} + n \rho_2^{'} + 1\right]^p}+ \frac{1}{ 2 \left[  \rho_1^{'} \left( \frac{1}{3} n \left(n^2-1\right) \right)^{\frac{1}{2}}   + n \rho_2^{'} + 1\right]^p} \right)^\frac{1}{p}.
$$

\subsubsection{Analytical Gradients for Parameter Estimation in MaGP}
\label{magpgrad}

Following the notations in Section~3 and based on (11), maximizing the likelihood is equivalent to minimizing the function:
$$
f = \log \vert \bm{\Phi} \vert + (\by - \widehat{\mu})^\TT \bm{\Phi}^{-1}(\by - \widehat{\mu}),
$$
where
$
\widehat{\mu} = (\textbf{1}^\TT\bm{\Phi}^{-1}\textbf{1})^{-1}\textbf{1}^\TT\bm{\Phi}^{-1}\textbf{y}
$.
For any parameter inside $\bm{\Phi}$, the expression of the analytical gradient given $\widehat{\mu}$ is:
$$
\frac{\partial f}{\partial \bullet} = tr(\bm{\Phi}^{-1}\frac{\partial \bm{\Phi}}{\partial \bullet}) - (\by - \widehat{\mu})^\TT \bm{\Phi}^{-1} \frac{\partial \bm{\Phi}}{\partial \bullet} \bm{\Phi}^{-1}(\by - \widehat{\mu}).
$$
In the MaGP model with covariance function in (10), for any $i,j = 1, \ldots, n$ and $h = 1, \ldots, k$,  we have:
$$
\frac{\partial \bm{\Phi}}{\partial \sigma_h^2} = \left( \frac{\partial \phi(\bw_i, \bw_j)}{\partial \sigma_h^2} \right)_{n \times n},
\frac{\partial \bm{\Phi}}{\partial \theta_h} = \left( \frac{\partial \phi(\bw_i, \bw_j)}{\partial \theta_h} \right)_{n \times n},
$$
$$
\frac{\partial \phi(\bw_i, \bw_j)}{\partial \sigma_h^2} = \hbox{exp} \big \{- \theta_h (x_{i,h} - x_{j,h})^2 \big \}
\hbox{exp} \big\{ -\sum_{l=1}^{t} (\tilde{o}_{i,h}^{(l)} - \tilde{o}_{j,h}^{(l)})^2 \big\},
$$
$$
\frac{\partial \phi(\bw_i, \bw_j)}{\partial \theta_h} = - \sigma_h^2 (x_{i,h} - x_{j,h})^2   \hbox{exp} \big \{- \theta_h (x_{i,h} - x_{j,h})^2 \big \}
\hbox{exp} \big\{ -\sum_{l=1}^{t} (\tilde{o}_{i,h}^{(l)} - \tilde{o}_{j,h}^{(l)})^2 \big\}.
$$
First, we consider the analytical gradients of  mapping parameters $\delta_1, \ldots, \delta_{2k-3}$ in the 2d-MaGP ($t=2$) model, where we rearrange the $2k-3$ mapping parameters in (5) by rows. For any $1 \leqslant s \leqslant 2k-3$, define two indicator functions ($h = 1, \ldots, k$):

\[
I_h^{(1)}(\bw_i, \bw_j) = \left\{
\begin{array}{ll}
1 \hbox{ when } o_{i,h} \neq o_{j,h} \hbox{ and } o_{i,h} = \lceil \frac{s-1}{2} \rceil +2 \nonumber \\
0 \hbox{ otherwise } \nonumber
\end{array}
\right.
,
\]
\[
I_h^{(2)}(\bw_i, \bw_j) = \left\{
\begin{array}{ll}
1 \hbox{ when } o_{i,h} \neq o_{j,h} \hbox{ and } o_{j,h} = \lceil \frac{s-1}{2} \rceil +2 \nonumber \\
0 \hbox{ otherwise } \nonumber
\end{array}
\right.
.
\]
It will be straightforward to prove that $\delta_s$ corresponds to the order-level $\lceil \frac{s-1}{2} \rceil +2$ where $\lceil x \rceil$ is the least integer that is no less than $x$. The indicator functions $I_h$ are used to judge where parameters $\delta_s$ are used. Thus, we have
$
\partial \bm{\Phi}/\partial \delta_s = \left( \partial \phi(\bw_i, \bw_j) / {\partial \delta_s} \right)_{n \times n}
$, and
\begin{eqnarray*}
	& & \frac{\partial \phi(\bw_i, \bw_j)}{\partial \delta_1} = \\
	& & -2 \sum_{h=1}^{k}  \sigma_h^2 (\delta_1 - \tilde{o}_{j,h}^{(1)}) \hbox{exp} \big \{- \theta_h (x_{i,h} - x_{j,h})^2 \big \}
	\hbox{exp} \big\{ -\sum_{l=1}^{2} (\tilde{o}_{i,h}^{(l)} - \tilde{o}_{j,h}^{(l)})^2 \big\} I_h^{(1)}(\bw_i, \bw_j)\\
	& & -2 \sum_{h=1}^{k}  \sigma_h^2 (\delta_1 - \tilde{o}_{i,h}^{(1)}) \hbox{exp} \big \{- \theta_h (x_{i,h} - x_{j,h})^2 \big \}
	\hbox{exp} \big\{ -\sum_{l=1}^{2} (\tilde{o}_{i,h}^{(l)} - \tilde{o}_{j,h}^{(l)})^2 \big\} I_h^{(2)}(\bw_i, \bw_j) \\
	& & = -2  \sum_{h=1}^{k}  \sigma_h^2 \hbox{exp} \big \{- \theta_h (x_{i,h} - x_{j,h})^2 -\sum_{l=1}^{2} (\tilde{o}_{i,h}^{(l)} - \tilde{o}_{j,h}^{(l)})^2 \big\} \big ( (\delta_1 - \tilde{o}_{j,h}^{(1)}) I_h^{(1)}(\bw_i, \bw_j) \\
	& & + (\delta_1 - \tilde{o}_{i,h}^{(1)}) I_h^{(2)}(\bw_i, \bw_j) \big ),
\end{eqnarray*}
\begin{eqnarray*}
	& & \frac{\partial \phi(\bw_i, \bw_j)}{\partial \delta_s} = \\
	& & -2 \sum_{h=1}^{k}  \sigma_h^2 (\delta_s - \tilde{o}_{j,h}^{(t(s))}) \hbox{exp} \big \{- \theta_h (x_{i,h} - x_{j,h})^2 \big \}
	\hbox{exp} \big\{ -\sum_{l=1}^{2} (\tilde{o}_{i,h}^{(l)} - \tilde{o}_{j,h}^{(l)})^2 \big\} I_h^{(1)}(\bw_i, \bw_j)\\
	& & -2 \sum_{h=1}^{k}  \sigma_h^2 (\delta_s - \tilde{o}_{i,h}^{(t(s))}) \hbox{exp} \big \{- \theta_h (x_{i,h} - x_{j,h})^2 \big \}
	\hbox{exp} \big\{ -\sum_{l=1}^{2} (\tilde{o}_{i,h}^{(l)} - \tilde{o}_{j,h}^{(l)})^2 \big\} I_h^{(2)}(\bw_i, \bw_j) \\
	& & = -2  \sum_{h=1}^{k}  \sigma_h^2 \hbox{exp} \big \{- \theta_h (x_{i,h} - x_{j,h})^2 -\sum_{l=1}^{2} (\tilde{o}_{i,h}^{(l)} - \tilde{o}_{j,h}^{(l)})^2 \big\} \big ( (\delta_s - \tilde{o}_{j,h}^{(t(s))}) I_h^{(1)}(\bw_i, \bw_j) \\
	& & + (\delta_s - \tilde{o}_{i,h}^{(t(s))}) I_h^{(2)}(\bw_i, \bw_j) \big ),
\end{eqnarray*}
where $s = 2, \ldots 2k-3$, and $t(s) = 1$ for even $s$ and $t(s) = 2$ for odd $s$.

Next, we consider 
the analytical gradients of mapping parameters $\delta_1, \ldots, \delta_{k(k-1)/2}$ in the
full-MaGP model ($t = k-1$), where we rearrange the $k(k-1)/2$ mapping parameters in (5) by rows. For any $1 \leqslant s \leqslant k(k-1)/2$, define two indicator functions ($h = 1, \ldots, k$):

\[
I_h^{(1)}(\bw_i, \bw_j) = \left\{
\begin{array}{ll}
1 \hbox{ when } o_{i,h} \neq o_{j,h} \hbox{ and } o_{i,h} = l_s \nonumber \\
0 \hbox{ otherwise } \nonumber
\end{array}
\right.
,
\]
\[
I_h^{(2)}(\bw_i, \bw_j) = \left\{
\begin{array}{ll}
1 \hbox{ when } o_{i,h} \neq o_{j,h} \hbox{ and } o_{j,h} = l_s \nonumber \\
0 \hbox{ otherwise } \nonumber
\end{array}
\right.
.
\]
where $\delta_s$ corresponds to the order-level $l_s$ such that $ (l_s-1)(l_s-2)/2 < s \leqslant l_s(l_s-1)/2$. Similar to the above, we can show that

\begin{eqnarray*}
	& & \frac{\partial \phi(\bw_i, \bw_j)}{\partial \delta_s} = \\
	& & -2 \sum_{h=1}^{k}  \sigma_h^2 (\delta_s - \tilde{o}_{j,h}^{(t(s))}) \hbox{exp} \big \{- \theta_h (x_{i,h} - x_{j,h})^2 \big \}
	\hbox{exp} \big\{ -\sum_{l=1}^{2} (\tilde{o}_{i,h}^{(l)} - \tilde{o}_{j,h}^{(l)})^2 \big\} I_h^{(1)}(\bw_i, \bw_j)\\
	& & -2 \sum_{h=1}^{k}  \sigma_h^2 (\delta_s - \tilde{o}_{i,h}^{(t(s))}) \hbox{exp} \big \{- \theta_h (x_{i,h} - x_{j,h})^2 \big \}
	\hbox{exp} \big\{ -\sum_{l=1}^{2} (\tilde{o}_{i,h}^{(l)} - \tilde{o}_{j,h}^{(l)})^2 \big\} I_h^{(2)}(\bw_i, \bw_j) \\
	& & = -2  \sum_{h=1}^{k}  \sigma_h^2 \hbox{exp} \big \{- \theta_h (x_{i,h} - x_{j,h})^2 -\sum_{l=1}^{2} (\tilde{o}_{i,h}^{(l)} - \tilde{o}_{j,h}^{(l)})^2 \big\} \big ( (\delta_s - \tilde{o}_{j,h}^{(t(s))}) I_h^{(1)}(\bw_i, \bw_j) \\
	& & + (\delta_s - \tilde{o}_{i,h}^{(t(s))}) I_h^{(2)}(\bw_i, \bw_j) \big ),
\end{eqnarray*}
where $s = 1, \ldots k(k-1)/2$ and $t(s) = s-(l_s-1)(l_s-2)/2$.
Finally, it will be straightforward to generalize the above results on analytical gradients to other  $t$-dimensional MaGP model ($t=3,\ldots, k-2$).

\subsubsection{Analytical Gradients for Optimizing  Expected Improvements}
\label{eigrad}
Following the notations and analytical results in Sections~3 and 4, the expected improvement in (16) at the input $\bw_* = (\bx_*, \bo_*) $ can be expressed in the closed form as
$$
E[I(\bw_*)] = (y_{\min }^{(n)}-\widehat{Y}(\bw_*)) \Phi \left( \frac{ y_{\min }^{(n)}-\widehat{Y}(\bw_*) }{s(\bw_*)} \right) + s(\bw_*) \varphi \left( \frac{ y_{\min }^{(n)}-\widehat{Y}(\bw_*) }{s(\bw_*)} \right)
$$
for the minimization scenario.   Using
the relations $\Phi^{'} = \phi$ and $\phi(t)^{'} = -t\phi(t)$, the gradient of $E[I(\bw_*)]$ can be derived as:
\begin{equation*}
\nabla E[I(\bw_*)] = - \nabla \widehat{Y}(\bw_*) \times \Phi \left( \frac{ y_{\min }^{(n)}-\widehat{Y}(\bw_*) }{s(\bw_*)} \right) + \nabla s(\bw_*) \times \varphi \left( \frac{ y_{\min }^{(n)}-\widehat{Y}(\bw_*) }{s(\bw_*)} \right).
\end{equation*}
Based on (14) and (15) and using the fact that $\nabla s^2(\bw_*)  = 2s(\bw_*) \nabla s(\bw_*)$, we have:
$$
\nabla \widehat{Y}(\bw_*) =
\nabla \bm{\gamma}^\TT \bm{\Phi}^{-1}(\by - \widehat{\mu} \textbf{1}),
$$ 
and
$$
\nabla s(\bw_*) = -\frac{1}{s(\bw_*)} \left( \nabla \bm{\gamma}^\TT \bm{\Phi}^{-1} \bm{\gamma} + \frac{(1 - \textbf{1}^\TT \bm{\Phi}^{-1}\bgamma)\nabla \bm{\gamma}^\TT \bm{\Phi}^{-1} \textbf{1} }{\textbf{1}^\TT \bm{\Phi}^{-1} \textbf{1}} \right).
$$
Given the sequence part $\bo_*$ in $\bw_*$ and all  estimated parameters in the MaGP model, we can derive that (for $h = 1, \ldots, k$ and $i = 1, \ldots, n$)
$$
\frac{\nabla \phi(\bw_{*}, \bw_i)}{\nabla x_{*,h}} = -2\theta_h\sigma_h^2(x_{*,h} - x_{i,h})\hbox{exp} \big\{ - \theta_h (x_{*,h} - x_{i,h})^2 -\sum_{l=1}^{t} (\tilde{o}_{*,h}^{(l)} - \tilde{o}_{i,h}^{(l)})^2 \big\}.
$$
Thus, we have $$\frac{\nabla \bm{\gamma}^\TT}{\nabla x_{*,h}} = \left(\frac{\nabla \phi(\bw_{*}, \bw_1)}{\nabla x_{*,h}}, \ldots, \frac{\nabla \phi(\bw_{*}, \bw_n)}{\nabla x_{*,h}} \right),$$ and 
$$\nabla \bm{\gamma}^\TT = \left((\frac{\nabla \bm{\gamma}^\TT}{\nabla x_{*,1}})^\TT, \ldots, (\frac{\nabla \bm{\gamma}^\TT}{\nabla x_{*,k}})^\TT \right)^\TT.$$
It will be straightforward to adapt this proof for the maximization scenario.

\subsection{On the Convergence for Learning QS Experiments}

Here, we look at the worst case scenario via studying the convergence rate.
We start from considering the setting for computer experiment, i.e. $\tau^2 =0$ in (10), and the \emph{QS-learning with naive EI}, where all possible sequences can be enumerated and the EI strategy is applied to the quantitative factors for each sequence. It is known that the EI strategy does not converge for all cases; see counter-examples in \cite{locatelli1997bayesian} and \cite{bull2011convergence}. \cite{bull2011convergence} showed that the EI strategy converges under Matern kernels when all parameters are fixed. The following Remark~\ref{remark1} directly follows Theorem 2 in \cite{bull2011convergence} under the same assumptions, since there is a finite number ($k!$) of sequences and the convergence rate (in terms of $n$) for only quantitative factors will keep the same as that in the \emph{QS-learning with naive EI}.

\begin{remark}
\label{remark1}
For QS experiments with $k$ components, the aforementioned {QS-learning with naive EI} under given model parameters converges at least at rate of $n^{- \text{min}(\nu, 1) / k}$ up to log factors, where $\nu$ is the smoothness parameter of Matern kernels in (8).
\end{remark}
Specifically, by Theorem 2 in \cite{bull2011convergence}, the rate is $O(n^{-1/d})$ for $\nu > 1$ or  $O(n^{-\nu/d} (\log n)^{\alpha})$ for $\nu \leq 1$ where $\alpha$ is some constant. In the fast QS learning procedure, if the estimated parameters can reach their true values and all possible sequences can be enumerated, it will converge to the global optimum by Remark~\ref{remark1} .


In practice, model parameters are estimated from the data and the smoothness parameter ($\nu$) specified in the Matern kernel may be different from the underlying smoothness ($\nu_f$) in the simulation model \citep{ wynne2021convergence}. In \cite{wenzel2021novel} and \cite{wynne2021convergence}, a $\gamma$-stabilized algorithm is developed for quantitative factors with domain $\mathcal{X}$ and any acquisition function (including EI) $F: \mathcal{X} \rightarrow \mathcal{R}$. When choosing the $(n+1)^{th}$ run based on current $n$ observations, it searches over $\mathcal{X}_{n, \gamma} = \left\{x \in \mathcal{X}: P_n(x) \geqslant \gamma \vert \vert P_n(x) \vert \vert_{L \infty} \right\}$, where $\gamma \in (0,1] $ and $P_n(x)$ is the square root of the posterior variance. That is, such a strategy only allows points to be selected from areas of non-trivial prediction variances; see \cite{wenzel2021novel} or \cite{wynne2021convergence} for detailed discussions. 
Here, we denote the \emph{QS learning with stabilized EI} as the method where all possible sequences are enumerated and this $\gamma$-stabilized algorithm using EI acquisition function is applied to the quantitative factors for each sequence. The posterior variance used in $P_n(x)$ is given by (15).  By Theorem~11 in \cite{wynne2021convergence}, it is straightforward to show the following Remark~\ref{remark2}.
 \begin{remark}
  \label{remark2}
For QS experiments with $k$ components, the aforementioned QS learning with stabilized EI using Matern kernels of smoothness $\nu > 1$ converges at least at rate of $n^{- \text{min}(\nu, \nu_f) / k}$, where $\nu_f$ is the underlying smoothness.
\end{remark}
In this QS learning for  $\bw = (\bx, \bo)$, the domain of $\bx$ (denoted as $[a,b]^k$) is a Lipschitz domain and the domain of $\bo$ is a finite set of points. For each sequence $\bo$, $(a,b)^k \times \bo$ is a Lipschitz domain. The union of all $(a,b)^k \times \bo$  is a Lipschitz domain, since different $\bo$'s are disjoint. Thus, Remark~2 hold under the same assumptions in \cite{wynne2021convergence}. 

Several techniques can be used to enumerate  sequences in the aforementioned QS-learning methods, e.g. the $\epsilon$-greedy approach \citep{sutton2018reinforcement}. With a probability of $\epsilon \in (0,1)$, uniform sampling is performed to draw a random sequence $\bo$ and then apply the EI to find the best $\bx$. With a probability of $1-\epsilon$, we apply a simplified Algorithm~1 in Section~4.1 where we keep its Step~1 but drop its Steps~2 and 3. That is, we select the sequence $\bo_c$ from the current best observations, and then apply the EI to find the best $\bx_c$. Clearly, such an $\epsilon$-greedy approach can enumerate all possible sequences and will converge to the global optimum, since there exist converged sub-sequences as stated in Remarks~\ref{remark1} and \ref{remark2}. Here, a good choice of $\epsilon$ should be small enough to not interfere with the EI algorithm, but large enough to prevent it from getting stuck in a local optimum.

It is worth to remarking that the small-sample performance is often more important and relevant than the rate of convergence in experimentation, since only a few number of trials are often allowed in practice \citep{fang2005}. The asymptotic guarantees may say little about the practical effectiveness of the methods \citep{sutton2018reinforcement}. Though the general EI does not converge for all cases, it outperforms many other convergence-guaranteed methods in real applications \citep{osborne2010bayesian}. In Sections~6 and 7 in the main manuscript and the following Sections~S5 and S6 in this Supplementary Materials, results from real and simulation experiments clearly illustrate the superior small-sample performances of our proposed QS-learning, though it may not converge without further restrictions as discussed above.

\subsection{Additional Implementation Details}

\subsubsection{Details on Model Estimation}

The minimization problem for the model estimation in (13) can be solved via some standard non-linear optimization algorithms in Matlab or R. We recommend to use a BFGS method in a hybrid algorithm from the R package ``rgenoud" \citep{Walter2011} which is a widely used package for GP estimation. Two key tuning parameters in ``rgenoud" package that control the computing time are ``pop.size" and ``max.generations". According to the available computing resources, we set ``pop.size" from 100 to 1000 and ``max.generations" from 10 to 1000, respectively. For large experiments, to further speed up the model estimation, we also recommend to use a low storage BFGS method implemented by the R package ``nloptr" (developed by Steven G. Johnson). We set its tuning parameter ``maxeval" (which controls the computing time) from 50 to 500 according to the practical needs. In both R packages, analytical gradients developed in  Section~S1.2 should be used to facilitate the computation.

\subsubsection{Details on the Algorithms~1 and 2}

We first discuss the parameter tuning in the proposed algorithms. In Algorithm~1, a recommended value for $N_{round}$ is from 10 to 100 based on the time constraint. To optimize the quantitative input $\bx$ given the sequence input $\bo_{c}$, we adopt a BFGS method  implemented by the R package ``rgenoud" \citep{Walter2011}, where we set ``pop.size" from 100 to 1000 and ``max.generations" from 10 to 1000. When the number of components $k$ is small or parallel computing is available, enumeration of all sequences $o$ given $x_c$ can be used in Algorithm 1; otherwise, Algorithm 2 (SFTA) should be adopted. 
In Algorithm 2, we typically set $n_{steps}^{(1)}$ from 100 to 5000, $n_{seq}$ from 500 to 2000, $n_{rounds}$ from 10 to 100 and $n_{steps}^{(2)}$ from 500 to 5000 according to the practical needs.


To illustrate the performance of Algorithm~2 (SFTA), we compare it to a random search method. Here, we consider the same traveling salesman problem (Example~3 in Section~7), and compare the SFTA to a random search for optimizing the sequence input $o$ given the quantitative input $x$ that will lead to the largest EI for selecting the first sequential run, where the estimated model parameters are $(\bm{\sigma^2}, \bm{\theta}, \bm{\delta}) = (8522.10, 4217.90, 6253.08, 5734.86, 4163.44, \\ 
3241.29, 4353.13, 1678.87,   15.83, 67.25,   20.31,   54.61,  
81.90,   62.45,    0.99,    0.26,   1.45,    1.49, -1.51, \\
1.37,   -1.45, 1.36,   -1.19,   -0.41,  -0.97,    1.04,  1.49, 0.74,  -0.13)$ and the given optimal quantitative input $x=(2.00, 2.66, 3.47, 2.01, 1.73, 1.82, 2.43, 3.63 )$. For $k=8$ components, there are $8! = 40320$ possible sequences in total and the true maximum EI is 5.34.

\begin{figure}[ht]
\vspace{-.2in}
	\centering
	\subfigure []{\includegraphics[scale=0.4]{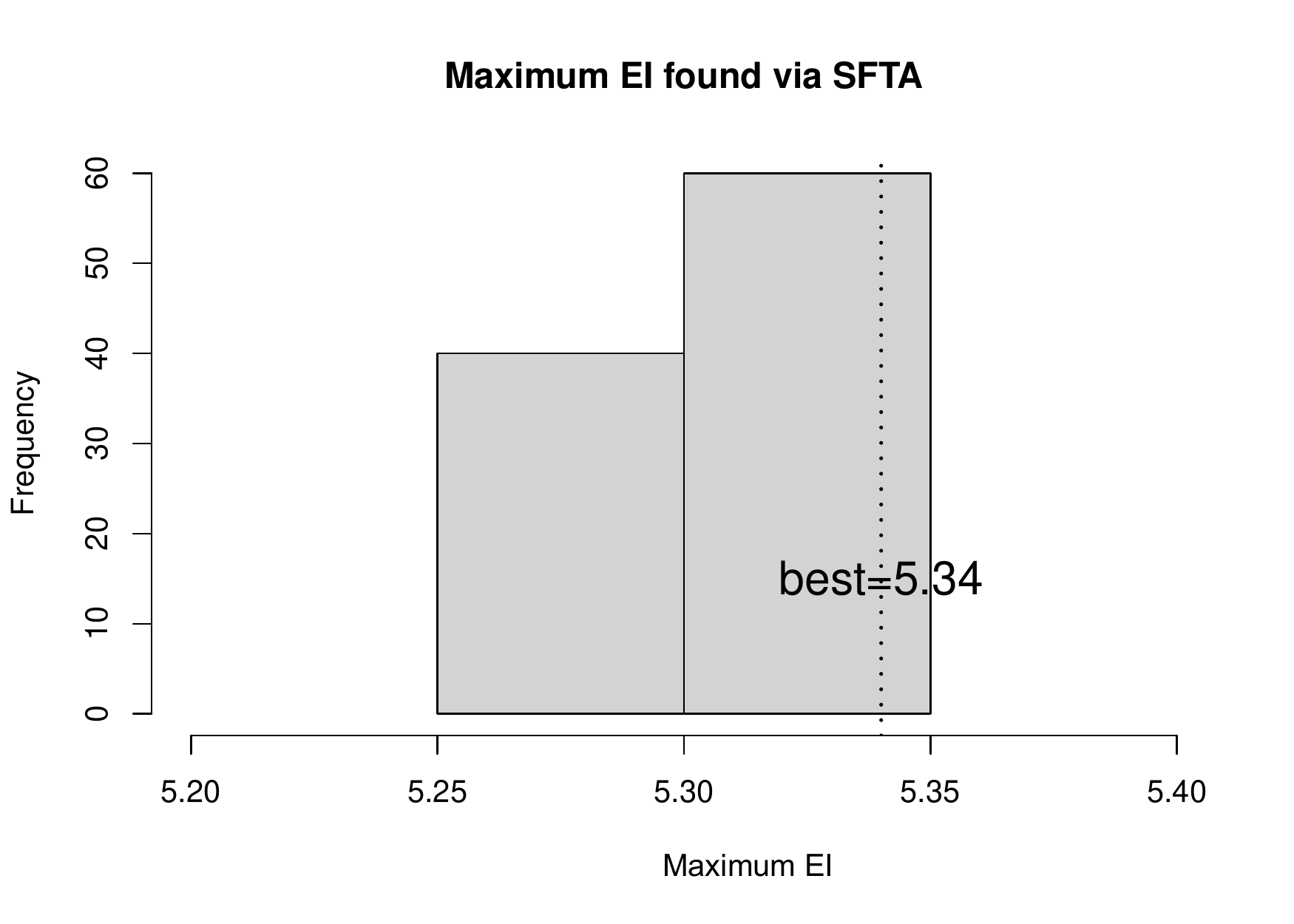}}\quad
	\subfigure [] {\includegraphics[scale=0.4]{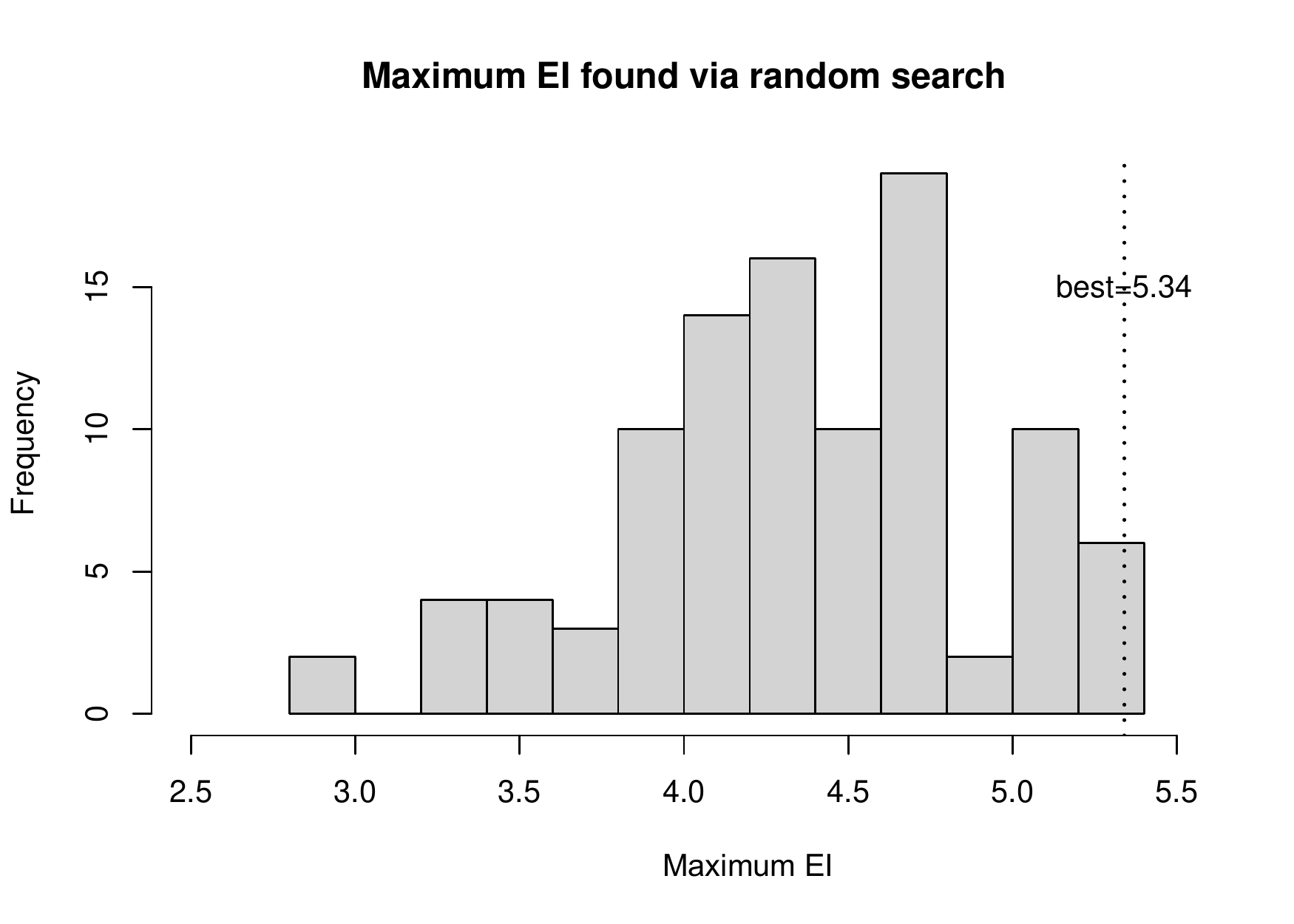}}
	\caption{Histograms of maximum EI values identified by (a)  SFTA and (b) a random search method.}
		\label{sfta}
\end{figure}

In SFTA, 600 sequences (100 sequences in Phase I and 500 sequences in Phase II) are compared, which costs less than a second on a regular laptop computer. We replicate it 100 times and show the maximum EIs found in Figure~\ref{sfta}(a). Its results are either 5.28 or 5.34. It finds the true maximum 5.34 in most replications.
As a comparison, we apply a random search method which randomly selects and compares 1000 sequences to find the maximum. The computing time is similar to that used in SFTA. We also replicate it 100 times and show its results in Figure~\ref{sfta}(b). Clearly, most of the results in Figure~\ref{sfta}(b) are not good. 

\subsubsection{Stopping Rules for the Fast QS-learning}
In Section~4.2, we set multiple stopping criteria for the fast QS-learning based on the obtained EI values, the budget constraint $N$ and the time constraint $T$. If the last three EIs are all less than $\alpha$ ($\alpha \in [0.1\%, 1\%]$) of the current best output, the algorithm will stop. In addition, if the maximum number of runs $N$ or the time constraint $T$ is reached, the algorithm will also stop. Here, as the computing time for model estimation is updated every batch of runs, we allow some flexibility in meeting with the time constraint. The algorithm will stop updating the parameters in  MaGP when the latest estimation exceeds the time $\alpha^{'} T$ ($\alpha^{'} \in (0,1]$ is a tuning parameter). Then, it will only update sequential runs until the time budget is used up.
Considering that updating parameters in the MaGP model ($O(n^3)$) requires much more computing time than updating the model having fixed parameters with a new run ($O(n^2)$), setting $\alpha^{'} = 0.95$ often suffices in practice.

\subsection{Additional Details on Numerical Studies}

\subsubsection{Details on the Case Study}

Here we provide some additional details on the case study in Section~6 of the main paper. We first provide a summary of drugs' names, levels and their corresponding doses in Table~\ref{tab:dld}, and then show the complete data in Table~\ref{tab:drugdata}.

\begin{table}[htp]
  \centering
  \caption{Drug levels and doses.}
    \begin{tabular}{cccccc}
    \toprule
    Drug       & \multicolumn{2}{c}{A: paclitaxel} & \multicolumn{2}{c}{B: doxorubicin} & C: mitoxantrone   \\
    \midrule
    Levels     & High: 1    & Low: 0     & High: 1    & Low: 0     & Fixed \\
    \midrule
    Doses       & 3.75 $\mu$M    & 2.8 $\mu$M     & 95 nM      & 70 nM      & 0.16 $\mu$M \\
    \bottomrule
    \end{tabular}%
    \label{tab:dld}
\end{table}%

\begin{table}[htp]
	\centering
	\small
	\caption{A Real Data on Lymphoma Treatment.}
	{\footnotesize
	\begin{tabular}{|cc|cc|c|r|c|cc|cc|c|r|}
		\cline{1-6}\cline{8-13}
		\multicolumn{2}{|c|}{Drug $A$} & \multicolumn{2}{c|}{Drug $B$} & \multicolumn{1}{c|}{Drug $C$} &     &       & \multicolumn{2}{c|}{Drug $A$} & \multicolumn{2}{c|}{Drug $B$} & \multicolumn{1}{c|}{Drug $C$} &  \\
		\multicolumn{1}{|c}{Dose} & \multicolumn{1}{c|}{Order} & \multicolumn{1}{c}{Dose} & \multicolumn{1}{c|}{Order} & \multicolumn{1}{c|}{Order} & \multicolumn{1}{c|}{{$y$}} & & \multicolumn{1}{c}{Dose} & \multicolumn{1}{c|}{Order} & \multicolumn{1}{c}{Dose} & \multicolumn{1}{c|}{Order} & \multicolumn{1}{c|}{Order} & \multicolumn{1}{c|}{{$y$}} \\
		\cline{1-6}\cline{8-13}
		1          & 1          & 1          & 2          & 3          & 39.91  &     & 1          & 1          & 0          & 2          & 3          & 44.33 \\
		1          & 1          & 1          & 3          & 2          & 44.38  &     & 1          & 1          & 0          & 3          & 2          & 38.18 \\
		1          & 2          & 1          & 1          & 3          & 17.08  &     & 1          & 2          & 0          & 1          & 3          & 22.26 \\
		1          & 2          & 1          & 3          & 1          & 20.88  &     & 1          & 2          & 0          & 3          & 1          & 31.40 \\
		1          & 3          & 1          & 1          & 2          & 34.68  &     & 1          & 3          & 0          & 1          & 2          & 38.91 \\
		1          & 3          & 1          & 2          & 1          & 37.37  &     & 1          & 3          & 0          & 2          & 1          & 42.30 \\
		0          & 1          & 0          & 2          & 3          & 30.00  &     & 0          & 1          & 1          & 2          & 3          & 44.87 \\
		0          & 1          & 0          & 3          & 2          & 47.18  &     & 0          & 1          & 1          & 3          & 2          & 43.93 \\
		0          & 2          & 0          & 1          & 3          & 25.10  &     & 0          & 2          & 1          & 1          & 3          & 26.02 \\
		0          & 2          & 0          & 3          & 1          & 33.60  &     & 0          & 2          & 1          & 3          & 1          & 22.56 \\
		0          & 3          & 0          & 1          & 2          & 35.04  &     & 0          & 3          & 1          & 1          & 2          & 31.15 \\
		0          & 3          & 0          & 2          & 1          & 35.04  &     & 0          & 3          & 1          & 2          & 1          & 37.19 \\
		\cline{1-6}\cline{8-13}
	\end{tabular}%
}
	\label{tab:drugdata}
\end{table}%

Below, we show the QS-design along with the sequential runs in QS-learning. We use bold font to highlight the optimal setting throughout this Supplementary Materials.

\footnotesize
$$
\stackrel{\mbox{QS-design}}{
\kbordermatrix{
	&X_A & O_A & X_B & O_B & O_C & Y \\
	& 0 &   3 &   0 &   2 &   1 & 35.04 \\
	&  1 &   2 &   0 &   1 &   3 & 22.26 \\
	&  0 &   1 &   1 &   3 &   2 & 43.93 \\
	&  1 &   2 &   1 &   3 &   1 & 20.88 \\
	&  0 &   1 &   0 &   2 &   3 & 30.00 \\
	&  1 &   1 &   0 &   3 &   2 & 38.18 \\
	&  0 &   2 &   1 &   1 &   3 & 26.02 \\
	&  1 &   3 &   1 &   1 &   2 & 34.68 \\
}
}
\stackrel{\mbox{Sequential runs}}{
\kbordermatrix{
	&X_A & O_A & X_B & O_B & O_C & Y \\
	&  1 &   2 &   1 &   1 &   3 & 17.08 \\
	&  1 &   1 &   1 &   2 &   3 & 39.91 \\
	& 1 &   2 &   0 &   3 &   1 & 31.40 \\
	&  1 &   1 &   0 &   2 &   3 & 44.33 \\
	&  1 &   1 &   1 &   3 &   2 & 44.38 \\
	&  1 &   3 &   1 &   2 &   1 & 37.37 \\
	&  \textbf{0} &   \textbf{1} &   \textbf{0} &   \textbf{3} &   \textbf{2} & \textbf{47.18} \\
}
}
$$
\normalsize


\subsubsection{Details on the Traveling Salesman Problem}

In Example~3 of Section 7, the sampled matrix for $s_{i,j}$ (rounding to 1 decimal place) is
	$$\big(s_{i,j}\big)_{\substack{ i=0,\ldots,k \\
			j=1,\ldots,k}} =
	\kbordermatrix{
		& 1  & 2  & 3  & 4  & 5  & 6  & 7  & 8  \\
		0 & 0.6 & 2.2 & 1.8 & 2.6 & 1.8 & 1.7 & 2.6 & 1.0 \\
		1 & 0 & 0.8 & 1.5 & 1.4 & 2.8 & 1.1 & 3.0 & 0.9 \\
		2 & 0.7 & 0 & 1.2 & 2.4 & 2.3 & 1.4 & 2.2 & 2.6 \\
		3 & 1.5 & 1.2 & 0 & 1.8 & 1.3 & 1.5 & 2.0 & 2.5 \\
		4 & 1.2 & 2.4 & 1.7 & 0 & 1.7 & 2.1 & 1.6 & 1.0 \\
		5 & 2.7 & 2.4 & 1.3 & 1.7 & 0 & 0.9 & 1.4 & 2.3 \\
		6 & 1.1 & 1.3 & 1.4 & 2.3 & 0.9 & 0 & 1.6 & 0.8 \\
		7 & 2.7 & 2.0 & 1.5 & 1.9 & 1.2 & 1.4 & 0 & 0.5 \\
		8 & 0.7 & 3.0 & 2.7 & 0.9 & 2.1 & 0.8 & 0.5 & 0 \\
	}.
	$$

In the following Figure~\ref{fig:hist2}, we show the histogram for the responses of 4,032,000 randomly sampled observations .
\begin{figure}[H]
	\centering
	\includegraphics[width=0.7\linewidth]{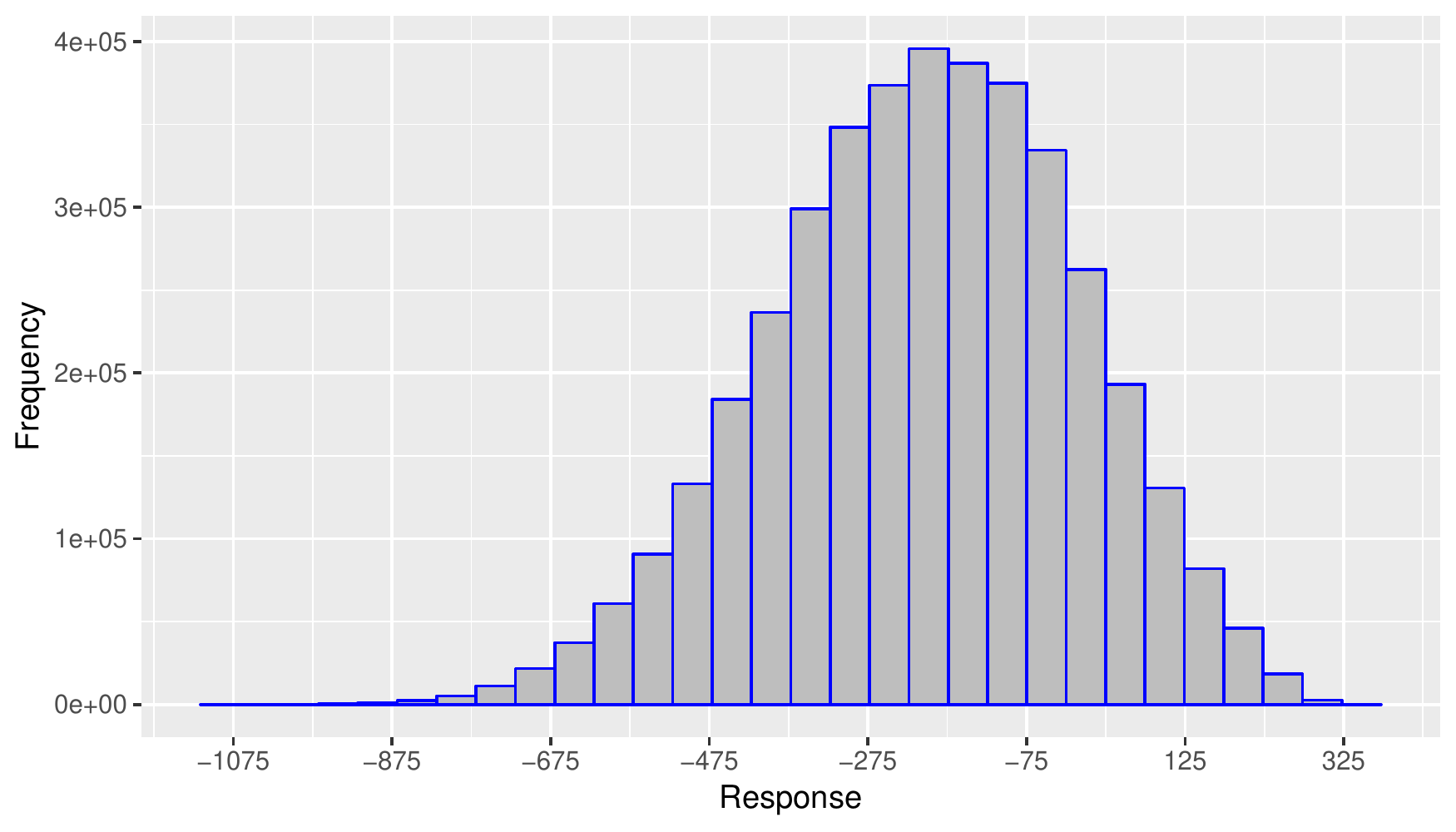} \vspace{-.2in}
	\caption{Histogram of the responses in a large random sample of 4,032,000 observations.}
		\label{fig:hist2}
\end{figure}

\begin{figure}[ht]
\vspace{-.2in}
	\centering
	\subfigure []{\includegraphics[scale=0.4]{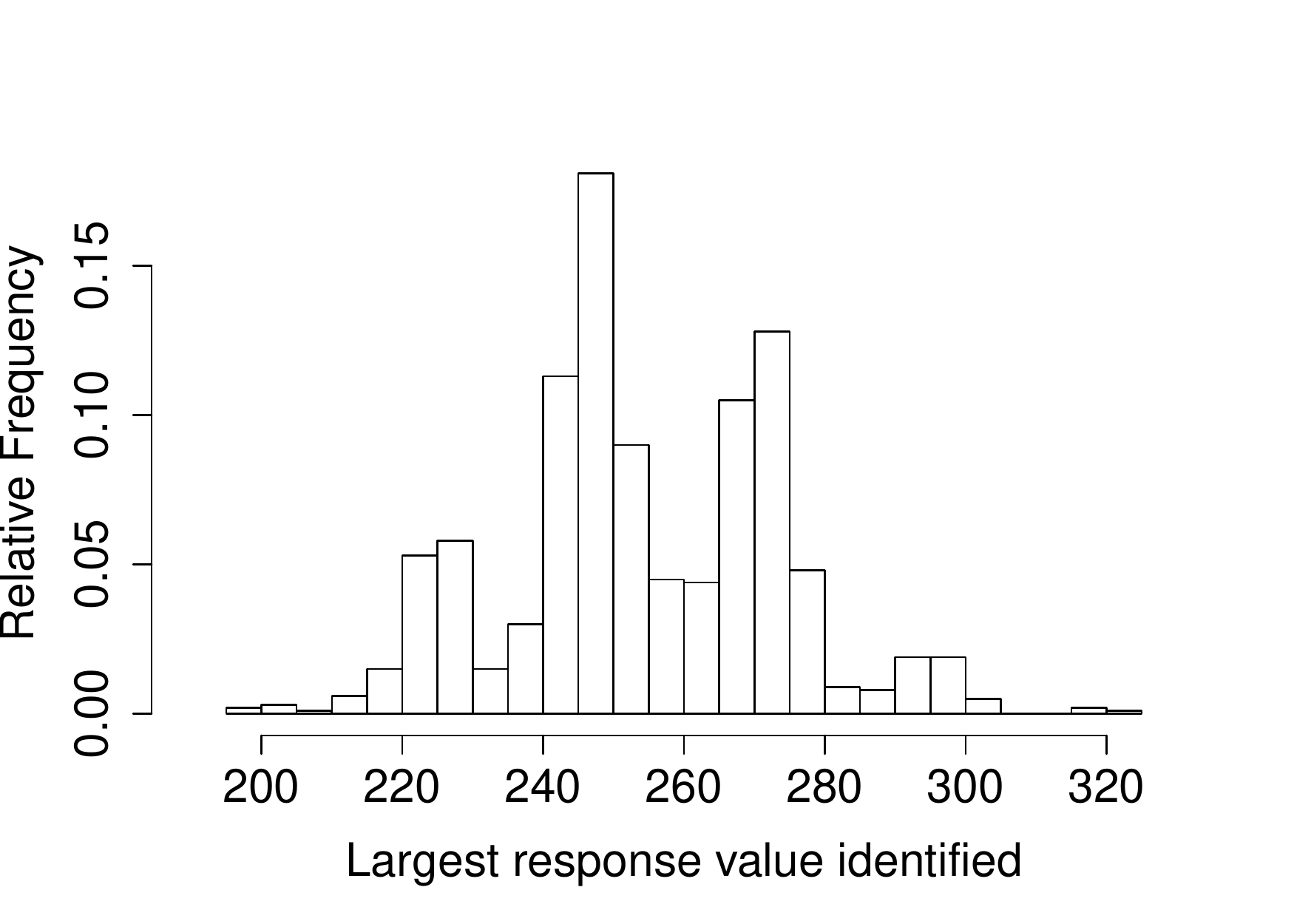}}\quad
	\subfigure [] {\includegraphics[scale=0.4]{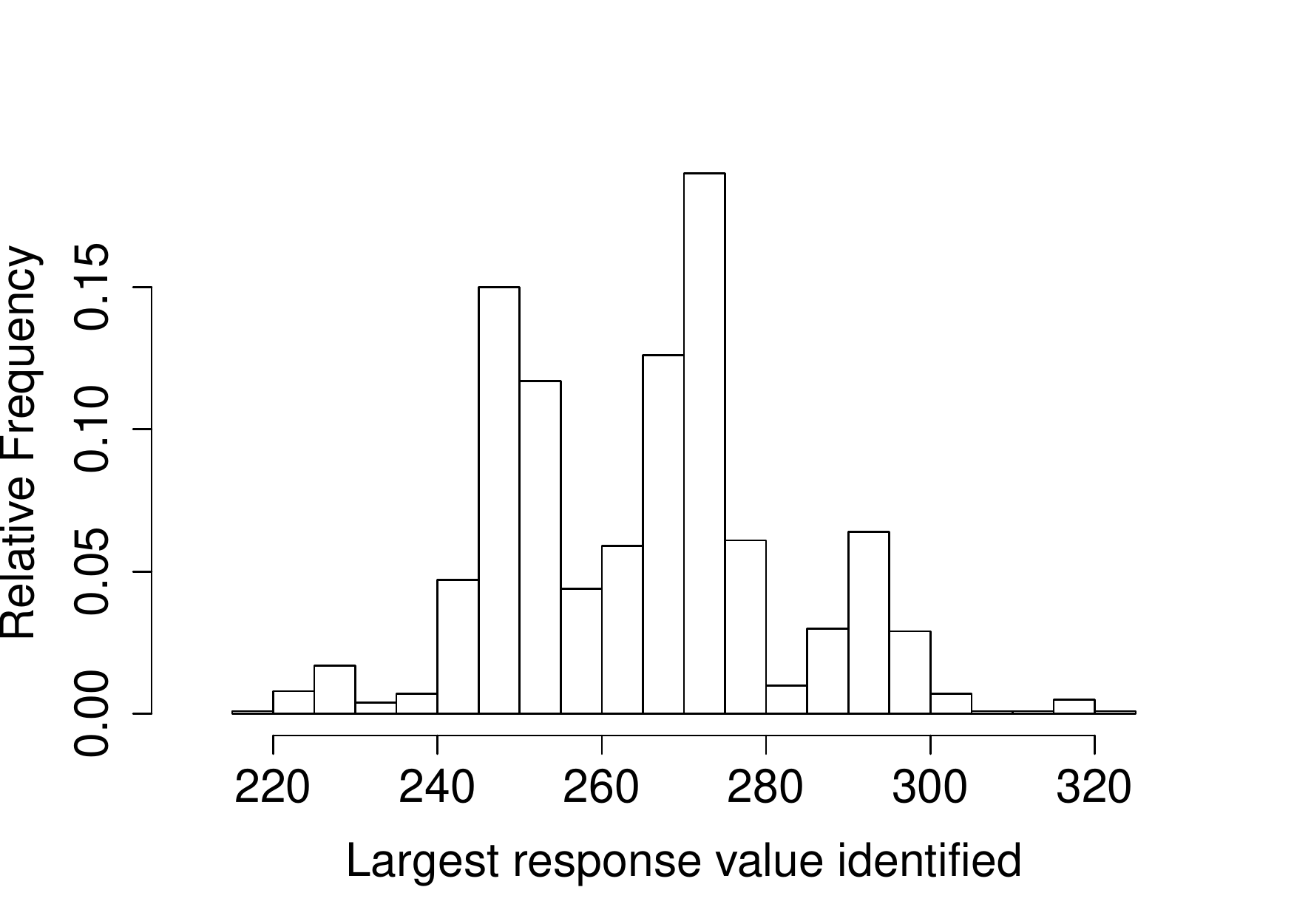}}
	\caption{Histograms of largest responses identified by (a) $BM_2$ and (b) $BM_3$.}
		\label{fig:tsp}
\end{figure}

In Figures~\ref{fig:tsp}(a) and (b), we show the histograms of largest responses identified by the sequential generalized PWO approach ($BM_2$) and sequential generalized CP approach ($BM_3$), respectively.

\newpage

Then, we show the 46-run QS-design along with its sequential runs of the QS-learning approach based on the 2d-MaGP model.

\scriptsize
$$
\stackrel{\mbox{QS-design}}{
	\kbordermatrix{
		& x_1 & o_1 & x_2 & o_2 & x_3 & o_3 & x_4 & o_4 & x_5 & o_5 & x_6 & o_6 & x_7 & o_7 & x_8 & o_8 & Y \\
		& 1.10 & 8 & 3.51 & 1 & 3.38 & 7 & 2.73 & 4 & 2.40 & 2 & 3.58 & 3 & 1.95 & 5 & 1.49 & 6 & -97.33 \\
		& 3.32 & 8 & 2.47 & 3 & 2.34 & 2 & 2.08 & 1 & 1.82 & 5 & 3.84 & 6 & 3.71 & 4 & 3.77 & 7 & -367.75 \\
		& 3.25 & 2 & 3.18 & 3 & 3.18 & 1 & 3.18 & 7 & 2.14 & 5 & 2.14 & 8 & 1.03 & 6 & 1.10 & 4 & -341.80 \\
		& 1.29 & 6 & 1.95 & 3 & 2.99 & 7 & 1.29 & 4 & 1.42 & 8 & 2.01 & 1 & 2.53 & 2 & 3.64 & 5 & 70.21 \\
		& 3.38 & 1 & 2.79 & 2 & 2.21 & 4 & 3.64 & 3 & 3.77 & 8 & 3.18 & 5 & 3.90 & 7 & 2.92 & 6 & -220.43 \\
		& 1.55 & 7 & 2.21 & 2 & 3.97 & 3 & 3.05 & 5 & 2.60 & 4 & 3.64 & 1 & 3.25 & 6 & 3.32 & 8 & -152.48 \\
		& 2.47 & 4 & 3.25 & 8 & 2.86 & 1 & 3.58 & 7 & 1.10 & 2 & 2.99 & 6 & 3.77 & 3 & 2.27 & 5 & -568.18 \\
		& 2.79 & 1 & 1.55 & 6 & 3.58 & 8 & 3.90 & 5 & 3.64 & 2 & 2.08 & 4 & 2.14 & 3 & 2.21 & 7 & -426.28 \\
		& 2.08 & 1 & 2.86 & 4 & 1.29 & 5 & 1.49 & 6 & 2.73 & 2 & 3.90 & 3 & 1.82 & 7 & 3.38 & 8 & -58.55 \\
		& 1.36 & 3 & 1.42 & 5 & 1.68 & 8 & 2.66 & 4 & 3.12 & 7 & 3.77 & 6 & 3.12 & 1 & 2.34 & 2 & -19.84 \\
		& 3.84 & 8 & 1.88 & 2 & 3.05 & 6 & 1.16 & 4 & 2.92 & 5 & 1.42 & 7 & 2.08 & 1 & 2.08 & 3 & -160.51 \\
		& 1.82 & 5 & 3.64 & 7 & 2.27 & 8 & 1.10 & 3 & 2.53 & 1 & 1.95 & 4 & 1.23 & 2 & 1.75 & 6 & -235.68 \\
		& 3.90 & 5 & 2.53 & 2 & 3.25 & 1 & 3.25 & 6 & 2.27 & 7 & 1.75 & 3 & 2.92 & 8 & 3.84 & 4 & 98.93 \\
		& 1.75 & 8 & 1.29 & 7 & 3.12 & 3 & 1.42 & 5 & 2.66 & 6 & 3.25 & 2 & 1.29 & 1 & 1.95 & 4 & -145.71 \\
		& 3.97 & 6 & 1.82 & 7 & 1.88 & 3 & 3.45 & 1 & 1.95 & 8 & 1.68 & 4 & 2.73 & 5 & 1.36 & 2 & -277.53 \\
		& 1.03 & 1 & 3.77 & 8 & 1.49 & 6 & 2.40 & 2 & 1.88 & 7 & 2.53 & 4 & 3.38 & 5 & 3.18 & 3 & -15.17 \\
		& 3.12 & 7 & 2.99 & 3 & 1.42 & 4 & 1.95 & 5 & 1.75 & 1 & 1.62 & 6 & 1.75 & 8 & 3.97 & 2 & -20.89 \\
		& 3.58 & 4 & 3.32 & 6 & 3.77 & 7 & 2.47 & 3 & 2.86 & 1 & 2.34 & 2 & 3.58 & 5 & 1.55 & 8 & -400.69 \\
		& 2.73 & 2 & 2.14 & 8 & 1.62 & 7 & 2.34 & 6 & 3.97 & 1 & 2.66 & 5 & 1.49 & 4 & 1.16 & 3 & -210.02 \\
		& 3.05 & 3 & 3.58 & 1 & 1.82 & 2 & 2.99 & 7 & 3.71 & 6 & 1.16 & 4 & 2.21 & 8 & 2.66 & 5 & -50.76 \\
		& 2.86 & 2 & 3.84 & 5 & 3.84 & 4 & 1.62 & 1 & 1.55 & 6 & 2.79 & 7 & 2.27 & 3 & 3.25 & 8 & -613.34 \\
		& 1.23 & 6 & 1.49 & 3 & 2.60 & 2 & 3.84 & 8 & 1.62 & 1 & 2.21 & 7 & 2.60 & 5 & 1.68 & 4 & -158.14 \\
		& 2.66 & 2 & 1.75 & 6 & 1.03 & 3 & 3.97 & 4 & 2.21 & 1 & 2.60 & 8 & 2.99 & 7 & 3.45 & 5 & -463.02 \\
		& 3.18 & 5 & 1.68 & 1 & 2.79 & 4 & 3.32 & 6 & 2.47 & 3 & 3.97 & 8 & 2.66 & 2 & 1.23 & 7 & -382.50 \\
		& 2.53 & 8 & 3.38 & 5 & 2.47 & 2 & 2.14 & 6 & 1.16 & 4 & 1.03 & 1 & 2.86 & 7 & 1.82 & 3 & -126.88 \\
		& 2.40 & 6 & 3.90 & 4 & 1.36 & 1 & 2.53 & 5 & 2.79 & 7 & 3.12 & 8 & 3.05 & 2 & 1.03 & 3 & -323.96 \\
		& 1.16 & 3 & 2.27 & 8 & 3.64 & 4 & 1.75 & 1 & 3.05 & 7 & 1.23 & 2 & 2.79 & 6 & 1.62 & 5 & -259.77 \\
		& 1.68 & 4 & 2.08 & 7 & 2.53 & 5 & 2.86 & 1 & 3.18 & 2 & 1.29 & 8 & 3.64 & 6 & 3.90 & 3 & -393.77 \\
		& 3.77 & 7 & 2.60 & 4 & 2.14 & 6 & 1.68 & 3 & 1.36 & 2 & 3.51 & 1 & 1.55 & 8 & 2.01 & 5 & -51.79 \\
		& 3.71 & 1 & 2.73 & 3 & 1.23 & 8 & 1.23 & 2 & 2.99 & 4 & 2.40 & 7 & 3.51 & 6 & 2.40 & 5 & -147.90 \\
		& 1.95 & 4 & 2.66 & 3 & 3.45 & 8 & 2.01 & 6 & 3.58 & 5 & 1.88 & 2 & 1.16 & 7 & 3.71 & 1 & 270.55 \\
		& 1.88 & 4 & 2.92 & 2 & 1.10 & 3 & 2.92 & 1 & 1.29 & 6 & 2.47 & 5 & 1.42 & 8 & 1.88 & 7 & -81.61 \\
		& 1.62 & 6 & 3.71 & 7 & 3.51 & 4 & 3.71 & 2 & 2.34 & 3 & 1.49 & 5 & 2.40 & 1 & 3.05 & 8 & -553.06 \\
		& 2.92 & 3 & 1.03 & 1 & 3.32 & 6 & 1.88 & 8 & 3.38 & 2 & 2.73 & 5 & 3.84 & 7 & 2.53 & 4 & -138.01 \\
		& 1.42 & 5 & 1.62 & 4 & 1.16 & 2 & 1.55 & 8 & 3.45 & 7 & 1.55 & 1 & 2.34 & 3 & 2.73 & 6 & -2.74 \\
		& 3.45 & 5 & 3.05 & 4 & 3.71 & 7 & 2.60 & 2 & 3.51 & 8 & 3.71 & 6 & 1.62 & 1 & 2.79 & 3 & -387.33 \\
		& 1.49 & 5 & 3.12 & 6 & 2.08 & 3 & 3.77 & 8 & 3.84 & 4 & 3.05 & 7 & 1.88 & 2 & 2.86 & 1 & -318.83 \\
		& 2.60 & 2 & 1.10 & 8 & 1.55 & 6 & 1.82 & 5 & 1.23 & 4 & 1.82 & 3 & 3.32 & 1 & 2.47 & 7 & -326.37 \\
		& 2.99 & 1 & 1.36 & 4 & 3.90 & 8 & 2.21 & 7 & 1.03 & 3 & 2.27 & 2 & 2.47 & 5 & 2.14 & 6 & 5.11 \\
		& 2.01 & 8 & 3.45 & 6 & 2.92 & 5 & 1.36 & 7 & 3.90 & 3 & 2.86 & 1 & 3.18 & 4 & 2.99 & 2 & -162.49 \\
		& 2.34 & 2 & 1.23 & 1 & 2.40 & 5 & 2.79 & 3 & 2.01 & 7 & 1.10 & 6 & 1.10 & 4 & 2.60 & 8 & -136.24 \\
		& 3.51 & 6 & 3.97 & 1 & 1.75 & 5 & 3.51 & 8 & 2.08 & 4 & 3.32 & 2 & 2.01 & 3 & 3.12 & 7 & -168.74 \\
		& 3.64 & 2 & 1.16 & 5 & 1.95 & 7 & 2.27 & 8 & 3.25 & 3 & 2.92 & 4 & 1.68 & 6 & 3.51 & 1 & -15.91 \\
		& 2.21 & 7 & 2.34 & 5 & 2.01 & 8 & 3.12 & 2 & 3.32 & 6 & 1.36 & 3 & 3.97 & 4 & 1.29 & 1 & 55.81 \\
		& 2.14 & 7 & 2.40 & 6 & 2.66 & 2 & 1.03 & 4 & 1.68 & 8 & 3.38 & 5 & 3.45 & 3 & 1.42 & 1 & -203.72 \\
		& 2.27 & 3 & 2.01 & 8 & 2.73 & 1 & 3.38 & 2 & 1.49 & 5 & 3.45 & 7 & 1.36 & 4 & 3.58 & 6 & -560.99 \\ }
}
$$

$$
\stackrel{\mbox{Sequential runs under the 2d-MaGP}}{
	\kbordermatrix{
		& x_1 & o_1 & x_2 & o_2 & x_3 & o_3 & x_4 & o_4 & x_5 & o_5 & x_6 & o_6 & x_7 & o_7 & x_8 & o_8 & Y \\
		&2.00 & 8 & 2.66 & 6 & 3.47 & 2 & 2.01 & 1 & 1.73 & 3 & 1.82 & 5 & 2.43 & 4 & 3.63 & 7 & 258.39 \\
  &1.97 & 8 & 2.61 & 2 & 1.46 & 6 & 2.02 & 1 & 2.65 & 7 & 2.53 & 3 & 2.43 & 5 & 4.00 & 4 & 85.06 \\
  &1.03 & 1 & 1.33 & 6 & 3.42 & 2 & 1.43 & 8 & 1.83 & 5 & 1.86 & 4 & 1.34 & 3 & 4.00 & 7 & 221.56 \\
  &1.59 & 1 & 2.98 & 8 & 3.05 & 6 & 1.95 & 5 & 3.29 & 2 & 1.79 & 4 & 1.01 & 3 & 3.78 & 7 & 165.24 \\
  &1.17 & 1 & 2.67 & 8 & 3.41 & 2 & 2.37 & 6 & 3.56 & 5 & 1.67 & 4 & 1.43 & 3 & 3.08 & 7 & 205.61 \\
  &1.81 & 8 & 1.37 & 6 & 1.49 & 2 & 2.92 & 1 & 3.56 & 5 & 1.78 & 4 & 1.36 & 3 & 3.80 & 7 & 280.84 \\
  &1.90 & 8 & 2.67 & 6 & 1.56 & 2 & 1.29 & 4 & 1.87 & 3 & 1.90 & 5 & 1.00 & 1 & 4.00 & 7 & 261.06 \\
  &1.76 & 2 & 2.67 & 6 & 1.51 & 8 & 2.93 & 1 & 1.80 & 5 & 1.78 & 4 & 1.44 & 3 & 3.97 & 7 & 271.20 \\
  &2.11 & 1 & 2.66 & 6 & 1.52 & 8 & 2.22 & 2 & 2.90 & 5 & 1.81 & 4 & 1.35 & 3 & 4.00 & 7 & 221.92 \\
  &1.96 & 8 & 2.94 & 6 & 1.55 & 2 & 2.01 & 1 & 1.03 & 4 & 4.00 & 5 & 1.91 & 3 & 3.69 & 7 & 292.25 \\
  &1.92 & 8 & 2.92 & 2 & 1.54 & 1 & 2.37 & 6 & 1.03 & 4 & 4.00 & 5 & 1.83 & 3 & 3.85 & 7 & 251.23 \\
  &1.98 & 8 & 2.95 & 6 & 1.61 & 2 & 2.91 & 1 & 1.03 & 4 & 1.94 & 5 & 2.59 & 3 & 3.50 & 7 & 287.63 \\
  &1.93 & 6 & 2.93 & 2 & 1.57 & 8 & 2.91 & 1 & 3.56 & 4 & 3.66 & 5 & 2.42 & 3 & 3.77 & 7 & 271.75 \\
  &1.93 & 8 & 2.68 & 6 & 1.56 & 2 & 2.01 & 1 & 1.74 & 4 & 3.63 & 5 & 2.36 & 3 & 4.00 & 7 & 298.83 \\
  &1.83 & 8 & 1.36 & 6 & 2.85 & 2 & 2.92 & 1 & 1.71 & 4 & 3.78 & 5 & 2.46 & 7 & 3.82 & 3 & 306.34 \\
  &1.63 & 8 & 1.36 & 6 & 1.56 & 2 & 2.92 & 1 & 1.71 & 5 & 3.83 & 4 & 2.79 & 7 & 4.00 & 3 & 295.58 \\
  &1.78 & 8 & 1.38 & 6 & 1.80 & 2 & 2.93 & 1 & 1.71 & 4 & 3.76 & 5 & 1.50 & 7 & 3.92 & 3 & 290.53 \\
  &1.80 & 8 & 1.39 & 6 & 1.64 & 5 & 2.01 & 1 & 3.10 & 2 & 3.75 & 4 & 1.78 & 7 & 3.82 & 3 & 198.70 \\
  &1.76 & 8 & 1.36 & 6 & 1.64 & 2 & 2.93 & 1 & 1.73 & 4 & 3.72 & 5 & 1.00 & 7 & 3.77 & 3 & 283.66 \\
  &1.82 & 8 & 2.68 & 6 & 2.00 & 2 & 2.02 & 1 & 1.09 & 4 & 3.73 & 5 & 2.38 & 7 & 3.75 & 3 & 296.17 \\
  &1.82 & 8 & 2.69 & 2 & 2.80 & 6 & 2.01 & 1 & 1.07 & 4 & 3.73 & 5 & 2.40 & 7 & 3.77 & 3 & 286.73 \\
  &1.81 & 8 & 2.69 & 6 & 2.75 & 2 & 2.01 & 1 & 1.02 & 4 & 3.71 & 5 & 2.41 & 7 & 3.76 & 3 & 296.45 \\
  &1.84 & 8 & 2.69 & 6 & 4.00 & 5 & 2.01 & 1 & 4.00 & 2 & 3.76 & 4 & 2.56 & 7 & 3.85 & 3 & 114.76 \\
  &1.74 & 8 & 2.94 & 6 & 2.83 & 2 & 2.04 & 1 & 1.07 & 5 & 3.09 & 4 & 2.22 & 7 & 3.79 & 3 & 285.82 \\
  &3.06 & 1 & 2.67 & 6 & 2.83 & 2 & 2.38 & 8 & 3.57 & 4 & 3.81 & 7 & 3.09 & 5 & 3.87 & 3 & 133.94 \\
  &1.77 & 8 & 1.34 & 6 & 1.54 & 2 & 1.43 & 1 & 1.65 & 4 & 3.67 & 5 & 2.21 & 7 & 3.40 & 3 & 276.33 \\
  &1.67 & 8 & 2.68 & 6 & 2.82 & 2 & 2.93 & 1 & 1.04 & 4 & 4.00 & 5 & 2.75 & 3 & 4.00 & 7 & 310.63 \\
  &1.62 & 8 & 1.37 & 6 & 2.84 & 2 & 2.02 & 1 & 1.71 & 5 & 1.96 & 4 & 2.42 & 3 & 4.00 & 7 & 279.46 \\
  &1.86 & 8 & 2.95 & 6 & 1.93 & 2 & 2.93 & 1 & 1.01 & 4 & 3.42 & 5 & 3.24 & 3 & 4.00 & 7 & 310.34 \\
  &1.84 & 8 & 2.76 & 6 & 3.49 & 2 & 2.93 & 1 & 1.00 & 4 & 3.59 & 5 & 2.53 & 3 & 4.00 & 7 & 298.84 \\
  &1.97 & 8 & 2.03 & 6 & 2.42 & 2 & 2.91 & 1 & 3.57 & 4 & 3.24 & 5 & 3.59 & 3 & 1.39 & 7 & 308.59 \\
  &1.93 & 8 & 2.11 & 6 & 2.43 & 2 & 2.91 & 1 & 1.80 & 4 & 3.26 & 5 & 3.43 & 3 & 3.19 & 7 & 308.11 \\
  &1.91 & 8 & 3.08 & 6 & 2.48 & 2 & 2.90 & 1 & 3.58 & 4 & 4.00 & 5 & 3.21 & 3 & 4.00 & 7 & 332.32 \\
  &1.00 & 8 & 3.60 & 6 & 2.48 & 2 & 2.87 & 1 & 3.57 & 4 & 4.00 & 5 & 3.28 & 3 & 4.00 & 7 & 335.32 \\
  &1.82 & 8 & 4.00 & 6 & 2.48 & 2 & 2.90 & 1 & 3.57 & 4 & 4.00 & 5 & 3.24 & 3 & 1.93 & 7 & 300.09 \\
  &1.25 & 6 & 1.00 & 8 & 2.48 & 2 & 2.85 & 1 & 3.57 & 4 & 4.00 & 5 & 3.79 & 3 & 4.00 & 7 & 314.84 \\
  &1.00 & 8 & 1.88 & 6 & 2.48 & 2 & 2.59 & 1 & 3.56 & 4 & 4.00 & 5 & 3.72 & 3 & 4.00 & 7 & 325.39 \\
  &1.01 & 8 & 3.09 & 6 & 2.46 & 2 & 2.94 & 1 & 3.58 & 4 & 3.91 & 5 & 3.05 & 3 & 3.62 & 7 & 328.86 \\
  &4.00 & 8 & 3.41 & 6 & 2.55 & 2 & 2.68 & 1 & 3.55 & 4 & 4.00 & 5 & 3.15 & 3 & 4.00 & 7 & 281.17 \\
  &1.00 & 8 & 3.41 & 6 & 2.47 & 2 & 2.86 & 1 & 2.96 & 4 & 4.00 & 5 & 3.09 & 3 & 4.00 & 7 & 329.95 \\
  &\textbf{1.14} & \textbf{8} & \textbf{3.44} & \textbf{6} & \textbf{2.48} & \textbf{2} & \textbf{2.86} & \textbf{1} & \textbf{3.78} & \textbf{4} & \textbf{4.00} &\textbf{ 5} & \textbf{3.11} & \textbf{3} & \textbf{4.00} & \textbf{7} & \textbf{335.61} \\
  &2.70 & 8 & 3.46 & 6 & 2.47 & 2 & 2.92 & 1 & 3.63 & 4 & 3.96 & 5 & 3.12 & 3 & 4.00 & 7 & 307.91 \\
		 }
}
$$

\normalsize

\newpage

Below, we show the sequential runs of the fast QS-learning approach based on the 2d-MaGP model.
\scriptsize
$$
\stackrel{\mbox{Sequential runs under the 2d-MaGP}}{
	\kbordermatrix{
		& x_1 & o_1 & x_2 & o_2 & x_3 & o_3 & x_4 & o_4 & x_5 & o_5 & x_6 & o_6 & x_7 & o_7 & x_8 & o_8 & Y \\
  &2.21 & 8 & 2.66 & 6 & 3.44 & 2 & 2.01 & 1 & 3.58 & 5 & 1.88 & 4 & 2.53 & 3 & 3.64 & 7 & 274.32 \\
  &1.95 & 8 & 2.66 & 6 & 2.01 & 2 & 2.01 & 1 & 3.58 & 5 & 1.75 & 4 & 2.53 & 3 & 3.63 & 7 & 296.95 \\
  &1.95 & 8 & 2.53 & 2 & 3.44 & 6 & 2.01 & 1 & 3.58 & 5 & 1.88 & 4 & 2.53 & 3 & 3.61 & 7 & 232.20 \\
  &1.95 & 8 & 2.66 & 6 & 2.99 & 2 & 2.01 & 1 & 3.58 & 5 & 1.75 & 4 & 1.82 & 3 & 3.63 & 7 & 285.02 \\
  &3.64 & 8 & 2.66 & 6 & 2.99 & 2 & 2.01 & 1 & 3.58 & 5 & 1.75 & 4 & 2.53 & 7 & 3.65 & 3 & 254.89 \\
  &2.73 & 8 & 2.66 & 6 & 2.99 & 2 & 2.01 & 1 & 3.58 & 5 & 3.90 & 4 & 2.53 & 3 & 3.84 & 7 & 260.25 \\
  &2.99 & 6 & 2.66 & 8 & 1.99 & 3 & 2.01 & 1 & 3.57 & 5 & 3.87 & 4 & 3.97 & 2 & 3.97 & 7 & -120.48 \\
  &2.69 & 8 & 1.99 & 6 & 2.53 & 2 & 1.15 & 1 & 1.17 & 5 & 1.37 & 4 & 3.33 & 3 & 1.56 & 7 & 257.66 \\
  &2.89 & 4 & 2.35 & 1 & 1.34 & 8 & 1.56 & 6 & 2.95 & 7 & 2.77 & 3 & 3.46 & 5 & 2.77 & 2 & -116.97 \\
  &1.95 & 5 & 2.66 & 6 & 2.01 & 7 & 2.01 & 1 & 3.58 & 3 & 1.75 & 4 & 2.53 & 8 & 3.63 & 2 & -359.70 \\
  &1.89 & 8 & 2.66 & 6 & 2.01 & 2 & 1.68 & 1 & 1.17 & 5 & 1.75 & 4 & 1.82 & 3 & 4.00 & 7 & 271.80 \\
  &3.66 & 8 & 1.35 & 6 & 1.57 & 1 & 1.52 & 2 & 1.17 & 5 & 1.88 & 4 & 2.46 & 3 & 3.51 & 7 & 247.16 \\
  &1.95 & 3 & 2.66 & 5 & 2.01 & 4 & 2.01 & 1 & 2.29 & 8 & 1.75 & 2 & 2.53 & 6 & 3.63 & 7 & -238.69 \\
  &1.95 & 5 & 2.66 & 2 & 2.01 & 8 & 2.01 & 3 & 3.58 & 4 & 1.75 & 7 & 2.53 & 6 & 3.63 & 1 & -193.42 \\
  &2.99 & 6 & 3.38 & 7 & 2.01 & 2 & 3.25 & 8 & 1.17 & 5 & 3.86 & 4 & 2.46 & 3 & 3.54 & 1 & -140.20 \\
  &3.18 & 8 & 2.66 & 6 & 2.99 & 2 & 1.29 & 4 & 2.73 & 3 & 2.01 & 1 & 2.97 & 5 & 1.73 & 7 & 196.77 \\
  &3.47 & 3 & 1.21 & 8 & 2.21 & 6 & 3.93 & 5 & 3.03 & 7 & 3.02 & 1 & 3.45 & 2 & 2.68 & 4 & -214.36 \\
  &1.95 & 5 & 2.66 & 2 & 2.01 & 7 & 2.01 & 8 & 3.58 & 4 & 1.75 & 1 & 2.53 & 3 & 3.63 & 6 & -87.77 \\
  &1.57 & 8 & 1.35 & 6 & 2.01 & 2 & 3.77 & 5 & 2.73 & 3 & 3.85 & 4 & 3.64 & 1 & 2.34 & 7 & 253.49 \\
  &2.04 & 8 & 2.66 & 2 & 2.01 & 6 & 3.05 & 5 & 2.73 & 3 & 3.64 & 7 & 3.43 & 1 & 2.50 & 4 & 27.97 \\
  &1.95 & 6 & 2.66 & 8 & 3.97 & 3 & 1.50 & 2 & 3.58 & 4 & 1.75 & 1 & 3.07 & 5 & 3.22 & 7 & 113.82 \\
  &1.95 & 6 & 1.21 & 8 & 3.46 & 2 & 3.93 & 5 & 3.58 & 4 & 1.75 & 1 & 1.48 & 7 & 4.00 & 3 & 250.76 \\
  &1.95 & 6 & 3.45 & 8 & 3.46 & 2 & 3.63 & 3 & 3.56 & 5 & 1.75 & 4 & 3.56 & 1 & 1.22 & 7 & 86.14 \\
  &1.95 & 8 & 1.21 & 6 & 1.99 & 3 & 1.50 & 2 & 2.60 & 4 & 2.01 & 1 & 3.26 & 5 & 1.47 & 7 & 243.11 \\
  &1.95 & 8 & 2.40 & 3 & 2.01 & 2 & 2.01 & 6 & 2.60 & 4 & 1.75 & 1 & 3.11 & 5 & 1.46 & 7 & 159.79 \\
  &1.95 & 8 & 2.66 & 2 & 1.29 & 6 & 1.95 & 3 & 1.88 & 7 & 1.38 & 5 & 3.38 & 1 & 2.96 & 4 & 125.63 \\
  &3.52 & 6 & 2.32 & 4 & 2.02 & 7 & 2.21 & 2 & 3.39 & 3 & 1.39 & 1 & 4.00 & 5 & 4.00 & 8 & -330.90 \\
  &1.58 & 8 & 3.55 & 6 & 1.51 & 2 & 2.30 & 1 & 3.16 & 5 & 1.28 & 4 & 3.06 & 3 & 3.90 & 7 & 298.71 \\
  &1.58 & 8 & 3.55 & 6 & 1.51 & 2 & 2.30 & 1 & 3.16 & 5 & 1.28 & 4 & 3.06 & 3 & 3.90 & 7 & 298.72 \\
  &1.58 & 8 & 3.55 & 6 & 1.51 & 2 & 2.30 & 1 & 3.16 & 5 & 1.28 & 4 & 3.06 & 3 & 3.90 & 7 & 298.72 \\
  &\textbf{2.40} & \textbf{8} & \textbf{2.32} & \textbf{6} & \textbf{1.74} & \textbf{2} & \textbf{3.20} & \textbf{1} & \textbf{3.50} & \textbf{5}& \textbf{2.11} & \textbf{4} & \textbf{3.93} & \textbf{3} & \textbf{4.00} & \textbf{7} & \textbf{313.02} \\
  &2.40 & 8 & 2.32 & 6 & 1.74 & 2 & 2.43 & 1 & 2.59 & 5 & 2.11 & 4 & 3.93 & 3 & 4.00 & 7 & 308.22 \\
  &2.40 & 5 & 2.32 & 8 & 1.74 & 3 & 3.20 & 4 & 3.50 & 1 & 2.11 & 7 & 3.93 & 6 & 4.00 & 2 & -436.97 \\
  &1.95 & 8 & 2.53 & 2 & 3.44 & 6 & 2.01 & 1 & 3.84 & 5 & 1.88 & 4 & 1.87 & 3 & 1.72 & 7 & 209.86 \\
  &2.21 & 8 & 2.32 & 2 & 3.45 & 6 & 2.29 & 3 & 3.16 & 5 & 1.89 & 4 & 2.31 & 1 & 3.79 & 7 & 211.54 \\
  &2.20 & 8 & 2.38 & 3 & 3.45 & 2 & 2.04 & 6 & 2.60 & 4 & 1.88 & 5 & 3.97 & 1 & 3.99 & 7 & 115.94 \\
  &2.73 & 6 & 2.79 & 7 & 2.49 & 2 & 3.20 & 3 & 1.01 & 8 & 3.06 & 5 & 2.66 & 1 & 1.74 & 4 & -215.78 \\
  &2.40 & 3 & 2.32 & 1 & 1.74 & 8 & 3.20 & 7 & 3.50 & 5 & 2.11 & 2 & 3.93 & 4 & 4.00 & 6 & -357.09 \\
  &1.97 & 8 & 1.99 & 6 & 2.99 & 2 & 3.32 & 1 & 2.59 & 5 & 1.38 & 4 & 2.41 & 3 & 2.92 & 7 & 291.76 \\
  &1.95 & 8 & 1.21 & 6 & 2.99 & 2 & 2.03 & 1 & 2.60 & 5 & 2.11 & 4 & 3.93 & 3 & 1.50 & 7 & 282.56 \\
  &1.95 & 8 & 2.32 & 2 & 1.29 & 6 & 3.27 & 3 & 3.15 & 5 & 2.11 & 4 & 3.23 & 1 & 3.62 & 7 & 251.25 \\
  &1.58 & 8 & 1.35 & 6 & 1.74 & 2 & 1.82 & 1 & 2.59 & 5 & 1.89 & 4 & 2.52 & 3 & 4.00 & 7 & 275.88 \\
  &2.40 & 8 & 2.20 & 6 & 3.25 & 2 & 1.63 & 1 & 2.79 & 5 & 1.03 & 4 & 2.52 & 3 & 2.34 & 7 & 267.05 \\
  &1.84 & 7 & 3.36 & 1 & 3.72 & 3 & 1.39 & 2 & 2.57 & 5 & 3.13 & 4 & 3.35 & 8 & 2.38 & 6 & -581.79 \\
		 }
}
$$

\normalsize

\newpage

\subsection{Additional Simulation on Arranging Four Mathematical Operations}
\label{mo}

\cite{robert2018} described a problem on arranging four mathematical operations: addition, subtraction, multiplication and division, where only the sequence factors are considered.
Here, we generalize this problem by considering both the quantities and sequences for arranging the four mathematical operations. By this simulation study, We first show the efficiency of QS-learning for optimization, and then evaluate the accuracy of MaGP for prediction.

\begin{example}
	\label{oper1}
	Suppose that the four mathematical operations (i.e. components $c_1$, $c_2$, $c_3$ and $c_4$) have the following effects starting with $Y_0 = 20$ before any components: $c_1$: add $1+10sin(2\pi x_1)$; $c_2$: subtract $2+10(x_2-0.4)^2$; $c_3$: multiply by $3 + x_3$; $c_4$: divide by $4-x_4$, where $x_i \in [0,1]$ for $i = 1, \ldots, 4$. In this example, the analytical solution exists for maximizing the response $Y$ after the four operations, where the optimal settings are $(x_1, \ldots x_4) = (0.25, 0.4, 1, 1)$ and $(o_1, \ldots, o_4) = (2,  4, 3, 1)$ and the maximum response is $y_{max} =   (Y_0/c_4 + c_1) \times c_3 - c_2 = (Y_0/3 + 11) \times 4 -2 =  68.67$.
\end{example}

\begin{figure}[htbp]
\vspace{-.2in}
    \centering
  \subfigure []{\includegraphics[scale=0.4]{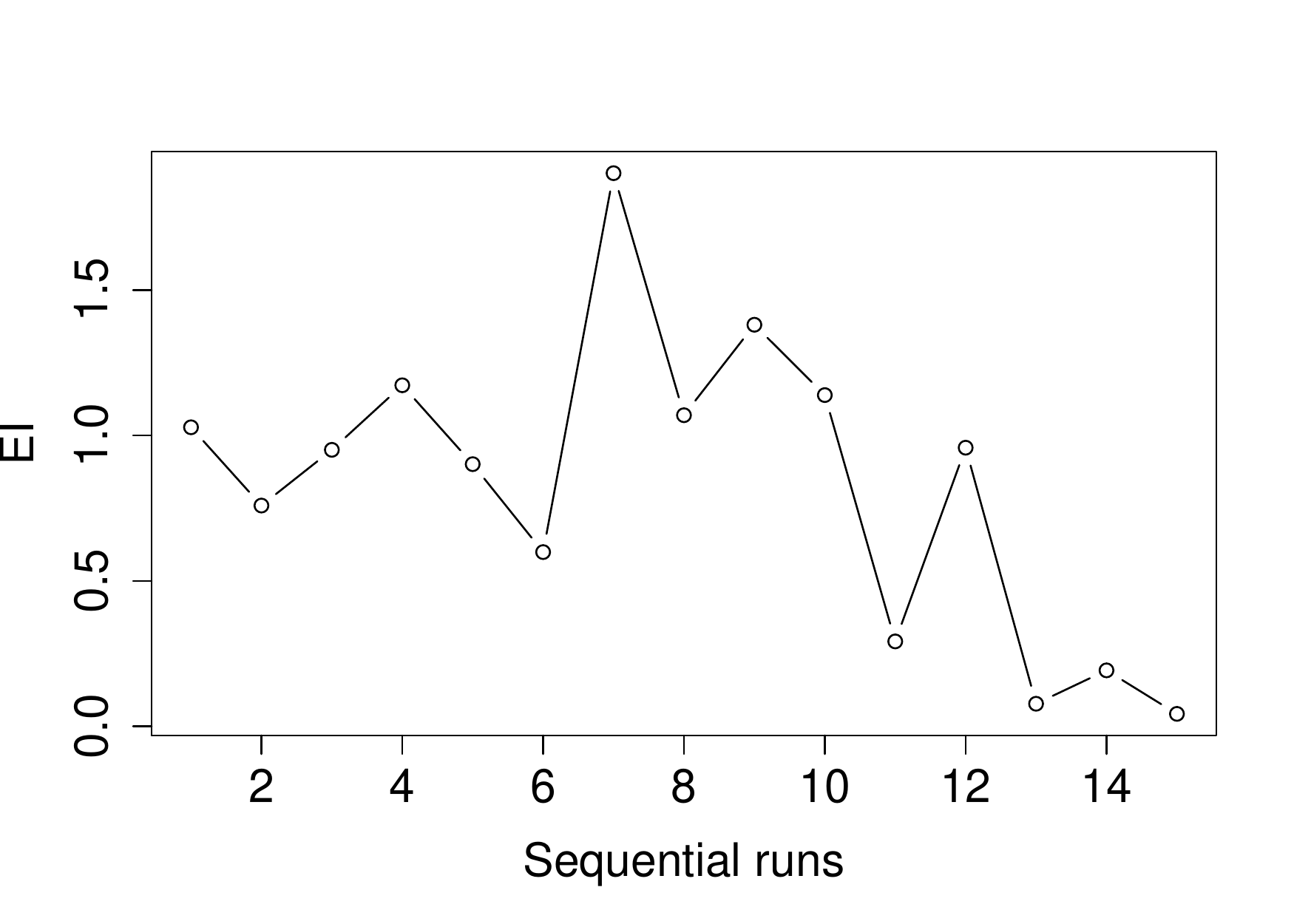}}\quad
  \subfigure [] {\includegraphics[scale=0.4]{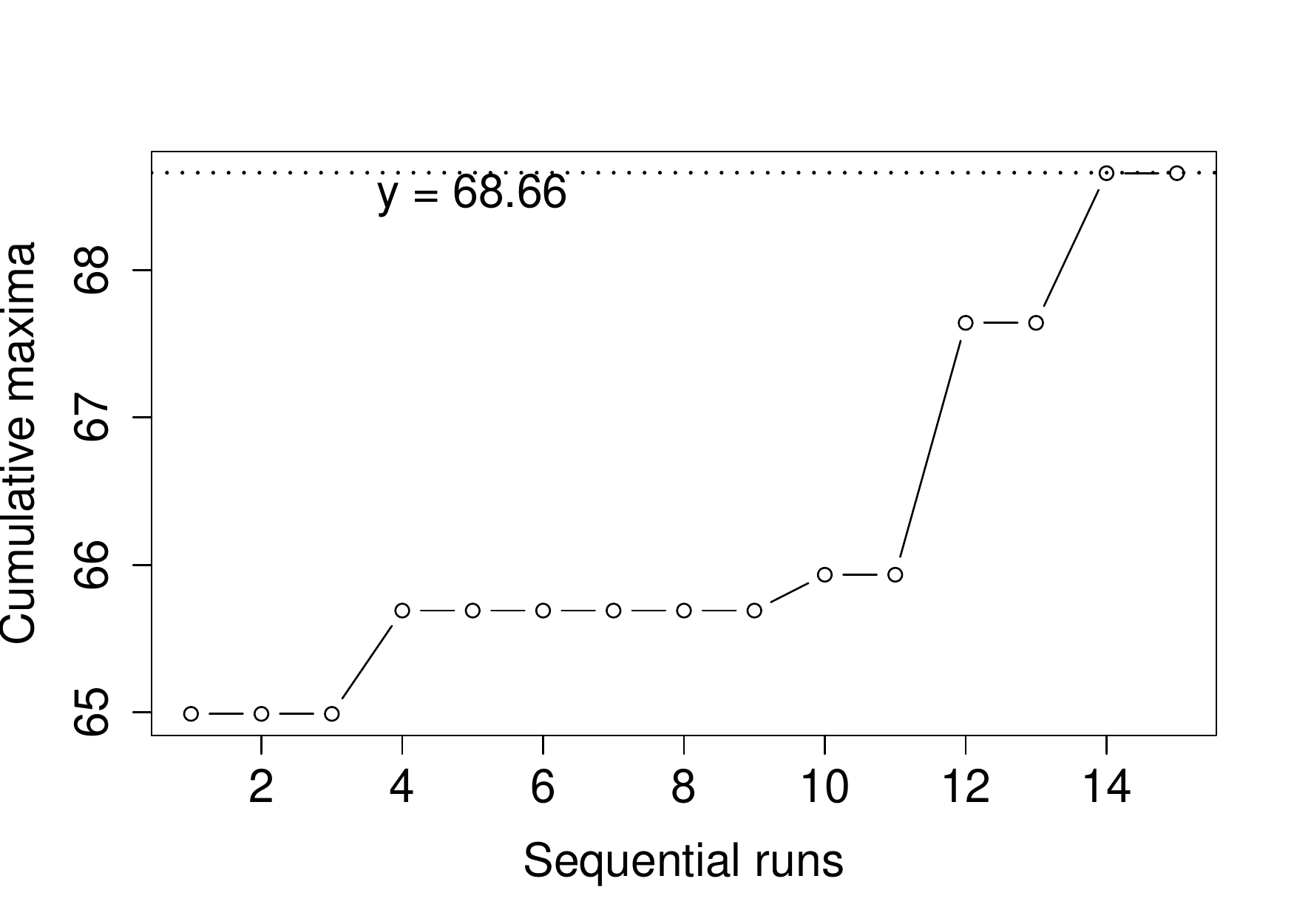}}
\caption{Plots for (a) expected improvements and (b) cumulative maximum responses from the QS-learning approach under the 2d-MaGP with 16-run QS-design.}
    \label{fig:ei2}
\end{figure}
\begin{figure}[ht]
\vspace{-.2in}
	\centering
	\subfigure []{\includegraphics[scale=0.4]{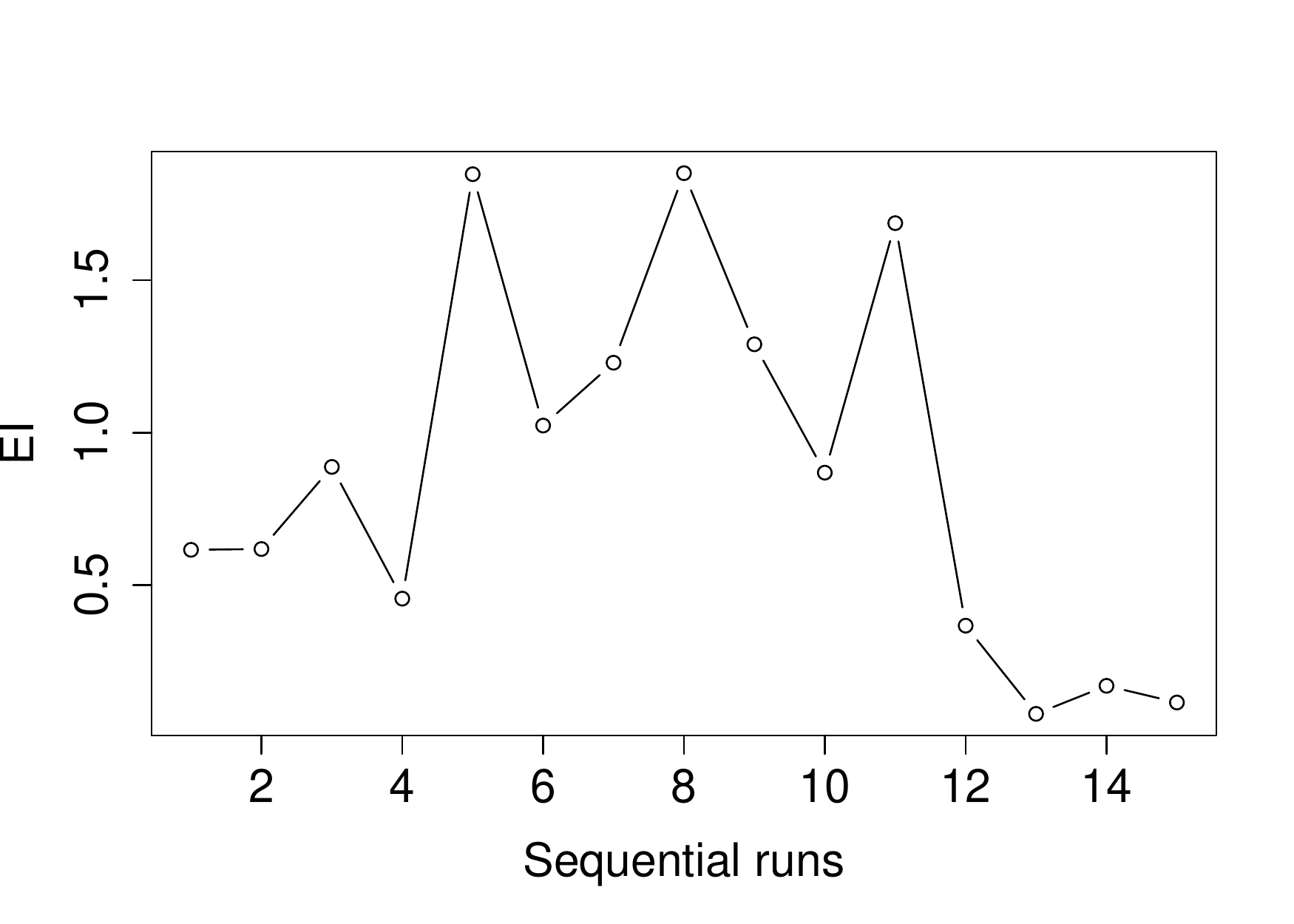}}\quad
	\subfigure [] {\includegraphics[scale=0.4]{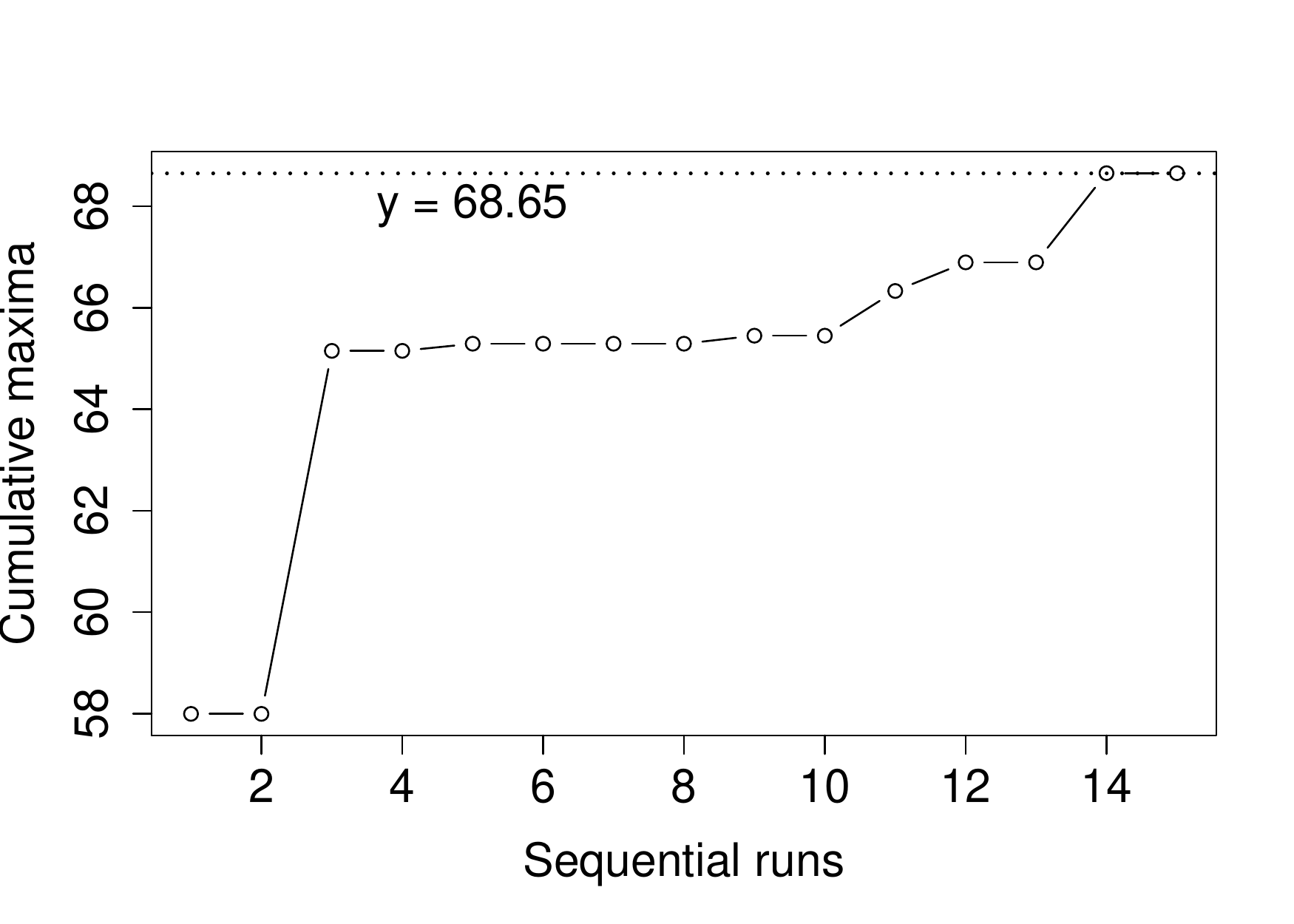}}
		\caption{Plots for (a) expected improvements and (b) cumulative maximum responses from the QS-learning approach under the full-MaGP with 16-run QS-design.}
			\label{fig:eifull1}
\end{figure}

Here we consider the QS-learning approach to identify the optimal setting for maximizing the response via experiment trials. We start from the QS-design with the rule-of-thumb run size of $2+k(k+3)/2)=16$ (as explained in Section~5), and then the QS-learning is to sequentially select the next design point.
When using the 2d-MaGP as the surrogate model, 15 sequential runs are selected until the stopping criterion is satisfied. The optimal settings of $(x_1, \ldots x_4) = (0.25, 0.40, 1.00, 1.00)$ (rounding to 2 decimal places) and $(o_1, \ldots, o_4) = (2,  4, 3, 1)$ are successfully identified in the $14^{th}$ sequential run. The maximum response found is $68.66$, where the minor difference from the truth ($68.67$) is due to rounding-off errors. Figure~\ref{fig:ei2} displays its expected improvements and cumulative maximum responses.
When using the full-MaGP, the QS-learning also stops after 15 sequential runs. The maximum response found is 68.65 and the best settings identified are $(x_1, \ldots, x_4) = (0.25, 0.39, 1, 1)$ and $(o_1, \ldots, o_4) = (2,4,3,1)$, which are still very close to the truth. Figure~\ref{fig:eifull1} displays its expected improvements and cumulative maximum responses.

Below, We show the QS-design with the rule-of-thumb run size of 16 along with the sequential runs in the QS-learning under the 2d-MaGP and the full-MaGP models, respectively.

\scriptsize
$$
\stackrel{\mbox{QS-design}}{
\kbordermatrix{
& x_1 & o_1 & x_2 & o_2 & x_3 & o_3 & x_4 & o_4 & Y \\
&0.03 & 1 & 0.78 & 2 & 0.22 & 3 & 0.84 & 4 & 19.88 \\
  &0.66 & 1 & 0.66 & 2 & 0.97 & 4 & 0.22 & 3 & 10.53 \\
  &0.47 & 1 & 0.34 & 3 & 0.47 & 2 & 0.03 & 4 & 19.55 \\
  &0.72 & 1 & 0.16 & 4 & 0.09 & 3 & 0.47 & 2 & 7.21 \\
  &0.34 & 2 & 0.91 & 1 & 0.53 & 3 & 0.16 & 4 & 22.74 \\
  &0.16 & 2 & 0.03 & 1 & 0.28 & 4 & 0.53 & 3 & 24.55 \\
  &0.84 & 4 & 0.47 & 1 & 0.72 & 2 & 0.97 & 3 & 14.71 \\
  &0.59 & 3 & 0.97 & 1 & 0.66 & 4 & 0.78 & 2 & 0.12 \\
  &0.78 & 2 & 0.72 & 4 & 0.16 & 1 & 0.59 & 3 & 12.93 \\
  &0.91 & 3 & 0.28 & 2 & 0.59 & 1 & 0.41 & 4 & 18.14 \\
  &0.09 & 3 & 0.22 & 4 & 0.84 & 1 & 0.28 & 2 & 24.90 \\
  &0.22 & 4 & 0.53 & 3 & 0.03 & 1 & 0.34 & 2 & 25.22 \\
  &0.28 & 2 & 0.59 & 3 & 0.91 & 4 & 0.72 & 1 & 56.75 \\
  &0.53 & 4 & 0.09 & 2 & 0.78 & 3 & 0.66 & 1 & 10.56 \\
  &0.41 & 3 & 0.41 & 4 & 0.34 & 2 & 0.91 & 1 & 26.17 \\
  &0.97 & 4 & 0.84 & 3 & 0.41 & 2 & 0.09 & 1 & 12.52 \\
}
}
\stackrel{\mbox{Sequential runs under 2d-MaGP}}{
\kbordermatrix{
  &x_1 & o_1 & x_2 & o_2 & x_3 & o_3 & x_4 & o_4 & Y \\
  &0.28 & 2 & 0.61 & 4 & 0.93 & 3 & 0.86 & 1 & 64.99 \\
  &0.27 & 2 & 0.61 & 4 & 1.00 & 1 & 1.00 & 3 & 27.86 \\
  &0.36 & 2 & 0.54 & 4 & 0.92 & 3 & 0.84 & 1 & 56.93 \\
  &0.25 & 2 & 0.28 & 4 & 0.91 & 3 & 0.85 & 1 & 65.69 \\
  &0.23 & 2 & 1.00 & 4 & 0.96 & 3 & 0.88 & 1 & 62.96 \\
  &0.26 & 2 & 0.00 & 3 & 0.89 & 4 & 0.91 & 1 & 53.86 \\
  &0.24 & 2 & 0.49 & 4 & 0.93 & 3 & 0.76 & 1 & 65.20 \\
  &0.25 & 4 & 0.28 & 2 & 0.93 & 3 & 0.78 & 1 & 26.99 \\
  &0.26 & 2 & 1.00 & 4 & 0.90 & 3 & 1.00 & 1 & 63.25 \\
  &0.26 & 2 & 0.00 & 4 & 0.94 & 3 & 1.00 & 1 & 65.93 \\
  &0.26 & 1 & 0.19 & 4 & 0.94 & 3 & 0.87 & 2 & 36.56 \\
  &0.25 & 2 & 0.36 & 4 & 0.94 & 3 & 1.00 & 1 & 67.64 \\
  &0.00 & 2 & 0.40 & 4 & 0.94 & 3 & 1.00 & 1 & 28.23 \\
  &\textbf{0.25} & \textbf{2} & \textbf{0.40} & \textbf{4} & \textbf{1.00} & \textbf{3} & \textbf{1.00} & \textbf{1} & \textbf{68.66} \\
  &0.25 & 2 & 0.40 & 4 & 0.33 & 3 & 1.00 & 1 & 56.78 \\
}
}
$$

$$
\stackrel{\mbox{Sequential runs under full-MaGP}}{
	\kbordermatrix{
  & x_1 & o_1 & x_2 & o_2 & x_3 & o_3 & x_4 & o_4 & Y \\
  &0.27 & 2 & 0.63 & 3 & 0.90 & 4 & 0.91 & 1 & 57.99 \\
  &0.26 & 2 & 0.33 & 1 & 0.90 & 4 & 0.91 & 3 & 36.47 \\
  &0.28 & 2 & 0.61 & 4 & 0.90 & 3 & 0.91 & 1 & 65.15 \\
  &0.25 & 2 & 0.61 & 4 & 0.35 & 1 & 1.00 & 3 & 23.54 \\
  &0.28 & 2 & 0.61 & 4 & 0.90 & 3 & 0.96 & 1 & 65.29 \\
  &0.28 & 3 & 1.00 & 4 & 0.91 & 2 & 1.00 & 1 & 31.27 \\
  &0.06 & 2 & 0.60 & 4 & 0.90 & 3 & 1.00 & 1 & 41.92 \\
  &0.29 & 2 & 0.61 & 4 & 0.90 & 3 & 0.90 & 1 & 64.57 \\
  &0.28 & 2 & 0.64 & 4 & 0.91 & 3 & 0.97 & 1 & 65.45 \\
  &0.28 & 2 & 1.00 & 4 & 0.90 & 3 & 1.00 & 1 & 62.61 \\
  &0.27 & 2 & 0.16 & 4 & 0.91 & 3 & 1.00 & 1 & 66.33 \\
  &0.27 & 2 & 0.36 & 4 & 0.91 & 3 & 1.00 & 1 & 66.89 \\
  &1.00 & 2 & 0.37 & 4 & 0.91 & 3 & 1.00 & 1 & 28.00 \\
  &\textbf{0.25} & \textbf{2} & \textbf{0.39} & \textbf{4} & \textbf{1.00} & \textbf{3} & \textbf{1.00} & \textbf{1} & \textbf{68.65} \\
  &0.27 & 2 & 0.24 & 3 & 0.00 & 4 & 1.00 & 1 & 46.05 \\
	}
}
$$
\normalsize	

\newpage


\begin{figure}[ht]
\vspace{-.2in}
    \centering
  \subfigure []{\includegraphics[scale=0.4]{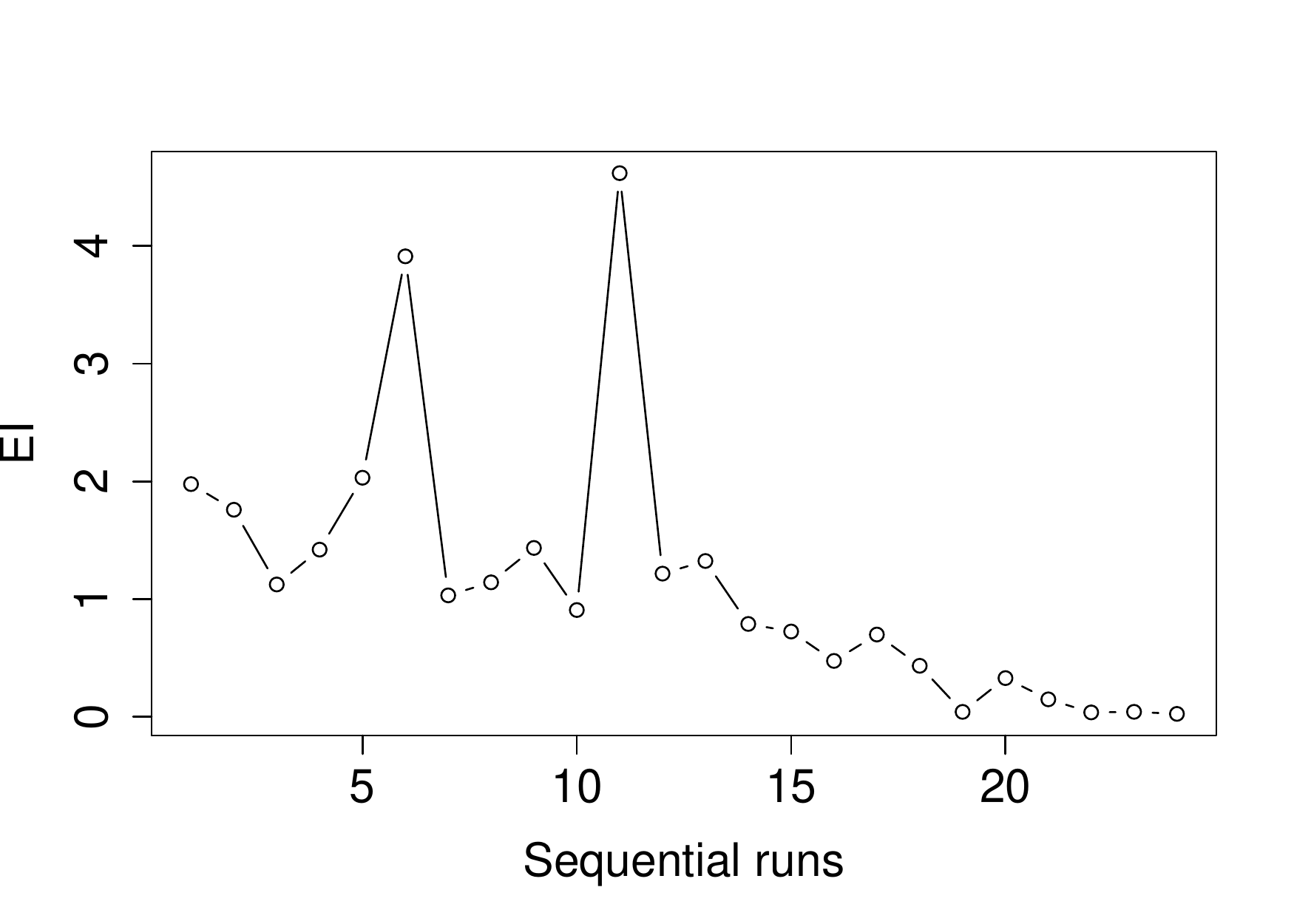}}\quad
  \subfigure [] {\includegraphics[scale=0.4]{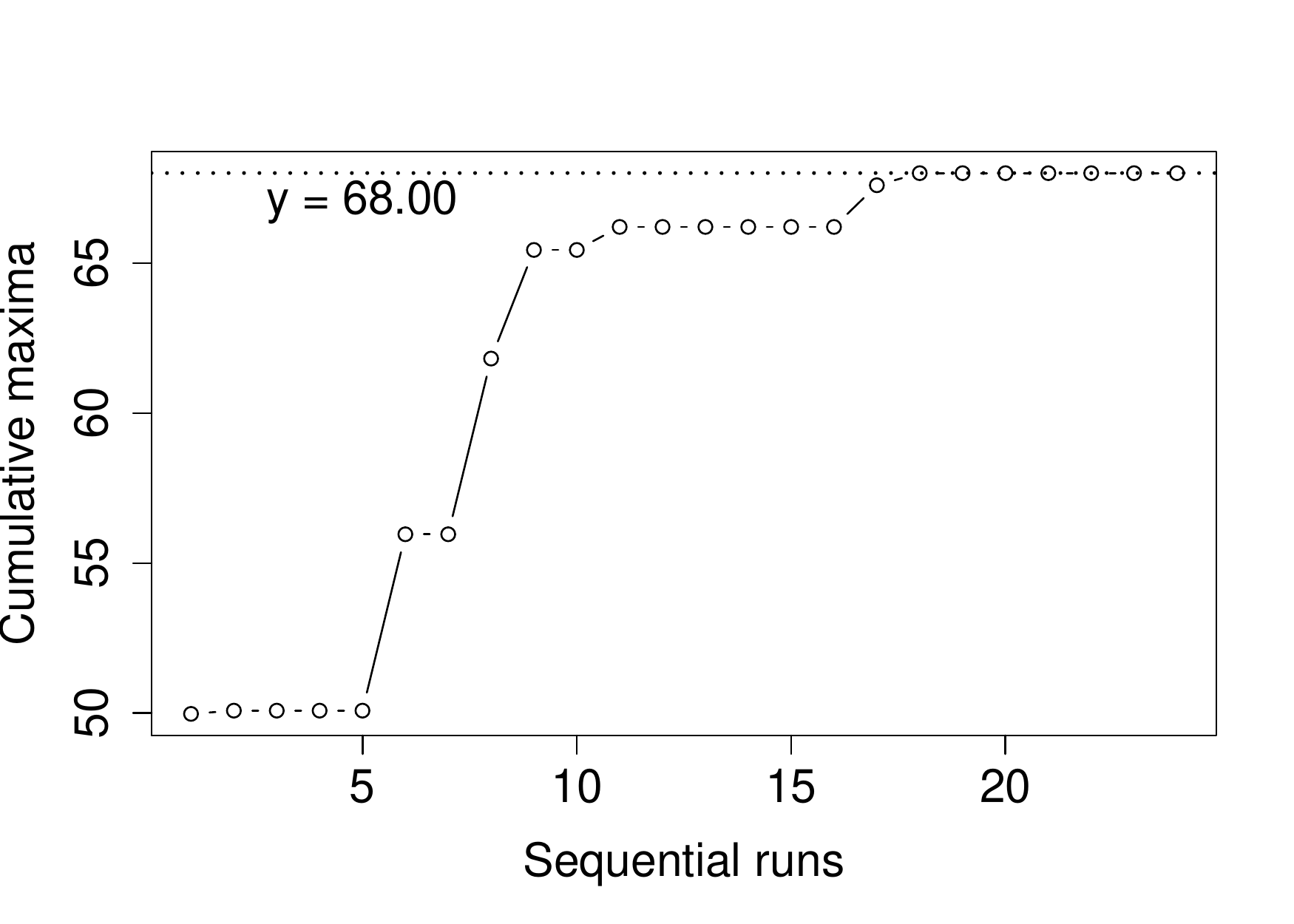}}
\caption{Plots for (a) expected improvements and (b) cumulative maximum responses from the QS-learning under the 2d-MaGP using the 4-run QS-design.}
    \label{fig:eifullfour}
\end{figure}

Next, we consider the QS-learning starting from the 4-run QS-design constructed by the algebraic method in Section~5.2. When using the 2d-MaGP model, 24 sequential runs are selected until the stopping criterion is satisfied. The best settings found are $(x_1, \ldots x_4) = (0.25, 0.40, 1.00, 1.00)$ and $(o_1, \ldots, o_4) = (3,1,4,2)$. Here, the identified settings for the quantitative inputs $(x_1, \ldots x_4)$ are optimal. Given these quantities, the identified setting for the sequence inputs $(3, 1, 4, 2)$ is the second best, which leads to the second largest response $68.00$ in this study.
When using the full-MaGP model, 22 sequential runs are selected and the identified maximum response is 67.99 with $(x_1, \ldots x_4) = (0.25, 0.42, 1.00, 1.00)$ and $(o_1, \ldots, o_4) = (3,1,4,2)$. As the QS-design from the proposed algebraic construction is small, the 2d-MaGP is often recommended for practical use. Figure~\ref{fig:eifullfour} displays its expected improvements and cumulative maximum responses.

\newpage

Below, we show the 4-run QS-design from the algebraic construction, along with the sequential runs in QS-learning under the 2d-MaGP and full-MaGP model, respectively.

\scriptsize
$$
\stackrel{\mbox{QS-design}}{
\kbordermatrix{
& x_1 & o_1 & x_2 & o_2 & x_3 & o_3 & x_4 & o_4 & Y \\
& 0.38 & 1 & 0.12 & 2 & 0.62 & 3 & 0.88 & 4 & 29.37 \\
& 0.88 & 2 & 0.38 & 4 & 0.12 & 1 & 0.62 & 3 & 14.71 \\
&  0.12 & 3 & 0.62 & 1 & 0.88 & 4 & 0.38 & 2 & 49.98 \\
&  0.62 & 4 & 0.88 & 3 & 0.38 & 2 & 0.12 & 1 & 7.09 \\
}
}
\stackrel{\mbox{Sequential runs under 2d-MaGP}}{
\kbordermatrix{
  & x_1 & o_1 & x_2 & o_2 & x_3 & o_3 & x_4 & o_4 & Y \\
  &0.12 & 3 & 1.00 & 1 & 0.00 & 4 & 0.38 & 2 & 35.99 \\
  &0.10 & 3 & 0.00 & 1 & 1.00 & 4 & 1.00 & 2 & 50.08 \\
  &0.50 & 2 & 1.00 & 1 & 1.00 & 4 & 1.00 & 3 & 20.91 \\
  &1.00 & 3 & 0.62 & 1 & 1.00 & 4 & 0.00 & 2 & 21.49 \\
  &0.04 & 3 & 0.00 & 2 & 0.84 & 4 & 0.77 & 1 & 24.46 \\
  &0.13 & 3 & 0.62 & 1 & 0.91 & 4 & 1.00 & 2 & 55.96 \\
  &0.16 & 1 & 0.53 & 3 & 0.85 & 4 & 1.00 & 2 & 29.55 \\
  &0.17 & 3 & 0.63 & 1 & 1.00 & 4 & 1.00 & 2 & 61.82 \\
  &0.20 & 3 & 0.61 & 1 & 1.00 & 4 & 1.00 & 2 & 65.44 \\
  &0.22 & 3 & 1.00 & 4 & 1.00 & 1 & 1.00 & 2 & 31.94 \\
  &0.20 & 3 & 0.30 & 1 & 1.00 & 4 & 1.00 & 2 & 66.20 \\
  &0.22 & 4 & 0.00 & 1 & 0.95 & 3 & 1.00 & 2 & 32.36 \\
  &0.22 & 3 & 0.00 & 1 & 0.97 & 4 & 0.92 & 2 & 63.92 \\
  &0.22 & 3 & 0.00 & 1 & 0.99 & 2 & 1.00 & 4 & 25.39 \\
  &0.22 & 3 & 0.00 & 2 & 1.00 & 4 & 1.00 & 1 & 55.74 \\
  &0.22 & 3 & 0.00 & 1 & 1.00 & 4 & 1.00 & 2 & 65.11 \\
  &0.23 & 3 & 0.39 & 1 & 1.00 & 4 & 1.00 & 2 & 67.59 \\
  &\textbf{0.25} & \textbf{3} & \textbf{0.40} & \textbf{1} & \textbf{1.00} & \textbf{4} & \textbf{1.00} & \textbf{2} & \textbf{68.00} \\
  &0.86 & 4 & 0.41 & 1 & 0.00 & 3 & 1.00 & 2 & 11.44 \\
  &0.25 & 3 & 0.41 & 1 & 0.00 & 2 & 0.00 & 4 & 16.25 \\
  &0.25 & 3 & 0.41 & 1 & 0.44 & 4 & 1.00 & 2 & 58.49 \\
  &0.25 & 3 & 0.57 & 2 & 1.00 & 4 & 0.00 & 1 & 54.88 \\
  &0.59 & 3 & 0.41 & 1 & 1.00 & 4 & 1.00 & 2 & 6.27 \\
  &0.53 & 4 & 0.39 & 1 & 1.00 & 3 & 1.00 & 2 & 22.88 \\
}
}
$$
$$
\stackrel{\mbox{Sequential runs under full-MaGP}}{
	\kbordermatrix{
  &0.05 & 3 & 0.54 & 1 & 0.00 & 4 & 1.00 & 2 & 29.39 \\
  &1.00 & 3 & 0.27 & 1 & 1.00 & 4 & 0.46 & 2 & 24.16 \\
  &0.05 & 3 & 0.72 & 1 & 0.82 & 4 & 1.00 & 2 & 36.46 \\
  &0.17 & 2 & 0.00 & 1 & 0.82 & 4 & 0.34 & 3 & 27.34 \\
  &0.15 & 3 & 1.00 & 2 & 0.88 & 4 & 0.87 & 1 & 38.48 \\
  &0.29 & 3 & 0.67 & 1 & 0.85 & 4 & 0.34 & 2 & 59.46 \\
  &0.39 & 3 & 0.73 & 1 & 0.80 & 4 & 0.35 & 2 & 45.53 \\
  &0.25 & 3 & 0.72 & 1 & 0.89 & 4 & 0.31 & 2 & 60.72 \\
  &0.25 & 3 & 0.00 & 1 & 0.89 & 4 & 0.00 & 2 & 58.80 \\
  &0.24 & 3 & 0.67 & 4 & 0.84 & 1 & 1.00 & 2 & 33.85 \\
  &0.25 & 4 & 1.00 & 1 & 0.87 & 3 & 0.00 & 2 & 24.92 \\
  &0.30 & 3 & 0.68 & 1 & 0.88 & 4 & 0.38 & 2 & 58.99 \\
  &0.26 & 1 & 1.00 & 3 & 0.88 & 2 & 0.24 & 4 & 30.43 \\
  &0.26 & 3 & 1.00 & 1 & 0.88 & 4 & 0.00 & 2 & 56.54 \\
  &0.26 & 3 & 0.00 & 1 & 0.92 & 4 & 0.39 & 2 & 60.89 \\
  &0.25 & 3 & 0.38 & 1 & 0.93 & 4 & 0.40 & 2 & 62.79 \\
  &0.25 & 3 & 0.40 & 1 & 1.00 & 4 & 0.49 & 2 & 64.51 \\
  &0.26 & 3 & 0.35 & 1 & 1.00 & 4 & 0.55 & 2 & 64.82 \\
  &0.25 & 3 & 0.58 & 1 & 1.00 & 4 & 1.00 & 2 & 67.56 \\
  &0.25 & 3 & 0.42 & 1 & 0.48 & 4 & 1.00 & 2 & 59.15 \\
  &\textbf{0.25} & \textbf{3} & \textbf{0.42} & \textbf{1} & \textbf{1.00} & \textbf{4} & \textbf{1.00} & \textbf{2} & \textbf{67.99} \\
  &0.26 & 3 & 0.39 & 1 & 0.00 & 2 & 0.88 & 4 & 20.82 \\
	}
}
$$

\normalsize

\begin{figure}[htbp]
	\centering
	\includegraphics[width=0.7\linewidth]{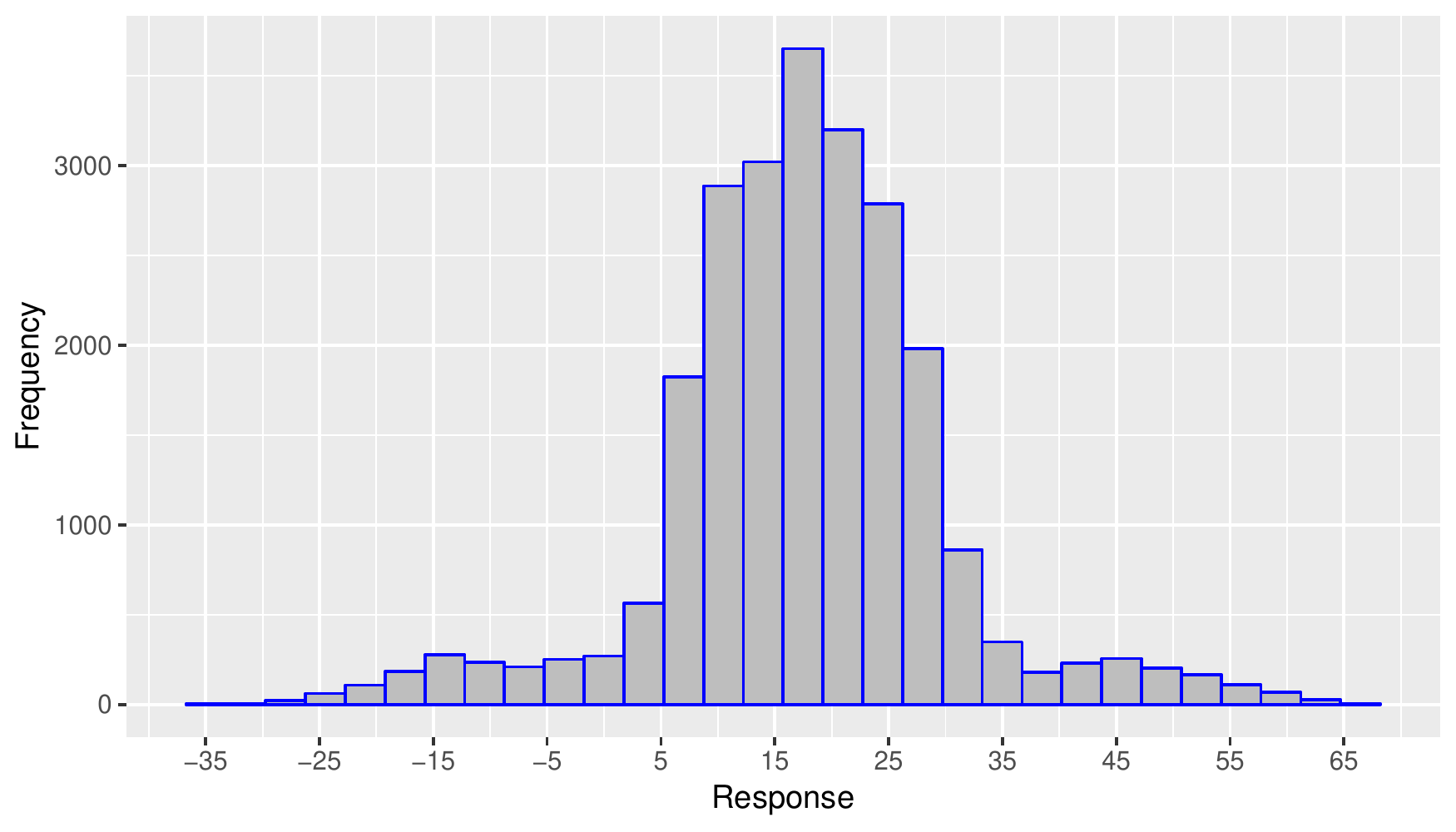} \vspace{-.2in}
	\caption{Histogram of largest responses identified by $BM_{1}$.}
	\label{fig:histfour}
\end{figure}

To compare the proposed QS-learning with some benchmark methods, we first consider a random sampling approach ($BM_1$) which is equivalent to using a large random one-shot experimental design. In $BM_1$, we draw a very large random sample consisting of 24,000 observations using a random design $D_{r}=(X_{r},O_{r})$ whose quantitative part $X_{r}$ is a random Latin hypercube design and sequence part $O_{r}$ includes a thousand replicates of all possible sequences.
The maximum response found by $BM_1$ is 66.85, which is clearly inferior compared to that by QS-learning. The histogram of largest responses identified by $BM_{1}$ is shown in Figure~\ref{fig:histfour}.

\begin{figure}[ht]
\vspace{-.2in}
	\centering
	\subfigure []{\includegraphics[scale=0.4]{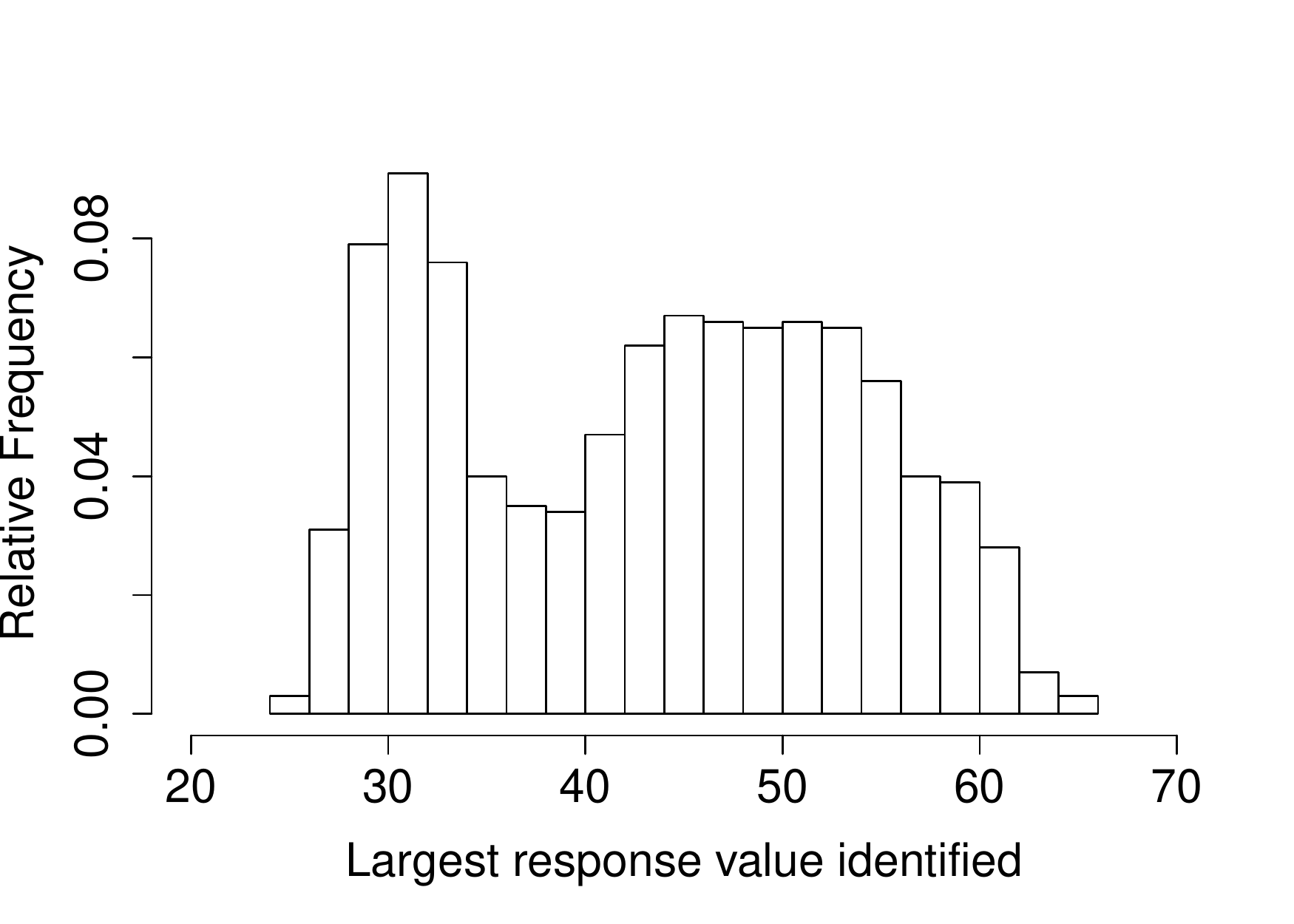}}\quad
	\subfigure [] {\includegraphics[scale=0.4]{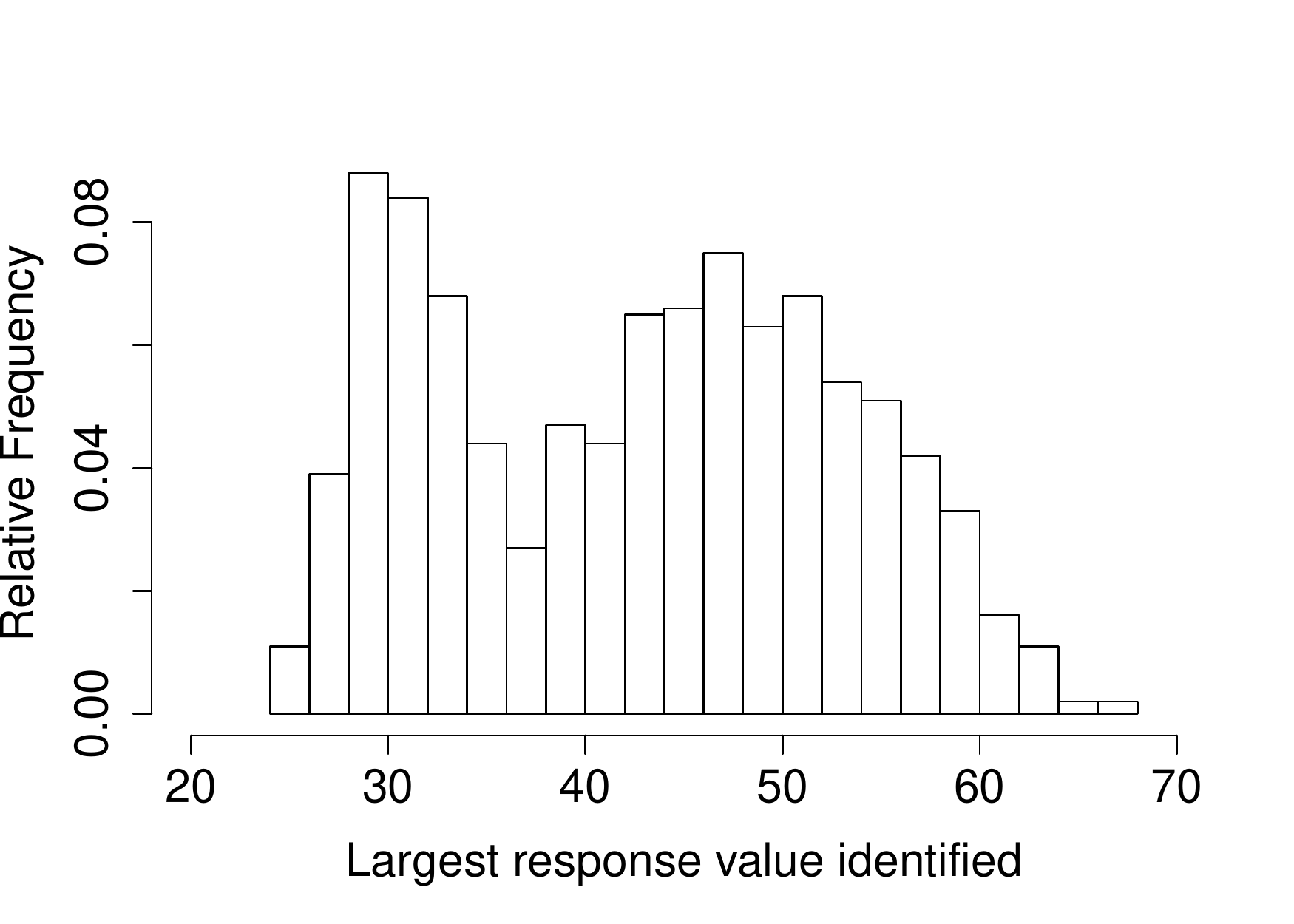}}
\caption{Histograms of largest responses identified by (a) $BM_{2}$ and (b) $BM_{3}$ approaches.}
	\label{fig:hist0}
\end{figure}

We also consider the sequential generalized PWO ($BM_2$) and CP ($BM_3$) approaches. Both methods start from random initial designs of 16 runs, where random Latin hypercube designs and random subsets of all possible sequences are used for the quantitative and sequence parts, respectively.
We replicate both methods 1000 times. Figure~\ref{fig:hist0} shows the histograms of their maximum responses found.
Even the best results in these 1000 replications (65.52 for $BM_{1}$ and 66.51 for $BM_{2}$) are worse than that by  QS-learning. Moreover, most results in Figure~\ref{fig:hist0} are not good.

Finally, we evaluate the prediction accuracy of the proposed MaGP model compared to the generalized PWO and CP models.
In Table~\ref{tab:fopre}, we report the prediction root mean square errors (RMSEs) of different models under various one-shot designs. Since the rule-of-thumb run size for the QS-design is 16 and the QS-learning discussed above includes 28 and 31 runs in total, here we consider random one-shot designs having run sizes of 16, 28 and 31 to form the training sets. In random designs, the quantitative parts are random Latin hypercube designs and the sequence parts include randomly selected sequences. The testing set is formed under a large 2400-run random design whose quantitative part is a random Latin hypercube design and sequence part includes 100 replicates of all possible sequences. We replicate the analysis 100 times and report both the median and mean of RMSEs in Table~\ref{tab:fopre}. In addition, we also list the results from all models using the training data formed by the proposed 16-run QS-design.

From Table~\ref{tab:fopre}, it is seen that both the 2d-MaGP and full-MaGP models have smaller RMSEs (thus stronger prediction power) compared to the generalized PWO and CP models for all cases.
Specifically, the proposed MaGP models perform very well under the proposed QS-design. The results from this 16-run design are even better than the median (or average) results from random 31-run designs. This justifies the necessity of using QS-designs to collect initial data in QS-learning.

\begin{table}[htbp]
  \centering
  \caption{The ``medians(means)" of the RMSEs from different models under various designs.}
    \begin{tabular}{lcccc}
    \toprule
          & PWO   & CP    & 2d-MaGP & full-MaGP \\
          \midrule
    Random 16-run & 14.81(15.44) & 20.57(22.30) & 11.86(12.11) & 11.59(11.99) \\
    Random 28-run & 11.76(12.12) & 13.19(13.60) & 10.86(11.47) & 11.01(11.29) \\
    Random 31-run & 11.08(11.29) & 12.19(12.61) & 10.70(10.84) & 10.69(10.90) \\
    QS-design 16-run & 15.25 & 15.63 & 9.41  & 10.01 \\
    \bottomrule
    \end{tabular}%
  \label{tab:fopre}%
\end{table}%

\subsection{Additional Simulations on  Single Machine Scheduling}
\label{smsp}

Single machine scheduling (SMS) is an NP-hard optimization problem in literature \citep{emmons1969one, allahverdi1999review, wan2013single}, where $k$ jobs are to be sequenced on a single machine. Here we consider a complex SMS problem whose object function is assumed to be black-box and expensive-to-evaluate. In the following Example~\ref{ex:SMS:BO}, we consider a classic SMS where only jobs' arrangement sequences are to be optimized. We show that the proposed QS-learning can also work well for only sequence inputs. In  Example~\ref{ex:SMS:BO2}, we further discuss a general SMS where both jobs' processing time and their arrangement sequences are to be optimized.
\begin{example}
	\label{ex:SMS:BO}
	Consider an SMS problem consisting of $k = 6$ jobs, indexed by Jobs $1, \ldots,6$. These jobs need to be processed on a single machine one after another, where Job $i$ takes a fixed processing time $x_i$ for $i=1,\ldots,6$. In this example, we randomly assign standardized $\bx = (x_1, \ldots , x_6) = (0.96, 0.74, 0.87, 0.43, 0.51, 0.64)$. Denote the sequence for arranging jobs as $\bm{\alpha} = (\alpha_1, \ldots, \alpha_k)$ and its corresponding order-sequence as $\bo = (o_1, \ldots, o_k)$, where Job $\alpha_i$  has the processing order $o_{\alpha_i} = i$. The completion time of Job $\alpha_i$ is $T(\alpha_i) = \sum_{j=1}^{i} x_{\alpha_j}$. Consider the quadratic cost of completion time by  \cite{townsend1978single}, which is defined as
	$$
	C(\bm{\alpha}) = \sum_{i=1}^{k}w_i T^2(\alpha_i),
	$$
	where $(w_1, \ldots, w_k) = (0.3, 0.6, 0.1, 0.9, 0.8, 0.5)$ are randomly assigned weights.
There are $6! = 720$ possible sequences in total and the cost function $C(\bm{\alpha})$ has a unique global minimum of $22.43$ at $\bm{\alpha}_* = (4, 5, 6, 2, 3, 1)$ or equivalently $\bo_* = (6,4,5,1,2,3)$.
\end{example}

\begin{figure}[htbp]
\vspace{-.2in}
	\centering
	\subfigure []{\includegraphics[scale=0.3]{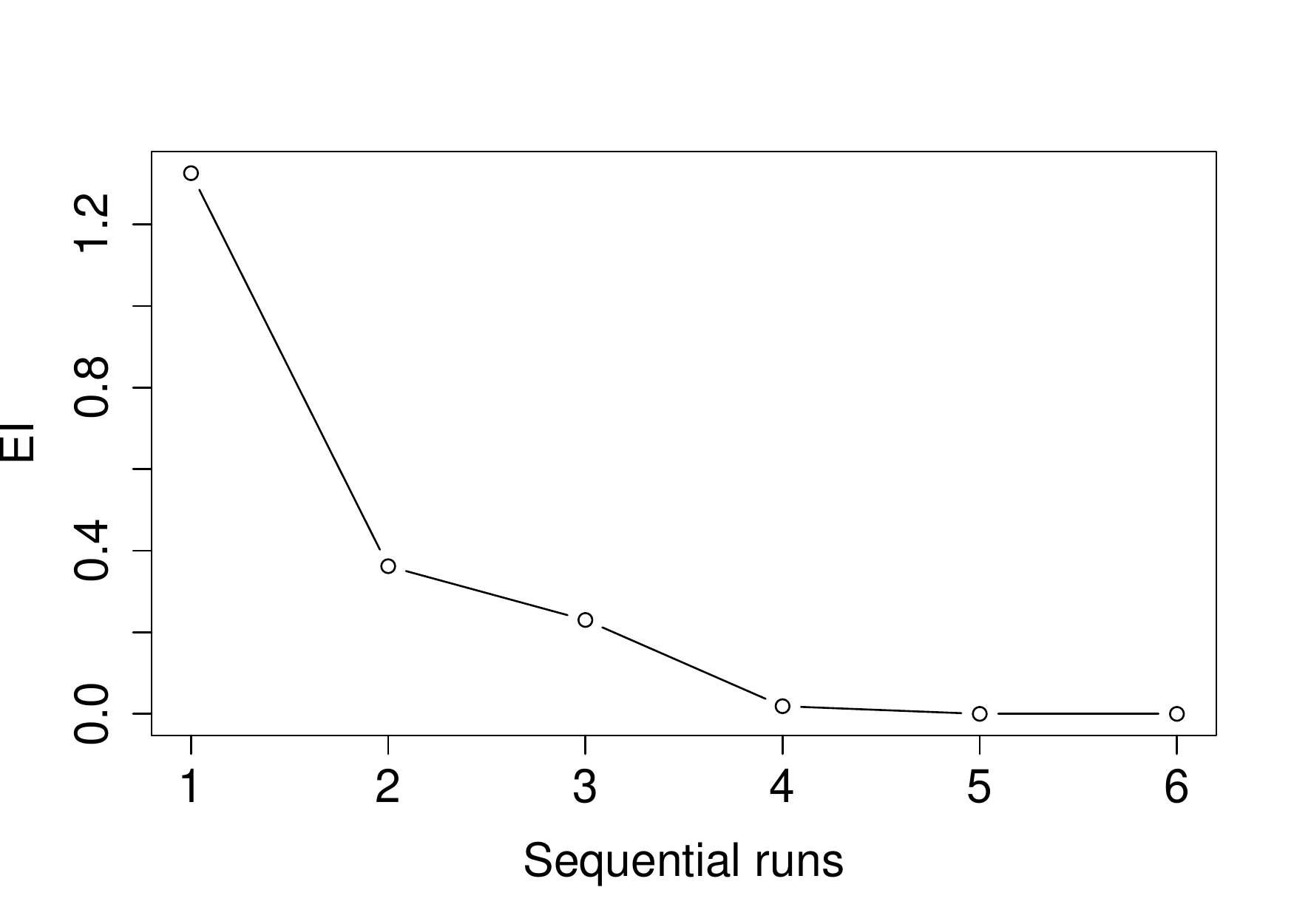}}\quad
	\subfigure [] {\includegraphics[scale=0.3]{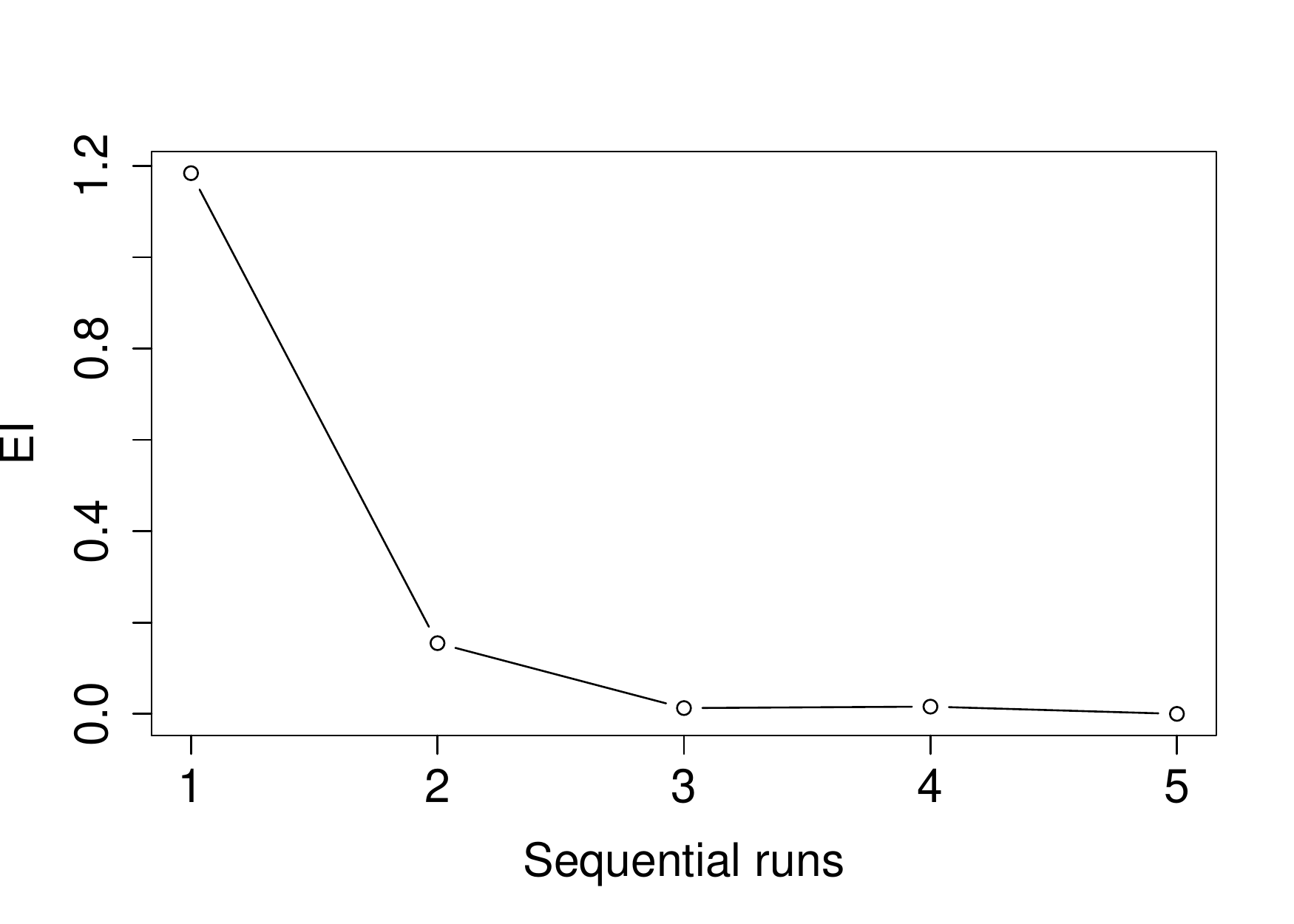}}
	\caption{Plots for expected improvements of sequential runs in QS-learning under (a) 2d-MaGP and (b) full-MaGP in Example~\ref{ex:SMS:BO}.}
		\label{fig:eifull2}
\end{figure}

\begin{figure}[htbp]
\vspace{-.2in}
	\centering
	\subfigure []{\includegraphics[scale=0.3]{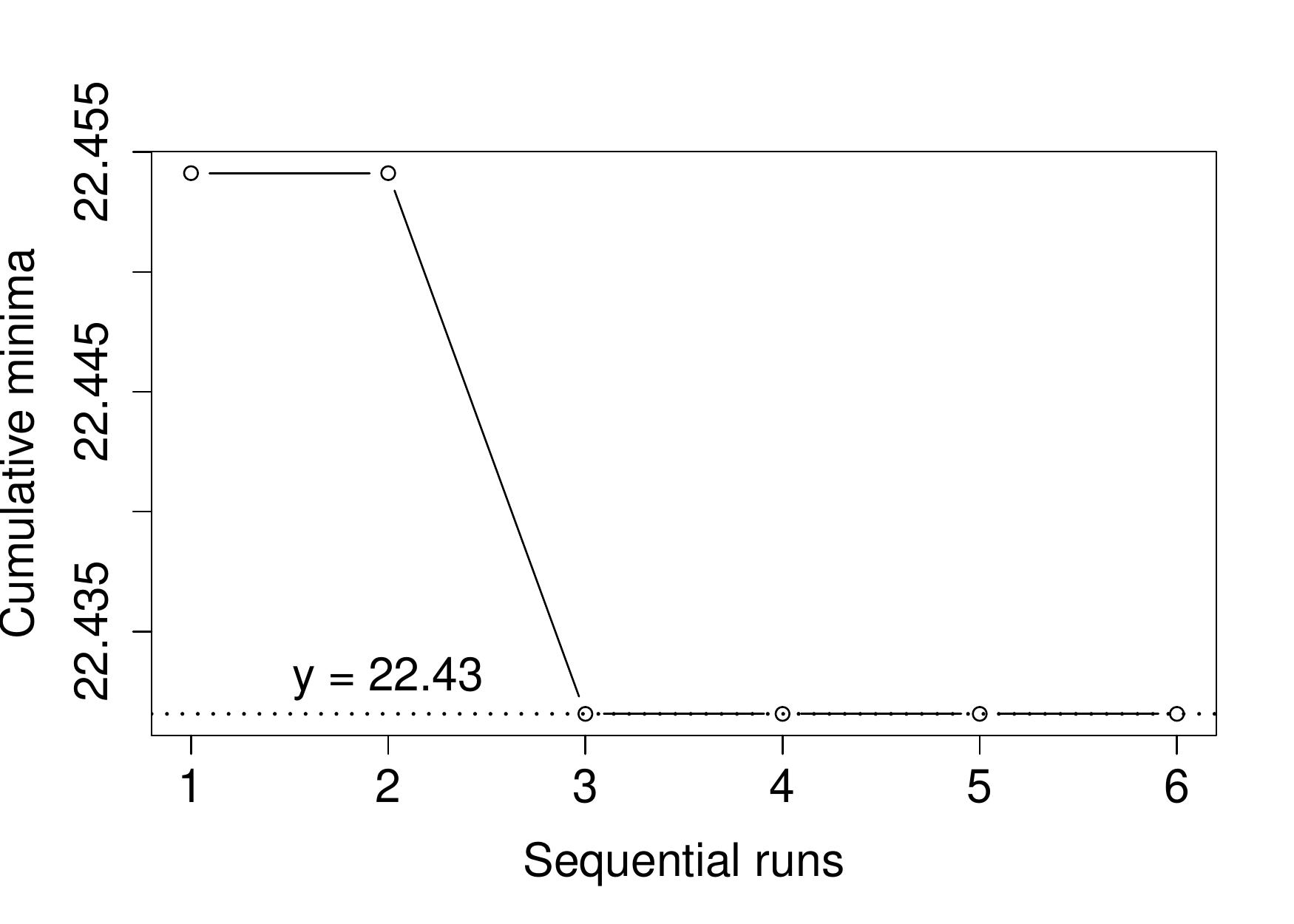}}\quad
	\subfigure [] {\includegraphics[scale=0.3]{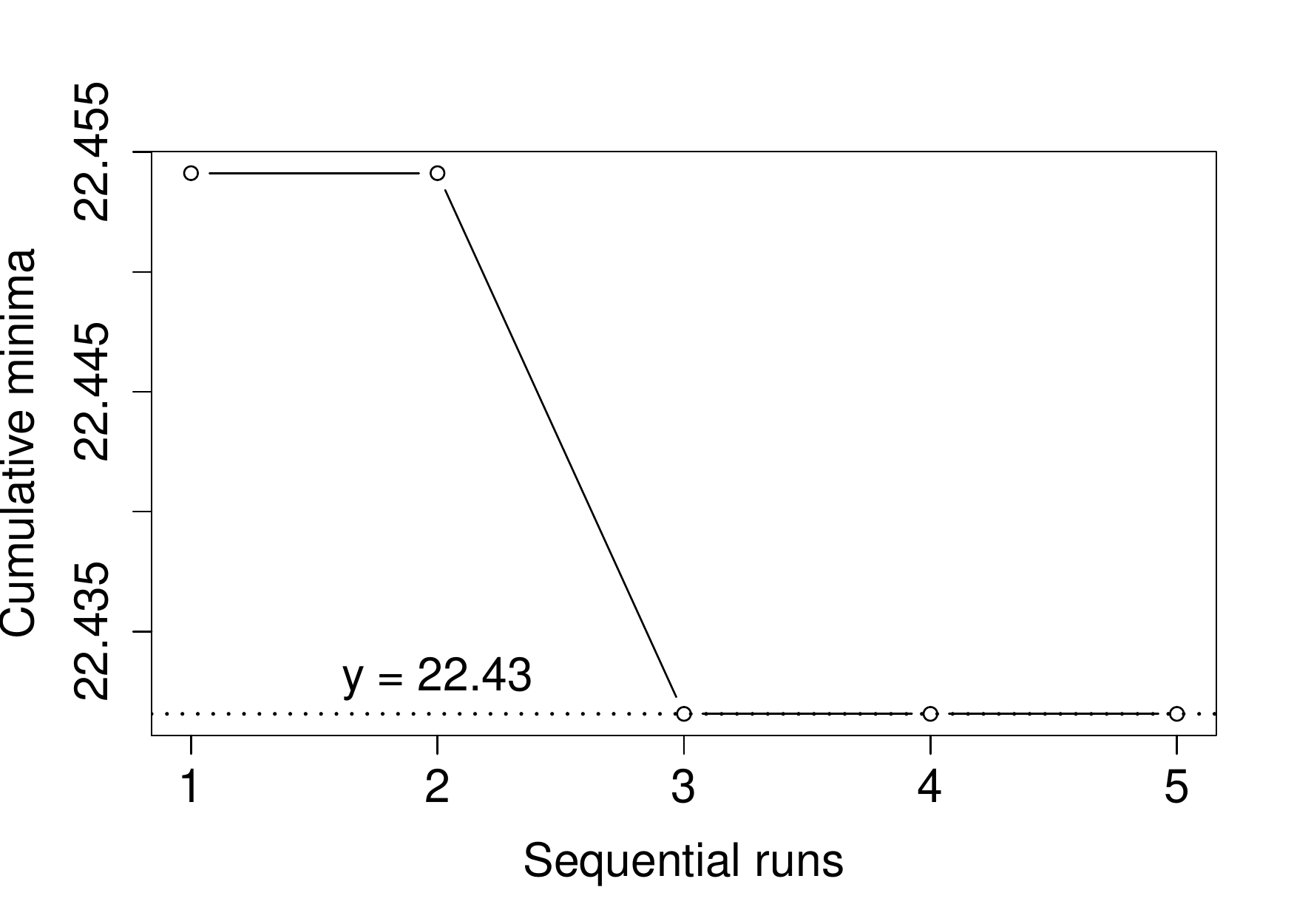}}
	\caption{Plots for cumulative maximum responses of sequential runs in  QS-learning under (a) the 2d-MaGP and (b) the full-MaGP in Example~\ref{ex:SMS:BO}.}
		\label{fig:cum3}
\end{figure}

We consider the proposed QS-learning to minimize the cost via experimental trials. As all processing time (quantitative inputs) are fixed, there are 15 and 21 parameters in the GP parts of the 2d-MaGP and full-MaGP models, respectively. Thus, we start from the QS-designs consisting of 15 and 21 runs (with fixed $X$ part), respectively.
When using the 2d-MaGP model as the surrogate, 6 sequential runs are selected before the stopping criterion is satisfied. When using the full-MaGP model, 5 sequential runs are selected. Figures~\ref{fig:eifull2} and \ref{fig:cum3} display the expected improvements and cumulative maximum responses of the selected sequential runs, respectively. It is seen that the QS-learning using both models can successfully identify the true minimum of $22.43$ and the optimal sequence $\bo_* = (6,4,5,1,2,3)$. In addition, we also try the QS-learning starting from the 6-run QS-design constructed via the algebraic construction (for part $O$ only). When using 2d-MaGP, 9 sequential runs are selected and the optimal sequence $\bo_* = (6,4,5,1,2,3)$ is found. 

Below, we first the 15-run QS-design along with its sequential runs from the QS-learning under 2d-MaGP, and then show the 21-run QS-design along with its sequential runs from the QS-learning under full-MaGP.

\scriptsize
$$
\stackrel{\mbox{QS-design}}{
	\kbordermatrix{
		& o_1 & o_2 & o_3 & o_4 & o_5 & o_6 & Y \\
		&4 & 2 & 6 & 5 & 3 & 1 & 26.15 \\
		&4 & 3 & 1 & 5 & 6 & 2 & 30.59 \\
		&1 & 5 & 3 & 2 & 6 & 4 & 28.73 \\
		&5 & 6 & 3 & 2 & 4 & 1 & 24.50 \\
		&1 & 2 & 5 & 4 & 3 & 6 & 27.24 \\
		&2 & 6 & 1 & 4 & 5 & 3 & 28.33 \\
		&2 & 3 & 4 & 6 & 5 & 1 & 31.16 \\
		&3 & 4 & 2 & 5 & 1 & 6 & 28.77 \\
		&1 & 4 & 3 & 6 & 5 & 2 & 31.38 \\
		&4 & 6 & 2 & 1 & 3 & 5 & 26.22 \\
		&3 & 2 & 1 & 6 & 4 & 5 & 30.66 \\
		&5 & 2 & 4 & 1 & 6 & 3 & 26.88 \\
		&6 & 4 & 5 & 2 & 1 & 3 & 22.45 \\
		&2 & 4 & 6 & 3 & 1 & 5 & 25.23 \\
		&6 & 1 & 2 & 3 & 5 & 4 & 25.35 \\
	}
}
\stackrel{\mbox{Sequential runs under 2d-MaGP}}{
	\kbordermatrix{
		& o_1 & o_2 & o_3 & o_4 & o_5 & o_6 & Y \\
		&5 & 1 & 6 & 4 & 3 & 2 & 23.73 \\
		&5 & 2 & 6 & 3 & 1 & 4 & 23.36 \\
		&\textbf{6} & \textbf{4} & \textbf{5} & \textbf{1} & \textbf{2} & \textbf{3} & \textbf{22.43} \\
		&6 & 4 & 5 & 1 & 3 & 2 & 22.59 \\
		&6 & 3 & 5 & 1 & 2 & 4 & 22.46 \\
		&6 & 5 & 4 & 1 & 2 & 3 & 22.99 \\
	}
}
$$
\normalsize

\scriptsize
$$
\stackrel{\mbox{QS-design}}{
	\kbordermatrix{
		& o_1 & o_2 & o_3 & o_4 & o_5 & o_6 & Y \\
  &2 & 1 & 6 & 4 & 5 & 3 & 26.57 \\
  &3 & 6 & 1 & 4 & 2 & 5 & 26.74 \\
  &1 & 4 & 3 & 5 & 2 & 6 & 29.13 \\
  &4 & 6 & 3 & 1 & 5 & 2 & 26.60 \\
  &4 & 3 & 2 & 5 & 6 & 1 & 30.48 \\
  &3 & 1 & 2 & 6 & 5 & 4 & 31.34 \\
  &2 & 3 & 4 & 1 & 5 & 6 & 28.24 \\
  &1 & 2 & 4 & 3 & 6 & 5 & 29.78 \\
  &6 & 4 & 2 & 1 & 3 & 5 & 24.00 \\
  &3 & 4 & 1 & 5 & 6 & 2 & 30.69 \\
  &6 & 2 & 4 & 5 & 1 & 3 & 24.98 \\
  &5 & 3 & 4 & 6 & 1 & 2 & 27.77 \\
  &5 & 2 & 3 & 1 & 4 & 6 & 25.61 \\
  &5 & 1 & 3 & 2 & 6 & 4 & 27.08 \\
  &5 & 4 & 6 & 3 & 2 & 1 & 23.23 \\
  &4 & 5 & 1 & 6 & 2 & 3 & 29.45 \\
  &4 & 2 & 5 & 6 & 3 & 1 & 28.61 \\
  &1 & 6 & 5 & 2 & 4 & 3 & 25.57 \\
  &2 & 5 & 3 & 6 & 4 & 1 & 29.94 \\
  &6 & 2 & 1 & 3 & 5 & 4 & 25.41 \\
  &6 & 1 & 5 & 3 & 4 & 2 & 23.23 \\
	}
}
\stackrel{\mbox{Sequential runs under full-MaGP}}{
	\kbordermatrix{
		& o_1 & o_2 & o_3 & o_4 & o_5 & o_6 & Y \\
		&6 & 4 & 5 & 2 & 1 & 3 & 22.45 \\
		&6 & 4 & 5 & 1 & 3 & 2 & 22.59 \\
		&\textbf{6} & \textbf{4} & \textbf{5} & \textbf{1} & \textbf{2} & \textbf{3} & \textbf{22.43 }\\
		&6 & 3 & 5 & 1 & 2 & 4 & 22.46 \\
		&6 & 3 & 5 & 2 & 1 & 4 & 22.49 \\
	}
}
$$
\normalsize

\begin{figure}[htbp]
\vspace{-.2in}
    \centering
  \subfigure []{\includegraphics[scale=0.3]{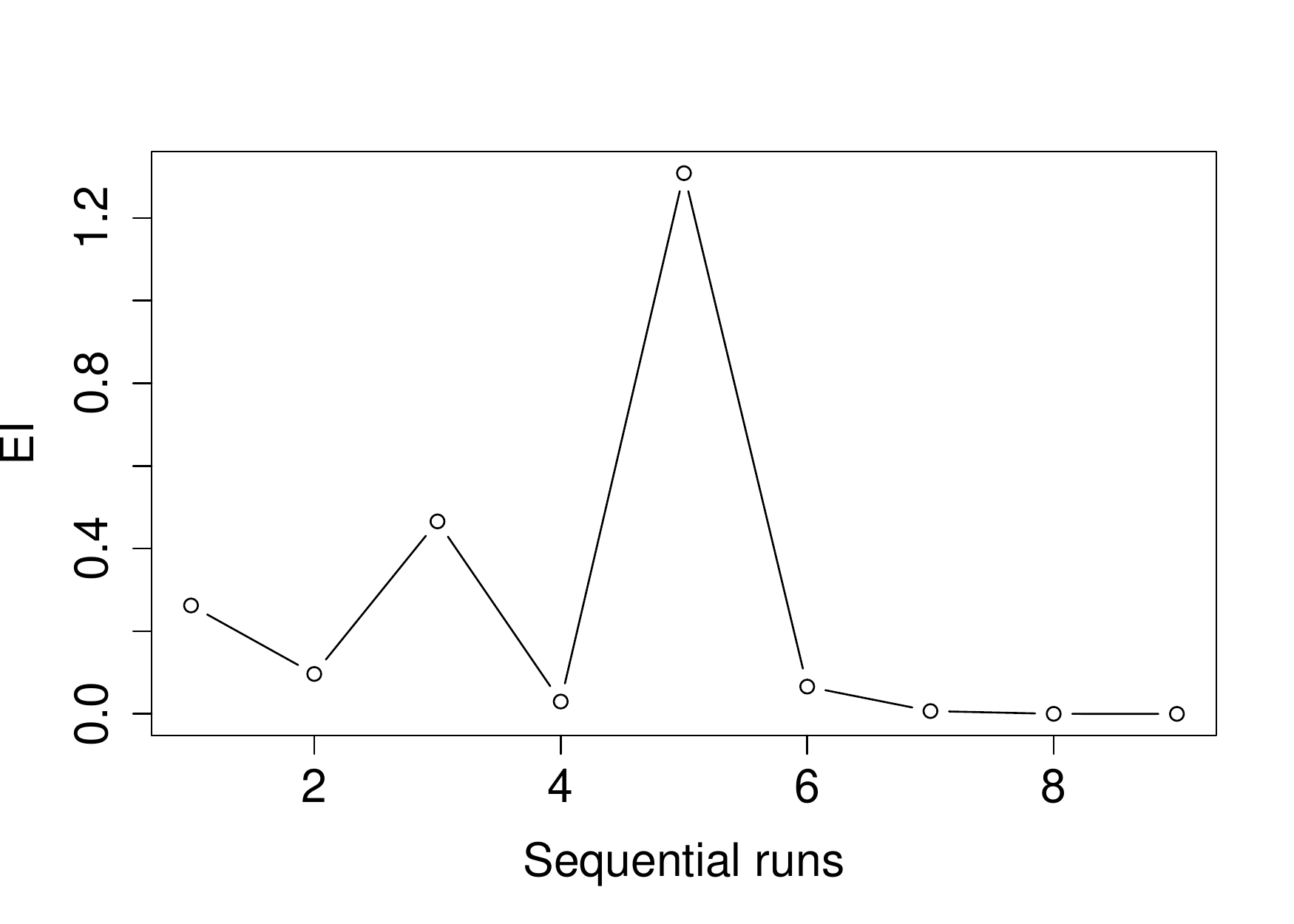}}\quad
  \subfigure [] {\includegraphics[scale=0.3]{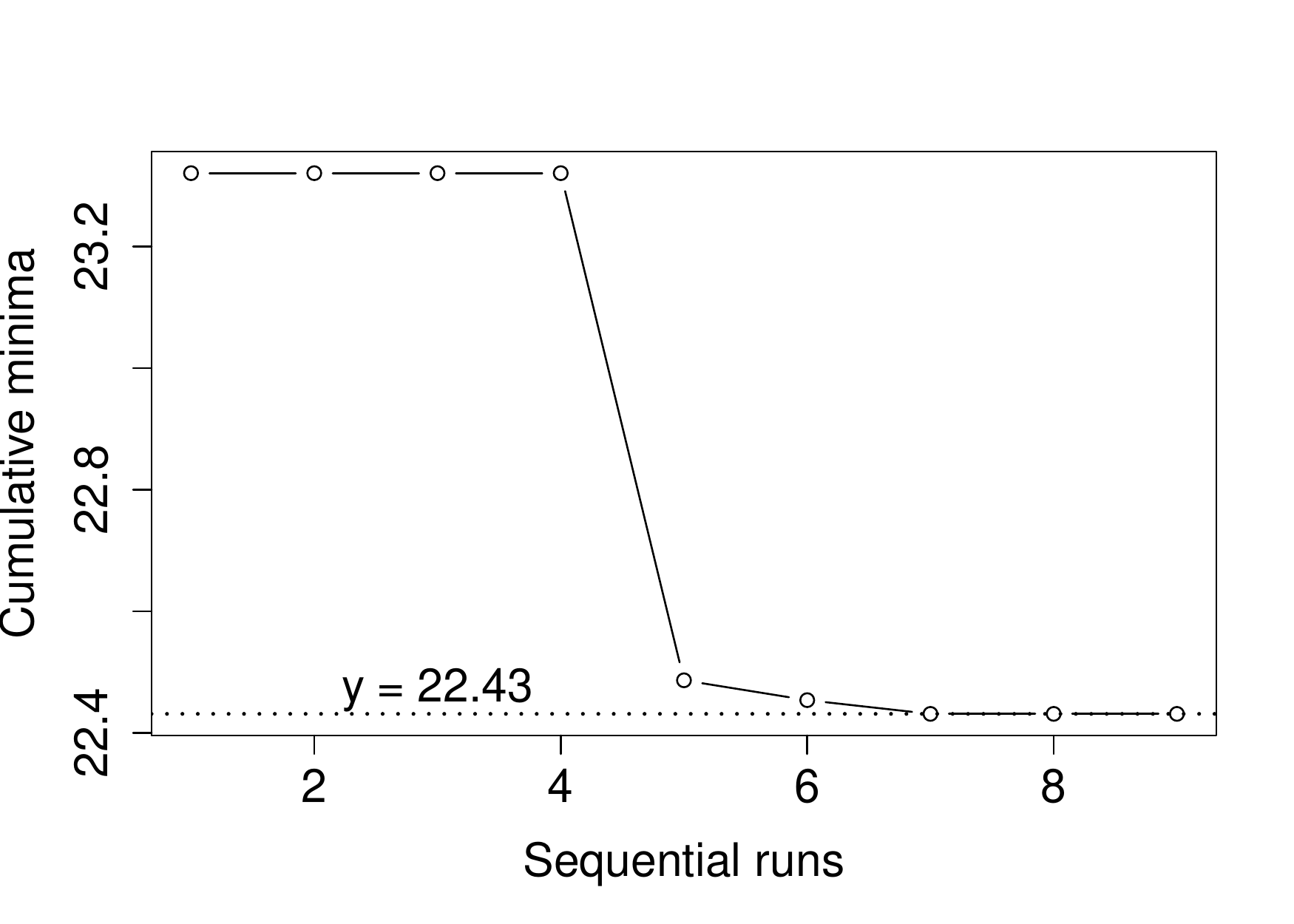}}
  \caption{Plots for (a) expected improvements and (b) cumulative maximum responses from QS-learning under 2d-MaGP using the 6-run QS-design in  Example~\ref{ex:SMS:BO}.}
      \label{fig:eisms6}
\end{figure}

Next, we consider using the $6$-run QS-design from the algebraic construction for the QS-learning under 2d-MaGP. Figure~\ref{fig:eisms6} displays the expected improvements and cumulative maximum responses of sequential runs. It is seen that only 15 runs in total (initial and sequential runs)  are needed to identify the optimum.

\newpage

Below, we list the QS-design along with its sequential runs.

\scriptsize
$$
\stackrel{\mbox{QS-design}}{
	\kbordermatrix{
		& o_1 & o_2 & o_3 & o_4 & o_5 & o_6 & Y \\
  &1 & 2 & 3 & 4 & 5 & 6 & 29.24 \\
  &2 & 4 & 6 & 1 & 3 & 5 & 25.07 \\
  &3 & 6 & 2 & 5 & 1 & 4 & 27.67 \\
  &4 & 1 & 5 & 2 & 6 & 3 & 27.43 \\
  &5 & 3 & 1 & 6 & 4 & 2 & 28.64 \\
  &6 & 5 & 4 & 3 & 2 & 1 & 23.32 \\
	}
}
\stackrel{\mbox{Sequential runs under 2d-MaGP}}{
	\kbordermatrix{
		& o_1 & o_2 & o_3 & o_4 & o_5 & o_6 & Y \\
  &5 & 3 & 4 & 2 & 6 & 1 & 26.81 \\
  &5 & 6 & 2 & 1 & 4 & 3 & 24.76 \\
  &5 & 1 & 6 & 3 & 4 & 2 & 23.70 \\
  &2 & 5 & 4 & 3 & 6 & 1 & 28.85 \\
  &6 & 3 & 5 & 2 & 1 & 4 & 22.49 \\
  &6 & 4 & 5 & 2 & 1 & 3 & 22.45 \\
  &\textbf{6} & \textbf{4} & \textbf{5} & \textbf{1} & \textbf{2} & \textbf{3} & \textbf{22.43} \\
  &6 & 3 & 5 & 1 & 2 & 4 & 22.46 \\
  &6 & 4 & 5 & 1 & 3 & 2 & 22.59 \\
	}
}
$$
\normalsize

\begin{figure}[htbp]
\vspace{-.2in}
	\centering
	\subfigure []{\includegraphics[scale=0.4]{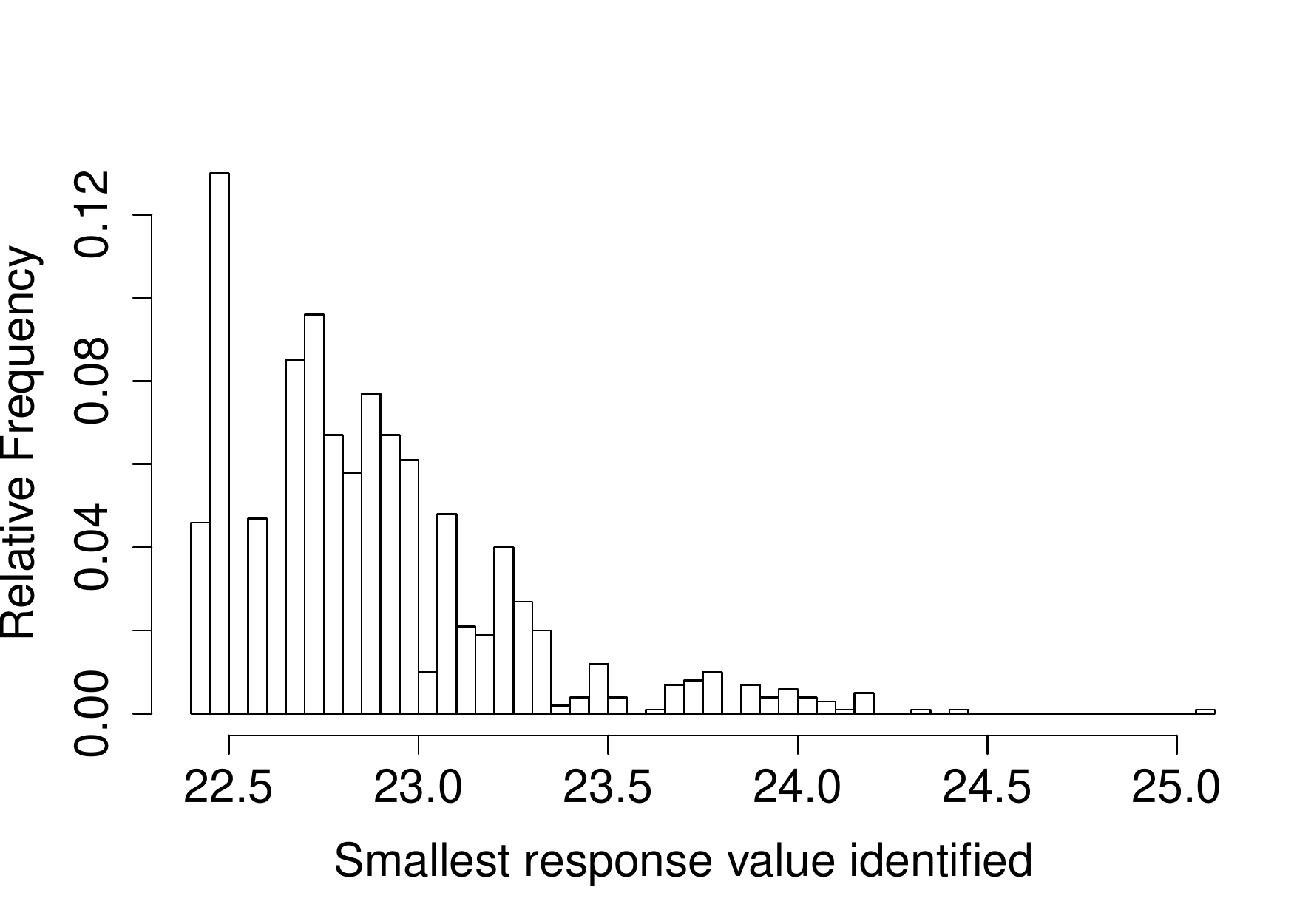}}\quad
	\subfigure [] {\includegraphics[scale=0.4]{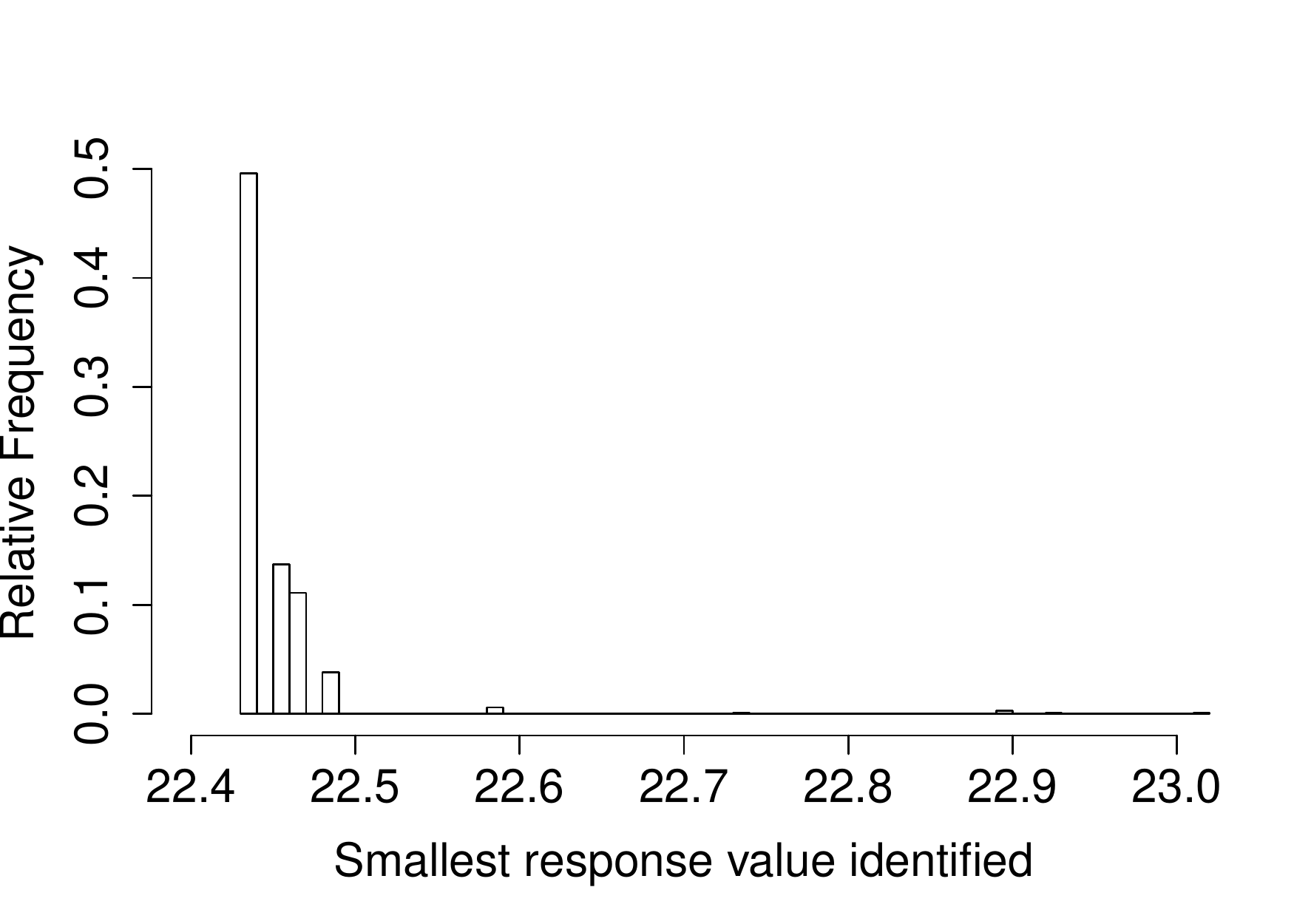}}
	\caption{Histograms of smallest response values identified by (a) $BM_{2}$ and (b) $BM_{3}$ methods in Example~\ref{ex:SMS:BO}.}
		\label{fig:smsf}
\end{figure}

To make a comparison, we consider the sequential generalized PWO ($BM_2$) and CP ($BM_3$) approaches.
The $BM_{2}$ and $BM_{3}$ start from random initial designs of 16 and 26 runs (random subsets of all possible sequences), respectively.
We replicate both methods 1000 times and show the histograms of their smallest response values identified in Figure~\ref{fig:smsf}.
It is seen that most results from $BM_{2}$ and $BM_{3}$ are not satisfactory.

To evaluate the prediction accuracy of the proposed MaGP model compared to the original PWO model \citep{van1995design, voelkel2019design} and CP model \citep{yang2018}, here we consider the cases where only sequence inputs are involved.
Random one-shot designs with run sizes of 30, 40, 50 and 60 are used to form the training data, and the testing data includes all 720 possible sequences. We replicate the analysis 100 times and report both the median and mean of prediction RMSEs in Table~\ref{tab:smspre}.
It is seen that both 2d-MaGP and full-MaGP have much smaller RMSEs,  thus stronger prediction power, compared to PWO and CP for all cases.
Note that both PWO and CP are good at capturing the interactions between the sequence orders \citep{van1995design, robert2018, yang2018}.
Considering that this SMS example contains significant interactions, it is clear that the MaGP model can better capture the interactions among sequence orders compared to PWO and CP.

\begin{table}[htbp]
  \centering
  \caption{The ``medians(means)" of RMSEs from different models under various designs.}
    \begin{tabular}{lcccc}
    \toprule

          & PWO   & CP    & 2d-MaGP & full-MaGP \\
          \midrule
    Random 30-run &  2.33(2.35) & 2.94(2.95) & 0.18(0.21) & 0.18(0.20) \\
    Random 40-run & 2.04(2.07) & 2.94(2.94) & 0.12(0.13) & 0.12(0.13) \\
    Random 50-run & 1.95(1.96) & 2.94(2.94) & 0.11(0.11) & 0.11(0.11) \\
    Random 60-run & 1.89(1.90) & 2.94(2.94) & 0.10(0.10)  & 0.10(0.10)  \\
    \bottomrule
    \end{tabular}%
  \label{tab:smspre}%
\end{table}%



\begin{figure}[ht]
\vspace{-.2in}
	\centering
	\subfigure []{\includegraphics[scale=0.4]{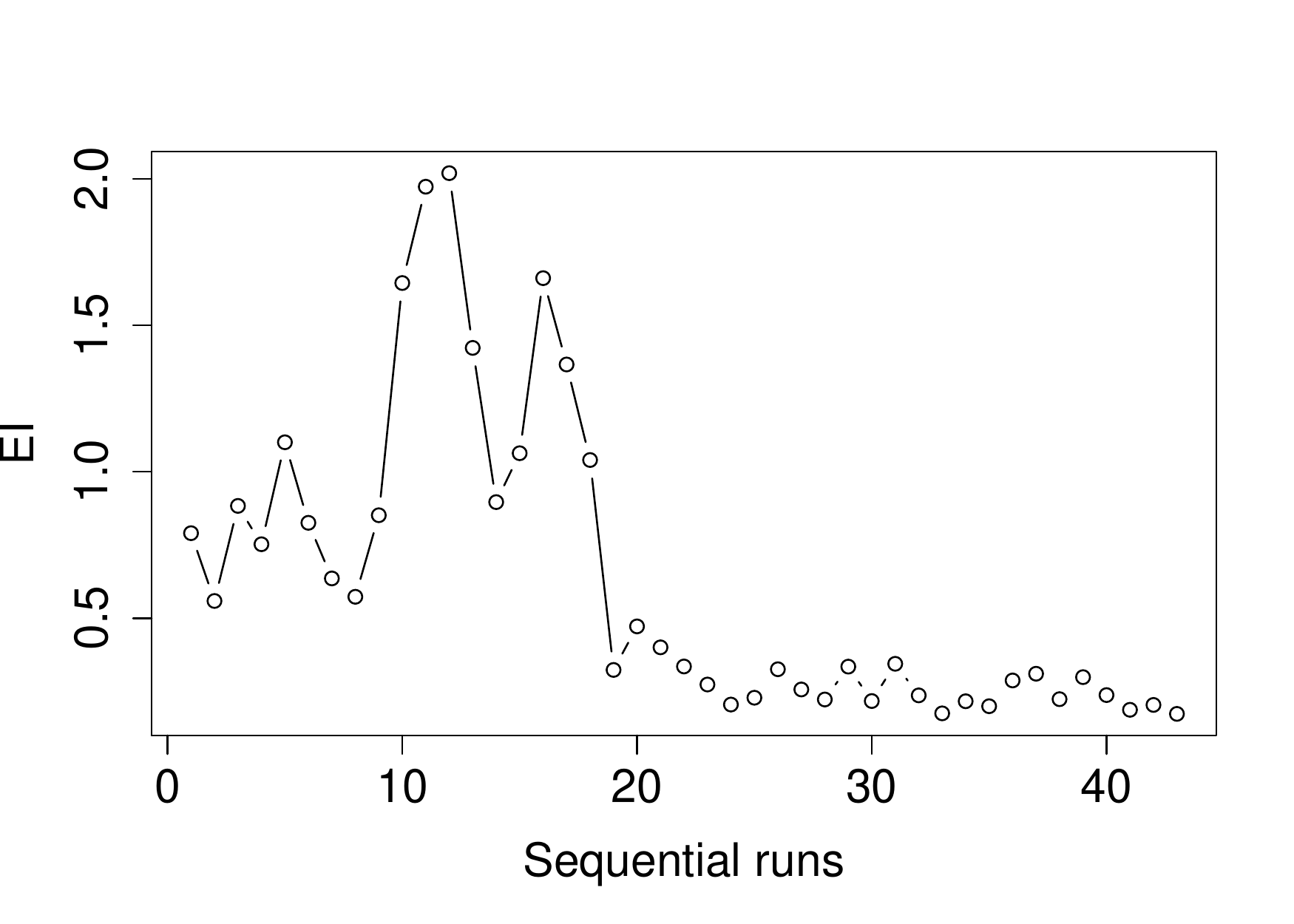}}\quad
	\subfigure [] {\includegraphics[scale=0.4]{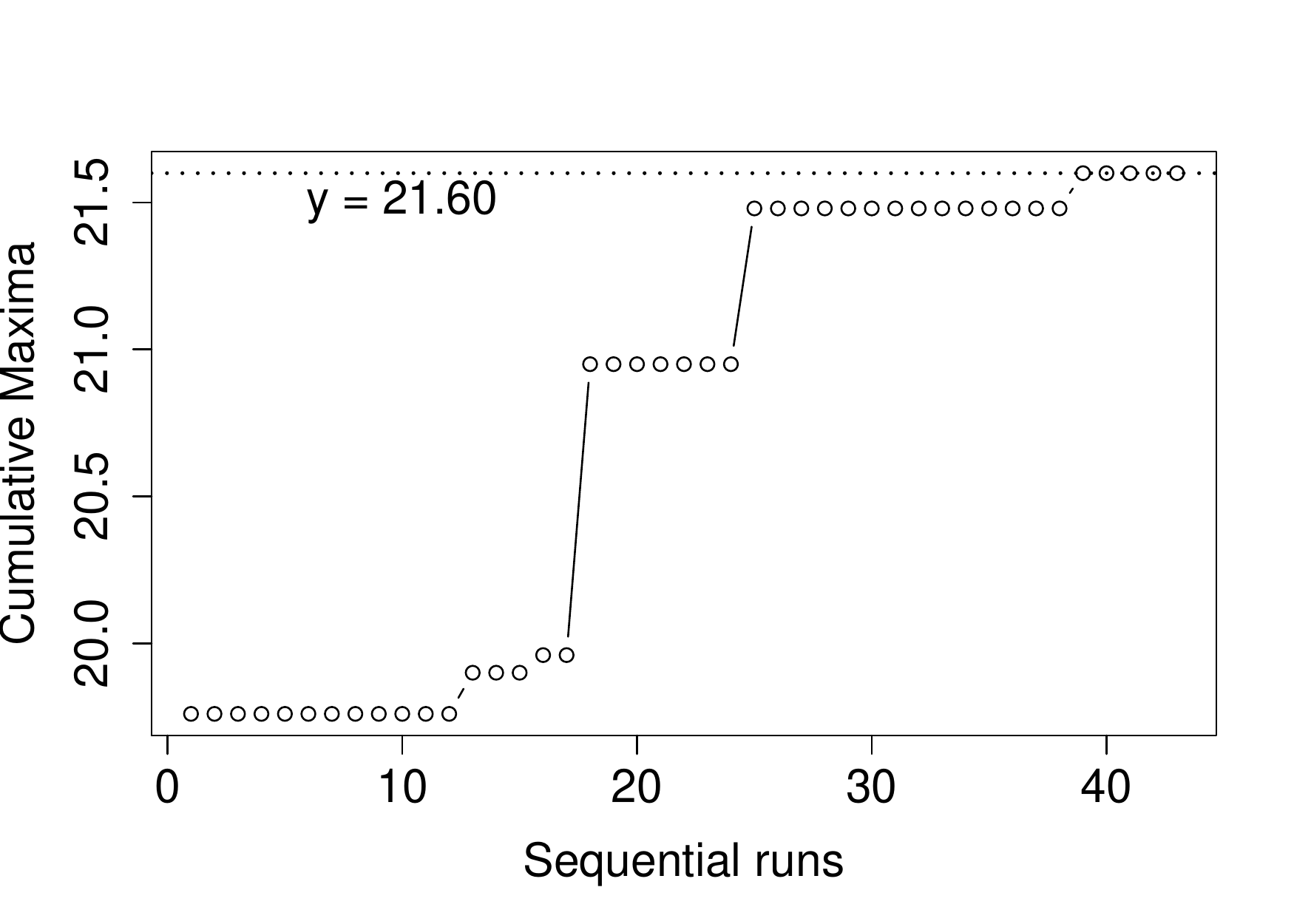}}
	\caption{Plots for (a) expected improvements and (b) cumulative maximum responses from the QS-learning approach under 2d-MaGP in Example \ref{ex:SMS:BO2}.}
	\label{fig:ei4}
\end{figure}

\begin{figure}[ht]
\vspace{-.2in}
	\centering
	\subfigure []{\includegraphics[scale=0.4]{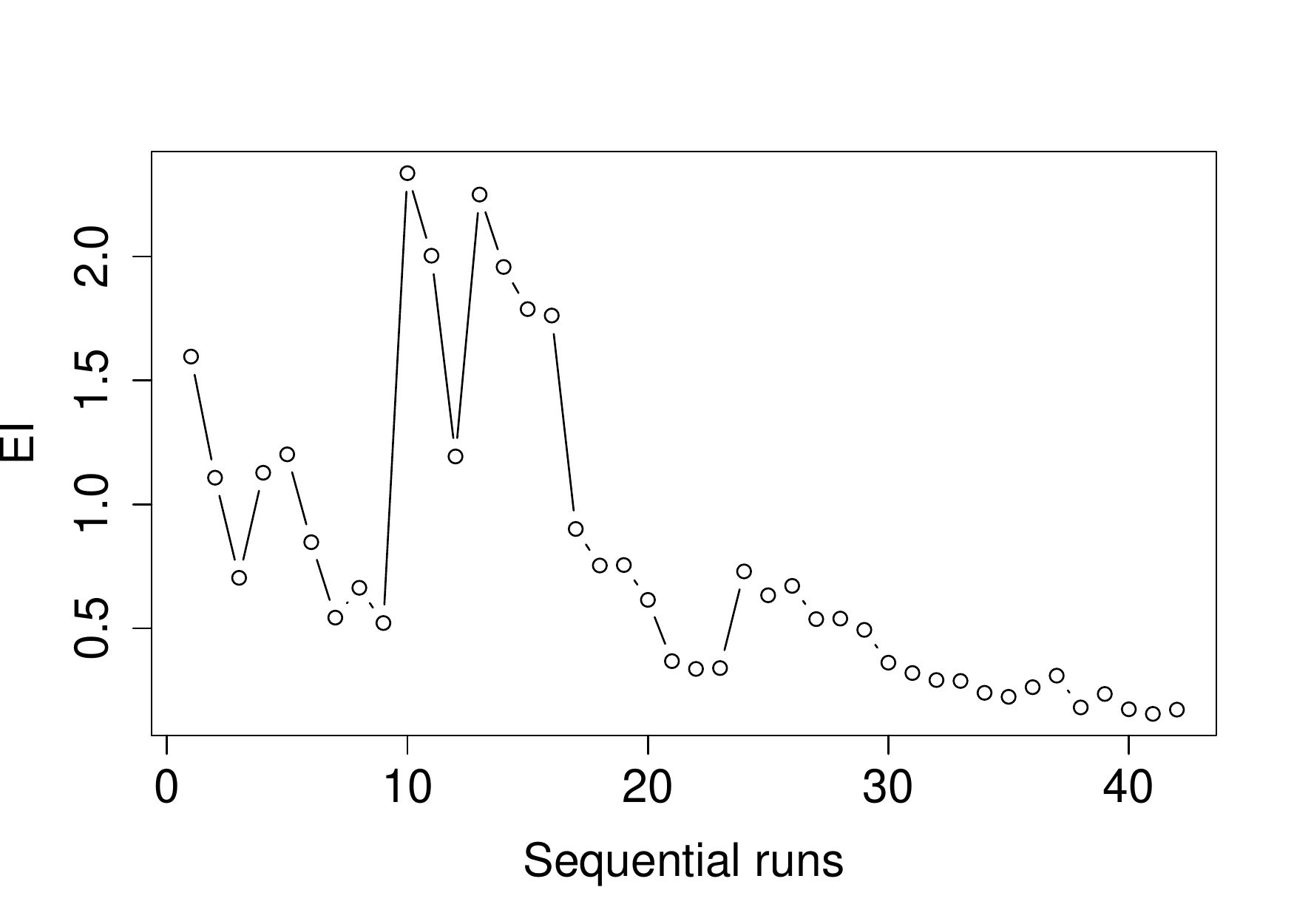}}\quad
	\subfigure [] {\includegraphics[scale=0.4]{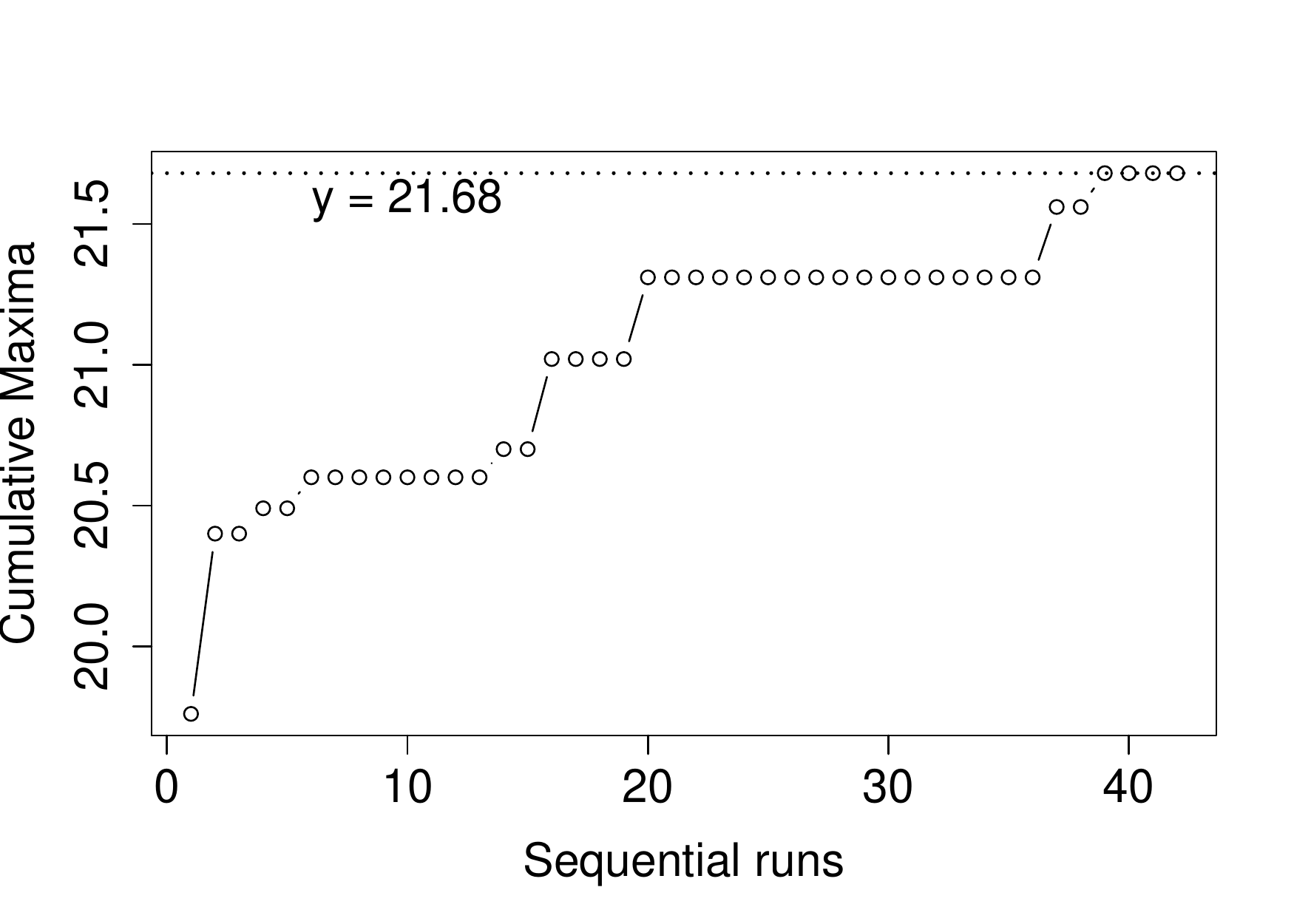}}
	\caption{Plots for (a) expected improvements and (b) cumulative maximum responses from the QS-learning approach under full-MaGP in Example \ref{ex:SMS:BO2}.}
		\label{fig:ei4full}
\end{figure}

\begin{example} \label{ex:SMS:BO2}
	Following the same way to define the cost function $C(\balpha)$ in Example \ref{ex:SMS:BO}, here we consider a general SMS problem where both the processing time $\bx = (x_1, \ldots , x_6)$ ($x_i \in (0,1)$) and job-sequence are to be optimized. Suppose the products manufactured by this single machine can sell for revenue of
	$
	R(\bx) = w_0 \sum_{i=1}^{k}x_i.
	$
	That is, products can sell for higher revenue if more time is spent on each job. The total profit made by this machine with processing time $\bx$ and job-sequence $\bm{\alpha}$ (or equivalently order-sequence $\bo$) is
	$$
	F(\bx,\balpha) = R(\bx) - C(\balpha, \bx) = w_0 \sum_{i=1}^{k}x_i - \sum_{h=1}^{k}w_h T^2(\alpha_h),
	$$
	where weights $w_1, \ldots, w_k$ are the same as those in Example \ref{ex:SMS:BO}, and we arbitrarily set $w_0 = 10$. The target of this study is to identify the optimal setting for maximizing the profit.
\end{example}
It is known that there is no analytical solution for this SMS problem. Here, we consider the proposed QS-learning to identify the optimal setting via experimental trials.
It starts from the 29-run (rule-of-thumb run size) QS-design, and then selects 43 and 42 sequential runs under the 2d-MaGP and full-MaGP models, respectively. When using 2d-MaGP, the maximum response found is 21.60 with  $\bx = (0.73, 0.46, 0.12, 0.97, 0.99, 0.14)$ and $\bo = (3,4,2,5,6,1)$ (or equivalently $\balpha = (6,3,1,2,4,5)$). When using full-MaGP, the maximum response found is 21.68 with $\bx = (0.52, 0.28, 0.03, 1.00, 1.00, 0.46)$ and $\bo = (4,2,1,5,6,3)$. Figures~\ref{fig:ei4} and \ref{fig:ei4full} show the plots for the expected improvements and cumulative maximum responses of sequential runs.


\newpage

Below, we show the 29-run QS-design along with the sequential runs in the QS-learning under 2d-MaGP and full-MaGP, respectively.

\scriptsize
$$
\stackrel{\mbox{QS-design}}{
	\kbordermatrix{
		& x_1 & o_1 & x_2 & o_2 & x_3 & o_3 & x_4 & o_4 & x_5 & o_5 & x_6 & o_6 &Y \\
		&0.24 & 1 & 0.14 & 2 & 0.28 & 6 & 0.83 & 3 & 0.76 & 5 & 0.83 & 4 & 15.76 \\
  &0.59 & 1 & 0.66 & 4 & 0.38 & 3 & 0.10 & 2 & 0.79 & 6 & 0.07 & 5 & 16.77 \\
  &0.38 & 5 & 0.48 & 4 & 1.00 & 3 & 0.72 & 6 & 0.07 & 2 & 0.62 & 1 & 17.25 \\
  &1.00 & 1 & 0.76 & 6 & 0.83 & 5 & 0.62 & 2 & 0.69 & 4 & 0.76 & 3 & 12.65 \\
  &0.07 & 5 & 0.72 & 6 & 0.24 & 2 & 0.34 & 3 & 0.90 & 4 & 0.55 & 1 & 16.42 \\
  &0.41 & 1 & 0.52 & 5 & 0.55 & 2 & 0.24 & 4 & 0.03 & 3 & 0.10 & 6 & 12.32 \\
  &0.90 & 2 & 0.59 & 6 & 0.69 & 4 & 0.86 & 5 & 0.14 & 3 & 0.14 & 1 & 17.78 \\
  &0.69 & 6 & 0.28 & 1 & 0.79 & 4 & 0.79 & 3 & 0.93 & 2 & 0.45 & 5 & 14.86 \\
  &0.03 & 3 & 0.55 & 5 & 0.90 & 4 & 0.52 & 2 & 1.00 & 6 & 0.69 & 1 & 19.02 \\
  &0.10 & 6 & 0.21 & 3 & 0.86 & 1 & 0.41 & 5 & 0.55 & 4 & 0.21 & 2 & 12.60 \\
  &0.48 & 2 & 0.34 & 1 & 0.93 & 5 & 0.14 & 6 & 0.48 & 3 & 0.86 & 4 & 14.59 \\
  &0.17 & 6 & 0.83 & 2 & 0.62 & 4 & 0.17 & 1 & 0.21 & 5 & 0.72 & 3 & 12.47 \\
  &0.14 & 6 & 0.79 & 5 & 0.21 & 3 & 0.76 & 2 & 0.24 & 1 & 0.34 & 4 & 14.42 \\
  &0.34 & 4 & 0.07 & 6 & 0.41 & 5 & 0.69 & 3 & 0.17 & 1 & 0.38 & 2 & 12.74 \\
  &0.62 & 4 & 0.90 & 2 & 0.97 & 3 & 0.21 & 5 & 0.52 & 1 & 0.31 & 6 & 10.81 \\
  &0.52 & 5 & 0.86 & 1 & 0.17 & 3 & 0.55 & 4 & 0.45 & 6 & 0.93 & 2 & 13.16 \\
  &0.66 & 3 & 0.69 & 6 & 0.48 & 1 & 0.07 & 5 & 0.97 & 2 & 0.79 & 4 & 13.51 \\
  &0.86 & 6 & 0.93 & 3 & 0.07 & 5 & 0.59 & 4 & 0.62 & 2 & 0.24 & 1 & 16.95 \\
  &0.83 & 2 & 1.00 & 4 & 0.59 & 1 & 0.48 & 3 & 0.10 & 5 & 0.59 & 6 & 13.02 \\
  &0.79 & 4 & 0.38 & 3 & 0.31 & 6 & 0.03 & 2 & 0.34 & 5 & 0.52 & 1 & 14.51 \\
  &0.72 & 3 & 0.41 & 2 & 0.03 & 1 & 0.97 & 6 & 0.38 & 4 & 0.48 & 5 & 19.76 \\
  &0.55 & 4 & 0.97 & 5 & 0.66 & 6 & 1.00 & 1 & 0.66 & 2 & 0.41 & 3 & 14.58 \\
  &0.93 & 3 & 0.10 & 2 & 0.72 & 5 & 0.38 & 1 & 0.41 & 6 & 0.17 & 4 & 16.61 \\
  &0.21 & 4 & 0.45 & 1 & 0.45 & 2 & 0.90 & 6 & 0.72 & 5 & 0.03 & 3 & 19.21 \\
  &0.28 & 1 & 0.24 & 3 & 0.10 & 4 & 0.28 & 5 & 0.31 & 2 & 0.90 & 6 & 16.58 \\
  &0.31 & 2 & 0.62 & 3 & 0.76 & 6 & 0.93 & 4 & 0.59 & 1 & 1.00 & 5 & 17.50 \\
  &0.45 & 5 & 0.03 & 4 & 0.34 & 6 & 0.31 & 1 & 0.86 & 3 & 0.28 & 2 & 14.76 \\
  &0.76 & 6 & 0.17 & 4 & 0.52 & 2 & 0.66 & 5 & 0.28 & 1 & 0.97 & 3 & 18.44 \\
  &0.97 & 2 & 0.31 & 5 & 0.14 & 3 & 0.45 & 1 & 0.83 & 4 & 0.66 & 6 & 15.47 \\ 		
	}
}
$$

$$
\stackrel{\mbox{Sequential runs under the 2d-MaGP}}{
	\kbordermatrix{
		& x_1 & o_1 & x_2 & o_2 & x_3 & o_3 & x_4 & o_4 & x_5 & o_5 & x_6 & o_6 &Y \\
	&0.75 & 3 & 0.79 & 5 & 0.75 & 6 & 0.76 & 2 & 0.26 & 1 & 0.35 & 4 & 18.12 \\
  &0.21 & 4 & 0.41 & 2 & 0.11 & 3 & 0.91 & 6 & 0.27 & 1 & 0.70 & 5 & 19.07 \\
  &0.72 & 3 & 0.19 & 5 & 0.11 & 4 & 0.96 & 6 & 0.31 & 2 & 0.54 & 1 & 18.18 \\
  &0.76 & 6 & 0.42 & 2 & 0.03 & 1 & 0.86 & 5 & 0.38 & 3 & 0.01 & 4 & 18.45 \\
  &0.74 & 3 & 0.10 & 4 & 0.45 & 2 & 0.88 & 6 & 0.27 & 1 & 0.22 & 5 & 17.82 \\
  &0.73 & 3 & 0.17 & 4 & 0.51 & 2 & 0.66 & 1 & 1.00 & 6 & 0.91 & 5 & 19.59 \\
  &0.04 & 3 & 0.45 & 5 & 0.75 & 2 & 0.97 & 6 & 0.38 & 4 & 1.00 & 1 & 17.26 \\
  &0.72 & 3 & 0.45 & 1 & 0.90 & 4 & 0.86 & 2 & 1.00 & 6 & 1.00 & 5 & 15.62 \\
  &0.21 & 4 & 0.61 & 6 & 0.03 & 1 & 0.52 & 2 & 1.00 & 3 & 0.49 & 5 & 17.20 \\
  &0.74 & 3 & 0.41 & 2 & 0.75 & 4 & 0.92 & 6 & 0.72 & 5 & 0.69 & 1 & 17.36 \\
  &0.91 & 2 & 0.55 & 5 & 0.75 & 6 & 0.66 & 1 & 0.38 & 4 & 0.01 & 3 & 16.94 \\
  &0.07 & 3 & 0.17 & 4 & 0.51 & 2 & 0.97 & 6 & 0.73 & 5 & 0.62 & 1 & 19.72 \\
  &0.73 & 4 & 0.17 & 3 & 0.04 & 1 & 0.96 & 2 & 1.00 & 6 & 0.61 & 5 & 19.90 \\
  &0.73 & 3 & 0.62 & 4 & 0.03 & 5 & 0.75 & 2 & 1.00 & 6 & 0.01 & 1 & 18.24 \\
  &0.73 & 3 & 0.41 & 6 & 0.45 & 2 & 0.93 & 4 & 0.27 & 5 & 0.63 & 1 & 13.22 \\
  &0.21 & 3 & 0.42 & 2 & 0.03 & 1 & 0.90 & 5 & 1.00 & 6 & 0.04 & 4 & 19.96 \\
  &0.21 & 4 & 0.57 & 3 & 0.04 & 1 & 0.52 & 2 & 1.00 & 6 & 0.48 & 5 & 19.65 \\
  &0.74 & 3 & 0.17 & 4 & 0.03 & 1 & 0.97 & 6 & 0.59 & 2 & 0.91 & 5 & 20.95 \\
  &0.73 & 4 & 0.17 & 3 & 0.51 & 1 & 0.90 & 6 & 0.60 & 2 & 0.48 & 5 & 18.55 \\
  &0.02 & 3 & 0.17 & 4 & 0.03 & 1 & 0.97 & 6 & 0.27 & 2 & 0.91 & 5 & 19.00 \\
  &0.75 & 4 & 0.78 & 5 & 0.51 & 2 & 0.97 & 6 & 0.59 & 1 & 0.04 & 3 & 19.86 \\
  &0.74 & 4 & 0.42 & 2 & 0.03 & 1 & 0.97 & 6 & 0.71 & 5 & 0.04 & 3 & 20.35 \\
  &0.21 & 4 & 0.17 & 3 & 0.03 & 1 & 0.96 & 6 & 0.07 & 2 & 0.91 & 5 & 18.99 \\
  &0.75 & 4 & 0.42 & 2 & 0.03 & 1 & 0.72 & 6 & 0.14 & 3 & 0.91 & 5 & 19.49 \\
  &0.73 & 4 & 0.41 & 2 & 0.03 & 1 & 0.97 & 5 & 1.00 & 6 & 0.04 & 3 & 21.48 \\
  &0.74 & 3 & 0.44 & 1 & 0.11 & 4 & 0.97 & 5 & 1.00 & 6 & 0.69 & 2 & 20.03 \\
  &0.21 & 4 & 0.42 & 2 & 0.04 & 1 & 0.96 & 6 & 0.99 & 5 & 0.04 & 3 & 20.17 \\
  &0.28 & 2 & 0.18 & 4 & 0.03 & 1 & 0.96 & 5 & 1.00 & 6 & 0.04 & 3 & 19.72 \\
  &0.73 & 4 & 0.40 & 2 & 0.04 & 1 & 0.97 & 6 & 0.74 & 5 & 0.03 & 3 & 20.44 \\
  &0.74 & 4 & 0.41 & 2 & 0.69 & 3 & 0.97 & 5 & 0.58 & 1 & 0.85 & 6 & 17.99 \\
  &0.74 & 3 & 0.63 & 4 & 0.51 & 2 & 0.97 & 5 & 1.00 & 6 & 0.16 & 1 & 20.57 \\
  &0.73 & 4 & 0.62 & 3 & 0.04 & 1 & 0.90 & 6 & 0.58 & 2 & 0.91 & 5 & 20.11 \\
  &0.73 & 4 & 0.42 & 1 & 0.76 & 5 & 0.96 & 6 & 0.60 & 2 & 0.04 & 3 & 20.08 \\
  &0.04 & 3 & 0.41 & 2 & 0.03 & 1 & 0.97 & 5 & 1.00 & 6 & 0.17 & 4 & 20.15 \\
  &0.75 & 3 & 0.16 & 4 & 0.51 & 2 & 0.96 & 5 & 1.00 & 6 & 0.69 & 1 & 19.50 \\
  &0.74 & 3 & 0.47 & 4 & 0.03 & 1 & 0.53 & 2 & 1.00 & 6 & 0.90 & 5 & 21.09 \\
  &0.74 & 3 & 0.66 & 4 & 0.04 & 1 & 0.50 & 2 & 1.00 & 6 & 0.91 & 5 & 20.87 \\
  &0.74 & 3 & 0.42 & 2 & 1.00 & 4 & 0.97 & 5 & 0.27 & 1 & 0.91 & 6 & 18.75 \\
  &\textbf{0.73} & \textbf{3} & \textbf{0.46} & \textbf{4} & \textbf{0.12} & \textbf{2} & \textbf{0.97} & \textbf{5} & \textbf{0.99} & \textbf{6} & \textbf{0.14} & \textbf{1} & \textbf{21.60} \\
  &0.06 & 4 & 0.47 & 3 & 0.11 & 2 & 0.38 & 1 & 1.00 & 6 & 0.70 & 5 & 19.90 \\
  &0.74 & 3 & 0.57 & 5 & 0.09 & 2 & 0.97 & 6 & 0.36 & 4 & 0.14 & 1 & 19.97 \\
  &0.32 & 2 & 0.65 & 3 & 0.04 & 1 & 0.97 & 5 & 0.99 & 6 & 0.05 & 4 & 21.15 \\
  &0.73 & 3 & 0.92 & 4 & 0.11 & 2 & 0.97 & 5 & 0.99 & 6 & 0.61 & 1 & 19.34 \\
	}
}
$$

$$
\stackrel{\mbox{Sequential runs under the full-MaGP}}{
	\kbordermatrix{
		& x_1 & o_1 & x_2 & o_2 & x_3 & o_3 & x_4 & o_4 & x_5 & o_5 & x_6 & o_6 &Y \\
		&0.03 & 3 & 0.19 & 4 & 0.07 & 5 & 0.94 & 6 & 0.30 & 2 & 0.67 & 1 & 16.20 \\
  &0.74 & 3 & 0.52 & 4 & 0.46 & 2 & 0.66 & 5 & 0.99 & 6 & 0.13 & 1 & 20.40 \\
  &0.79 & 6 & 0.56 & 1 & 0.47 & 2 & 0.66 & 5 & 0.37 & 4 & 0.05 & 3 & 18.48 \\
  &0.71 & 3 & 0.41 & 2 & 0.03 & 4 & 0.35 & 1 & 1.00 & 6 & 0.74 & 5 & 20.49 \\
  &0.71 & 4 & 0.42 & 2 & 0.01 & 3 & 0.76 & 5 & 1.00 & 6 & 1.00 & 1 & 18.81 \\
  &0.20 & 3 & 0.10 & 4 & 0.04 & 1 & 0.42 & 2 & 1.00 & 6 & 1.00 & 5 & 20.60 \\
  &0.03 & 4 & 0.56 & 5 & 0.50 & 2 & 0.93 & 6 & 0.32 & 3 & 0.00 & 1 & 18.22 \\
  &0.73 & 4 & 0.51 & 3 & 0.04 & 1 & 0.50 & 2 & 1.00 & 6 & 0.59 & 5 & 20.37 \\
  &0.73 & 3 & 0.55 & 4 & 0.47 & 2 & 1.00 & 1 & 1.00 & 6 & 1.00 & 5 & 16.05 \\
  &0.12 & 3 & 0.42 & 2 & 0.04 & 4 & 0.67 & 5 & 1.00 & 6 & 0.11 & 1 & 18.68 \\
  &0.73 & 3 & 0.57 & 5 & 0.91 & 4 & 0.45 & 2 & 1.00 & 6 & 0.22 & 1 & 19.32 \\
  &0.74 & 3 & 0.09 & 4 & 0.45 & 2 & 0.36 & 1 & 0.99 & 6 & 0.55 & 5 & 19.83 \\
  &0.73 & 3 & 0.50 & 4 & 0.04 & 1 & 0.88 & 6 & 0.63 & 2 & 0.78 & 5 & 19.83 \\
  &0.19 & 4 & 0.93 & 3 & 0.04 & 1 & 0.42 & 2 & 0.99 & 6 & 0.72 & 5 & 20.70 \\
  &0.71 & 3 & 0.51 & 4 & 0.08 & 1 & 0.45 & 2 & 0.91 & 5 & 0.90 & 6 & 20.56 \\
  &0.27 & 2 & 0.51 & 4 & 0.07 & 1 & 0.81 & 5 & 1.00 & 6 & 0.41 & 3 & 21.02 \\
  &0.25 & 3 & 0.39 & 1 & 0.04 & 2 & 0.79 & 4 & 1.00 & 6 & 0.46 & 5 & 19.87 \\
  &0.89 & 5 & 0.57 & 4 & 0.00 & 2 & 0.45 & 1 & 1.00 & 6 & 0.97 & 3 & 20.69 \\
  &0.51 & 4 & 0.68 & 3 & 0.01 & 1 & 0.45 & 2 & 0.71 & 5 & 0.96 & 6 & 20.53 \\
  &0.28 & 1 & 0.42 & 3 & 0.37 & 4 & 0.46 & 2 & 1.00 & 6 & 0.98 & 5 & 21.31 \\
  &0.24 & 3 & 1.00 & 5 & 0.01 & 1 & 0.44 & 2 & 0.30 & 4 & 0.99 & 6 & 21.09 \\
  &0.23 & 3 & 1.00 & 5 & 0.25 & 4 & 0.46 & 2 & 1.00 & 6 & 0.71 & 1 & 20.60 \\
  &0.29 & 2 & 0.37 & 3 & 0.00 & 1 & 0.69 & 5 & 0.31 & 4 & 0.72 & 6 & 17.81 \\
  &0.72 & 3 & 0.98 & 5 & 0.00 & 4 & 0.47 & 2 & 0.68 & 1 & 0.96 & 6 & 19.90 \\
  &0.26 & 3 & 0.49 & 4 & 0.00 & 1 & 0.44 & 2 & 0.98 & 6 & 0.95 & 5 & 21.23 \\
  &0.27 & 1 & 0.52 & 4 & 1.00 & 5 & 0.44 & 2 & 0.99 & 6 & 0.97 & 3 & 19.99 \\
  &0.26 & 3 & 1.00 & 5 & 0.00 & 2 & 0.49 & 1 & 0.31 & 4 & 0.90 & 6 & 20.56 \\
  &0.87 & 6 & 0.47 & 4 & 0.00 & 1 & 0.44 & 2 & 0.76 & 5 & 0.43 & 3 & 19.96 \\
  &0.27 & 1 & 0.61 & 3 & 0.00 & 4 & 1.00 & 5 & 0.65 & 2 & 0.97 & 6 & 20.86 \\
  &0.25 & 4 & 0.50 & 3 & 0.00 & 1 & 0.43 & 2 & 1.00 & 5 & 0.97 & 6 & 21.27 \\
  &0.75 & 4 & 0.71 & 3 & 0.00 & 1 & 0.41 & 2 & 1.00 & 6 & 0.95 & 5 & 21.16 \\
  &0.28 & 2 & 0.55 & 3 & 0.00 & 1 & 0.86 & 5 & 0.97 & 6 & 0.14 & 4 & 20.46 \\
  &0.88 & 3 & 0.31 & 2 & 0.01 & 4 & 0.47 & 1 & 1.00 & 5 & 0.98 & 6 & 20.93 \\
  &0.24 & 4 & 0.31 & 2 & 0.00 & 1 & 0.00 & 3 & 1.00 & 6 & 0.98 & 5 & 19.88 \\
  &0.24 & 4 & 0.39 & 3 & 0.00 & 1 & 0.79 & 5 & 0.63 & 2 & 0.98 & 6 & 20.58 \\
  &0.26 & 1 & 0.89 & 3 & 0.00 & 4 & 0.43 & 2 & 1.00 & 5 & 0.96 & 6 & 20.99 \\
  &0.52 & 4 & 0.21 & 2 & 0.00 & 1 & 1.00 & 5 & 1.00 & 6 & 0.41 & 3 & 21.56 \\
  &0.02 & 4 & 0.20 & 2 & 0.00 & 1 & 1.00 & 5 & 1.00 & 6 & 0.40 & 3 & 20.23 \\
  &\textbf{0.52} & \textbf{4} & \textbf{0.28} & \textbf{2} & \textbf{0.03} & \textbf{1} & \textbf{1.00 } & \textbf{5} & \textbf{1.00 }& \textbf{6} & \textbf{0.46} & \textbf{3} & \textbf{21.68} \\
  &0.53 & 4 & 0.28 & 2 & 0.01 & 1 & 1.00 & 5 & 1.00 & 6 & 0.89 & 3 & 21.53 \\
  &0.53 & 4 & 0.48 & 3 & 0.00 & 1 & 1.00 & 5 & 1.00 & 6 & 0.76 & 2 & 21.14 \\
  &0.91 & 3 & 0.30 & 2 & 0.00 & 1 & 1.00 & 6 & 0.26 & 4 & 0.97 & 5 & 21.57 \\
	}
}
$$
\normalsize

\begin{figure}[htbp]
	\centering
	\includegraphics[width=0.7\linewidth]{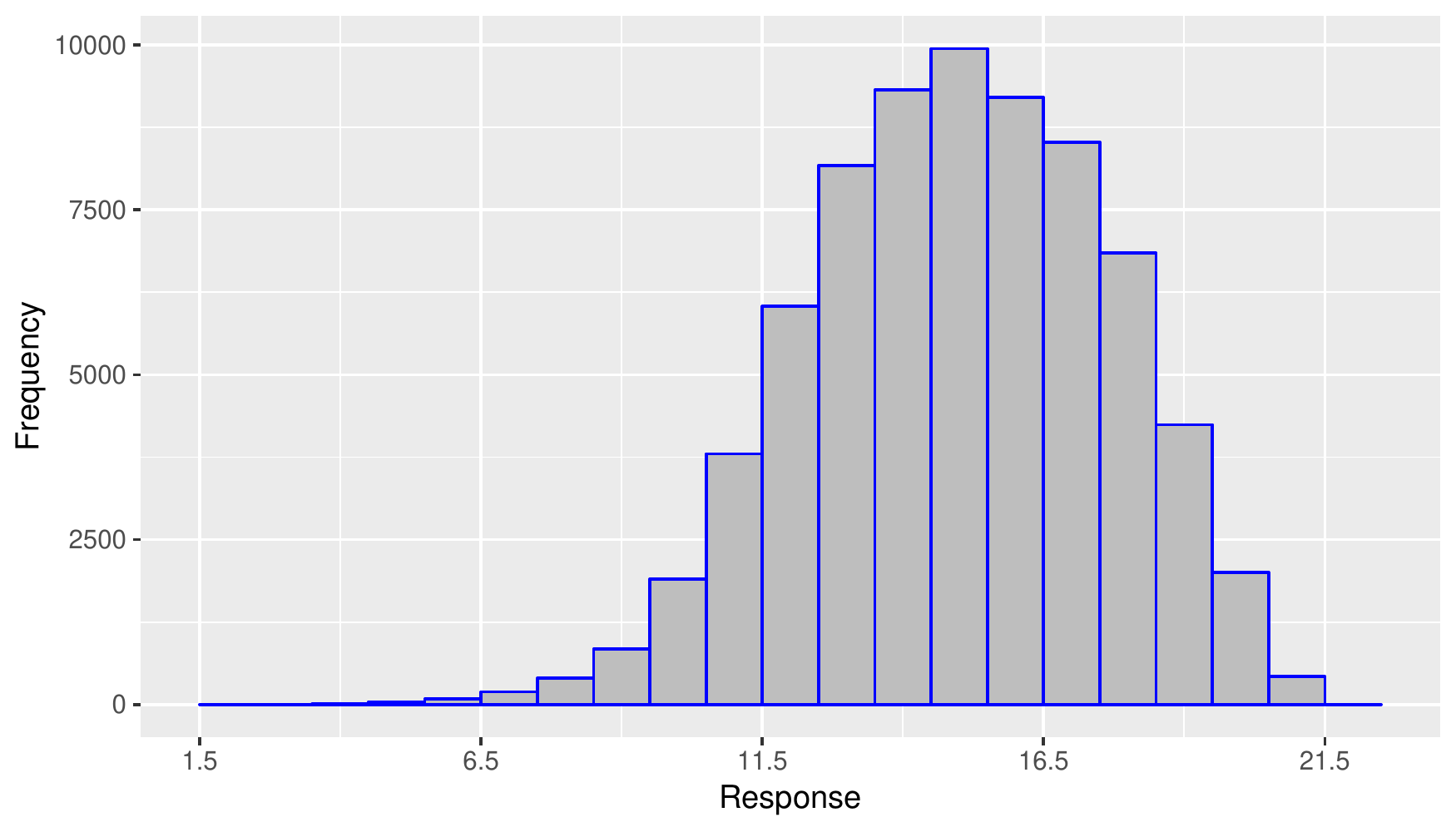} \vspace{-.2in}
	\caption{Histogram of the responses in a large random sample of 72,000 observations in Example~\ref{ex:SMS:BO2}.}
		\label{fig:hist}
\end{figure}

\begin{figure}[htbp]
\vspace{-.2in}
	\centering
	\subfigure []{\includegraphics[scale=0.4]{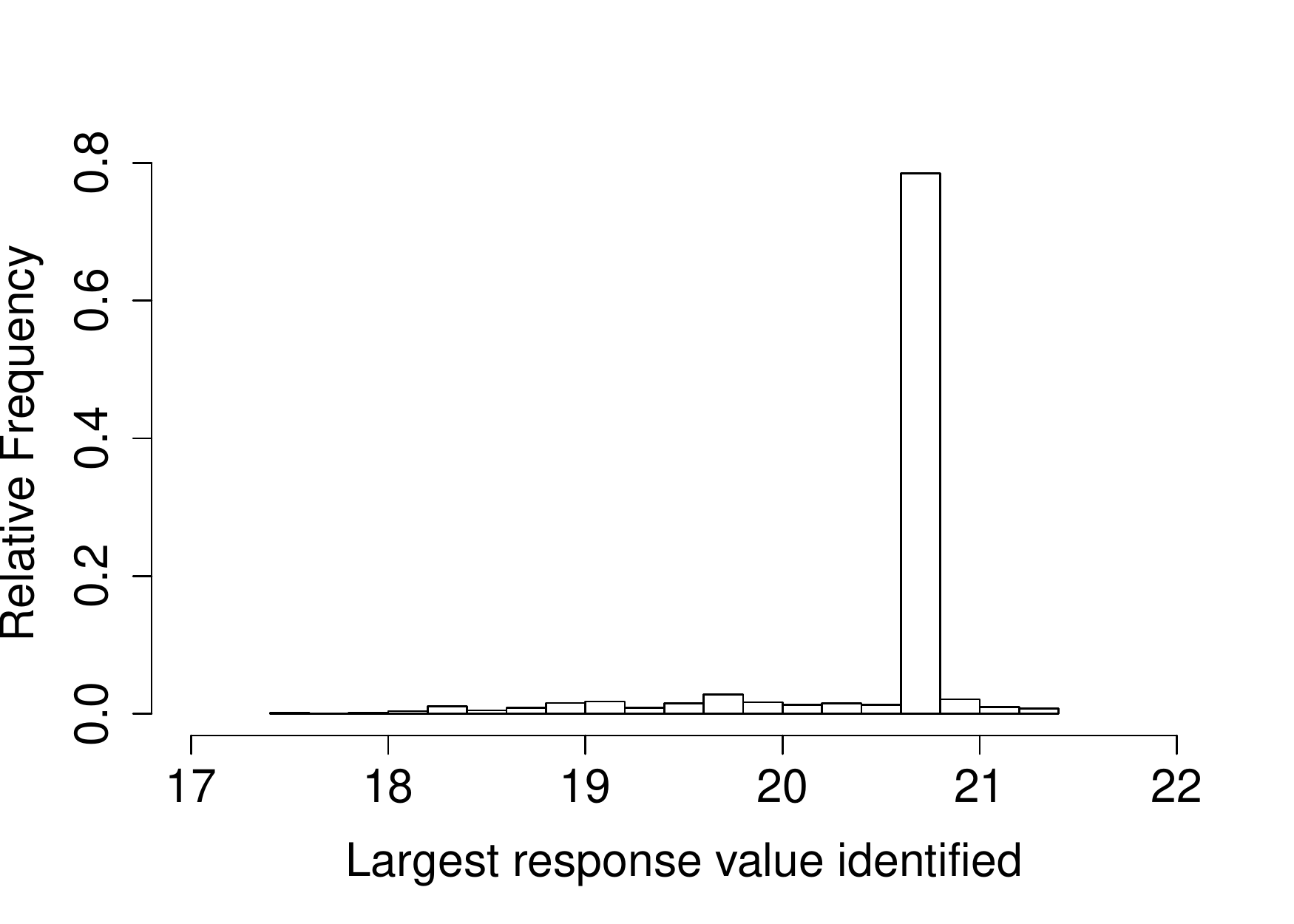}}\quad
	\subfigure [] {\includegraphics[scale=0.4]{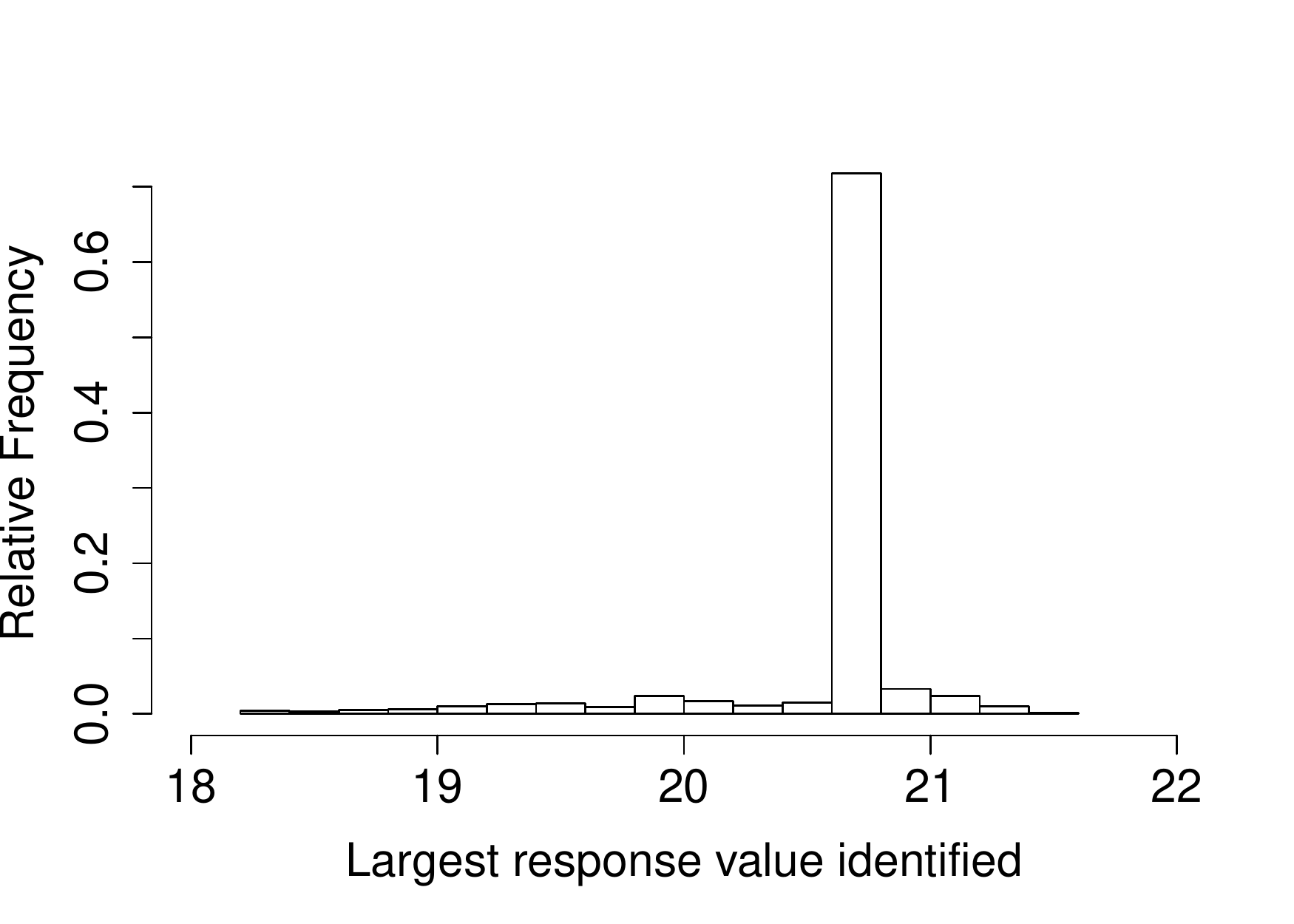}}
	\caption{Histograms of largest responses identified by (a) $BM_2$ and (b) $BM_3$ in Example~\ref{ex:SMS:BO2}.}
		\label{fig:smsqs}
\end{figure}

To make a comparison, we first consider a random sampling approach ($BM_1$).
Here, we draw a large random sample consisting of 72,000 observations using a random design whose quantitative part is a random Latin hypercube and sequence part includes a hundred replicates of all possible sequences. We show the histogram of these 72,000 observations in Figure~\ref{fig:hist}, and the maximum response found is 21.53 which is worse than that by the QS-learning using only around 70 runs.
In addition, we consider the sequential generalized PWO ($BM_2$) and CP ($BM_3$) approaches starting from randomly selected 22 and 32 runs, respectively.
We replicate the $BM_{2}$ and $BM_{3}$ methods 1000 times. In Figure~\ref{fig:smsqs}, we show the histograms of largest response values identified  by $BM_2$ and $BM_3$.
Their average results are 20.57 and 20.66, and their best results are 21.36 and 21.40, respectively. Their performances are clearly inferior.

\bibliographystyle{asa}
\bibliography{magp}

\end{document}